%% file: yarntopology.tex
\documentclass[acmtog]{acmart}
\acmSubmissionID{243}

\usepackage{booktabs} 

\usepackage{tabularx}
\usepackage[ruled]{algorithm2e} 

\SetAlFnt{\small}
\SetAlCapFnt{\small}
\SetAlCapNameFnt{\small}
\SetAlCapHSkip{0pt}

\usepackage{bm}
\usepackage{xcolor}
\usepackage{multirow,multicol,tabularx}
\usepackage{nicefrac}

\newcommand{\R}{\mathbb{R}}    
\newcommand{\rv}{\mathbf{r}}   
\newcommand{\atan}{\text{atan}}
\newcommand{\eq}{\!=\!}

\renewcommand{\vec}[1]{\mathbf{#1}}
\newcommand{\grad}{\bm{\nabla}}

\newcommand{\revision}{}
\newcommand{\revisionp}{}
\SetCommentSty{graycomments}

\acmJournal{TOG}
\acmYear{2021}
\acmVolume{40}
\acmNumber{4}
\acmArticle{106}
\acmMonth{8}

\setcopyright{acmcopyright}
\acmDOI{10.1145/3450626.3459778}

\begin{document}
\title{Fast Linking Numbers for Topology Verification of Loopy Structures}

\author{Ante Qu}
\orcid{0000-0002-1635-0721}
\email{antequ@cs.stanford.edu}
\author{Doug L. James}
\email{djames@cs.stanford.edu}
\affiliation{%
  \department{Computer Science}
  \institution{Stanford University}
  \streetaddress{353 Jane Stanford Way}
  \city{Stanford}
  \state{CA}
  \postcode{94305}
  \country{USA}
}

\include{abstract}

%
%


%
%


\begin{teaserfigure}
  \includegraphics[height=0.42\hsize]{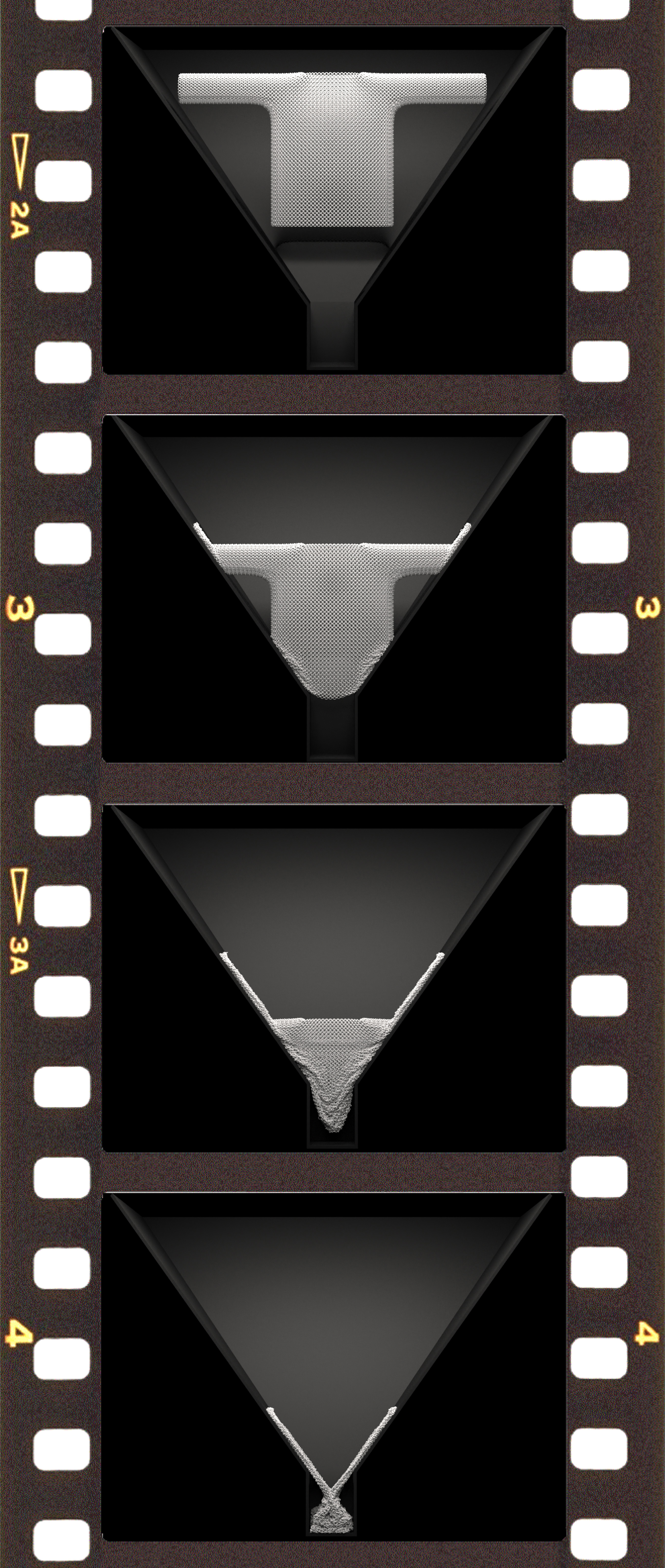}
  \hspace{-1.6mm}
  \includegraphics[height=0.42\hsize]{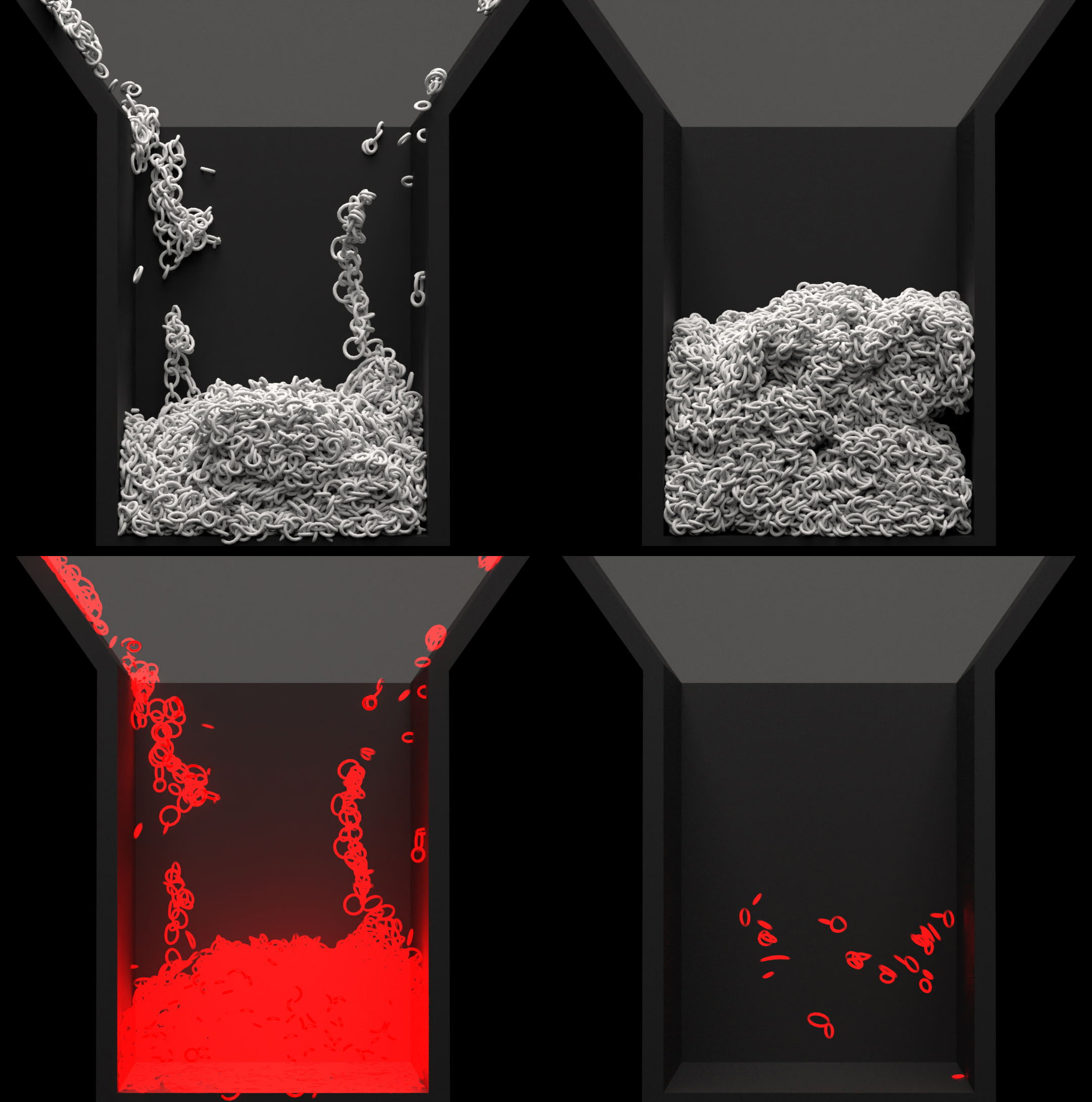} 
  \hspace{-1.6mm}  
  \includegraphics[height=0.42\hsize]{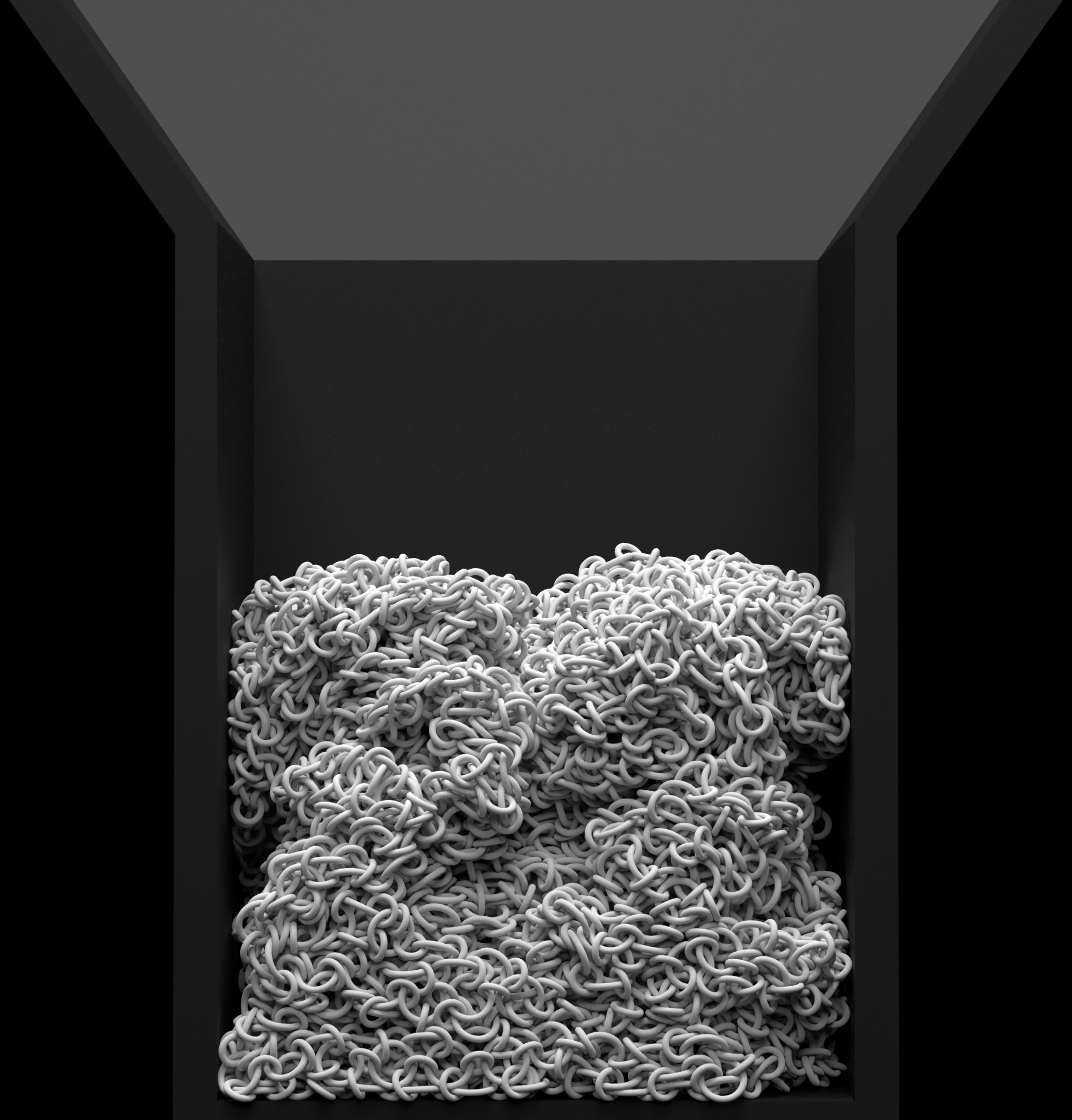}
  \\
  \begin{tabular}{cccc}
    \hspace{4mm} (a) Animation &
    \hspace{11mm} (b) $\nicefrac{1}{\Delta t} = 480 \, Hz$ &
    \hspace{11mm} (c) $\nicefrac{1}{\Delta t} = 1200 \, Hz$ &
    \hspace{28mm} (d) $\nicefrac{1}{\Delta t} = 2400 \, Hz$
  \end{tabular}
\caption{ \textbf{Topologically verified rigid-body simulation of fitted Kusari chainmail} (a) is used to pack the garment of 14112 curved rings with 18752 links into a box for 3D printing. Our fast linking number methods can detect pull-through between previously linked curved loops that result from large timesteps: (b) widespread failure when timestepped at $f=480\,Hz$ with \revision{4048} destroyed links shown in red; (c) timestepping at $f=1200\,Hz$ is much better and looks visually fine, but still has 16 failed links hiding inside the pile; whereas (d) using $f=2400\,Hz$ we can verify that there are no violated links (or spurious new ones) in the final result. Rapid topology verification also allows corrupted simulations to be detected and aborted early.}
\label{fig:teaser}
\end{teaserfigure}

\maketitle

\input{introduction}
\input{background}

\input{method}

\input{results}
\input{conclusions}
\input{acknowledgments}

\bibliographystyle{ACM-Reference-Format}
\bibliography{yarntopology-bibliography}

\input{appendix}

\end{document}

%% file: abstract.tex
\begin{abstract}

It is increasingly common to model, simulate, and process complex materials based on loopy structures, such as in yarn-level cloth garments, which possess topological constraints between inter-looping curves. While the input model may satisfy specific topological linkages between pairs of closed loops, subsequent processing may violate those topological conditions. In this paper, we explore a family of methods for efficiently computing and verifying linking numbers between closed curves, and apply these to applications in geometry processing, animation, and simulation, so as to verify that topological invariants are preserved during and after processing of the input models. Our method has three stages: (1) we identify potentially interacting loop--loop pairs, then (2) carefully discretize each loop's spline curves into line segments so as to enable (3) efficient linking number evaluation using accelerated kernels based on either counting projected segment--segment crossings, or by evaluating the Gauss linking integral using direct or fast summation methods (Barnes--Hut or fast multipole methods). We evaluate CPU and GPU implementations of these methods on a suite of test problems, including yarn-level cloth and chainmail, that involve significant processing: physics-based relaxation and animation, user-modeled deformations, curve compression and reparameterization. We show that topology errors can be efficiently identified to enable more robust processing of loopy structures.
  
\end{abstract}

%% file: introduction.tex
\section{Introduction}

It is increasingly common to model, simulate, and process complex materials based on loopy structures, such as in knitted yarn-level cloth garments, which possess topological constraints between inter-looping curves. In contrast to a solid or continuum model, the integrity of a loopy material depends on the preservation of its loop--loop topology. \revisionp{A failure to preserve these links, or an illegal creation of new links between unlinked loops, can result in an incorrect representation of the loopy material.}

Unfortunately, common processing operations can easily destroy the topological structure of loopy materials. For example, time-stepping dynamics with large time steps or contact solver errors that cause pull-through events, deformation processing that allows loops to separate or interpenetrate, or even applying compression and reparameterization on curves, can all cause topology errors. Furthermore, once the topology of a loopy material has been ruined, the results can be disastrous (an unraveling garment) or misleading (\revisionp{incorrect} yarn-level pattern design).

\begin{figure}[!htb]
  \centerline{
    \begin{tabular*}{0.5\textwidth}{cccc}
      \includegraphics[height=10.5mm]{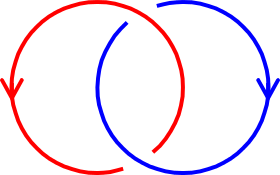} &
      \includegraphics[height=10.5mm]{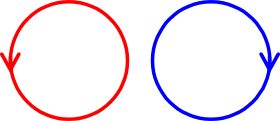} &  
      \includegraphics[height=10.5mm]{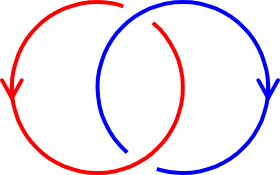} & 
      \includegraphics[height=10.5mm]{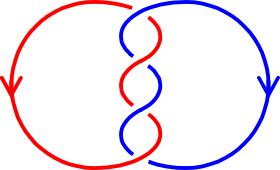} \\
      -1 & 0 & +1 & +2
    \end{tabular*}
  }
  \caption{{\bf Linking numbers} count the oriented linkages between two closed curves 
  (examples from \url{https://en.wikipedia.org/wiki/Linking_number}).
  \label{fig:whatIsALinkingNumber}}
\end{figure}

In this paper, we explore a family of methods for efficiently computing and verifying {\em linking numbers} between closed curves. Intuitively, if two loops are unlinked (i.e., they can be pulled apart), the linking number is zero; otherwise, the linking number is a nonzero signed integer corresponding to how many times they loop through one another, with a sign to disambiguate the looping direction (see Figure~\ref{fig:whatIsALinkingNumber}).  In mathematical terms, the linking number is a homotopical invariant that describes the ``linkage'' of two oriented closed curves in 3D space. While a closed curve can be either a ``knot'' or a simple ``loop,'' we will use the term ``loop'' to refer to all closed curves. Given two disjoint, oriented loops $\gamma_1$ and $\gamma_2$ where each maps $S^1 \to \R^3$,
the linking number is an integer-valued function, $\lambda(\gamma_1,\gamma_2)$, that counts the number of times each curve winds around the other \cite{rolfsen1976knots}. Curve deformations that result in one part of a curve crossing the other loop at a single point must change the linking number by $\pm 1$. By evaluating (or knowing) the linking numbers before a computation, then evaluating them afterwards, we can detect certain topological changes. \revision{A linking number checksum can therefore be useful as a sanity check for topology preservation---while we cannot rule out intermediate topology changes when two states have the same linking number, different linking numbers imply a topology change.} We efficiently and systematically explore this idea and apply it to several applications of loopy materials in computer graphics where it is useful to verify the preservation of topological invariants during and after processing (see Figure~\ref{fig:teaser} for a preview of our results).

Our method has three stages: (1) to avoid computing linking numbers between obviously separated loops, we first perform a pair search to identify potentially linked loops; (2) we discretize each loop's spline curves into line segments for efficient processing, being careful to ensure that the discretized \revisionp{collection of loops remain topologically equivalent to the input}; then (3) we evaluate all loop--loop linking numbers (and record the results in a sparse matrix) using one of several accelerated linking-number kernels: either (i) by counting projected segment--segment crossings, or (ii) by evaluating the Gauss linking integral using direct or fast summation methods (both Barnes--Hut and fast multipole methods).  An overview of our method is shown in Figure~\ref{fig:overview}.  We evaluate CPU and GPU implementations of these methods on a suite of test problems, including yarn-level cloth and chainmail examples, that involve significant processing: physics-based relaxation and animation, user-modeled deformations, curve compression and reparameterization. \revision{\revisionp{From our evaluation we conclude that counting crossings and the Barnes--Hut method are both efficient for computing the linking matrix.}} In addition to loopy structures, we also show an example where our method is part of an automatic procedure to topologically verify braids, which are structures with open curves that all attach to two rigid ends; in particular, we verify the relaxation of stitch patterns by attaching the ends of stitch rows to rigid blocks.  We show that \revision{many} topology errors can be efficiently identified to enable more robust processing of loopy structures.

%% file: background.tex
\section{Background}
\label{sec:background}

\subsection{Computing the Linking Number}
\label{sec:bgComputingLN}

There are several equivalent ways to calculate (and define) the linking number, all of which will be explored in this paper. The linking number, $\lambda(\gamma_1, \gamma_2)$, is a numerical invariant of two oriented curves, $\gamma_1, \gamma_2$. It intuitively counts the number of times one curve winds around the other curve (see Figure~\ref{fig:whatIsALinkingNumber}). The linking number is invariant to \revision{\emph{link homotopy}, a deformation in which curves can pass through themselves, but not through other curves \cite{milnor1954link}}.

\paragraph{Counting Crossings} One way to compute the linking number is to count crossings in a \emph{link diagram}. Suppose our curves are discretized into polylines. The link diagram is a 2D \emph{regular projection} of this set of 3D curves that also stores the above--belowness at every intersection; that is, if the projection plane is the XY plane, then the Z coordinate of each curve indicates which curve is above or below. A projection is deemed \emph{regular} if no set of 3 or more points, or 2 or more vertices, coincide at the same projected point \cite{rolfsen1976knots}.

\begin{figure}[!htb]
  \centering
   \includegraphics[width=0.35\textwidth]{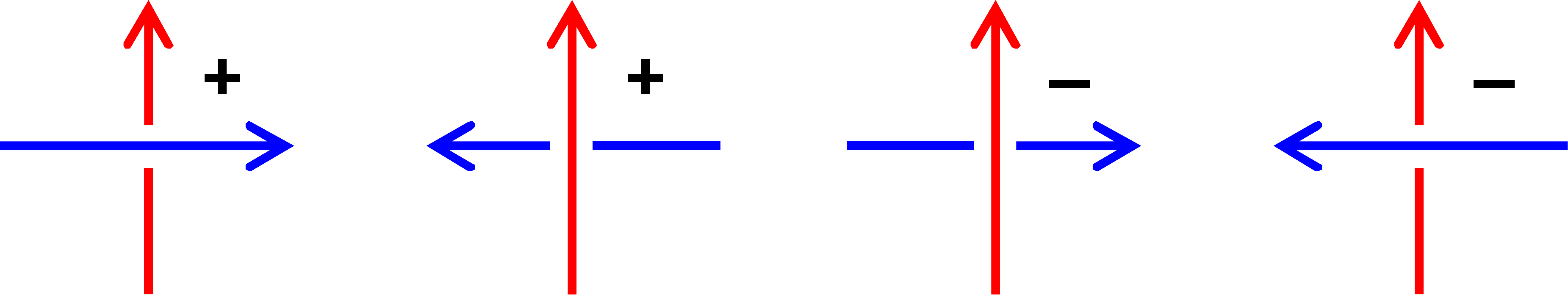}
  \caption{{\bf Crossing orientation} determines whether to increment or decrement the linking number.
  (From \url{https://en.wikipedia.org/wiki/Linking_number}).
  \label{fig:crossingOrientation}}
\end{figure}

To compute the linking number, start with $\lambda(\gamma_1, \gamma_2)=0$, and at every detected intersection, compare the orientation of the curves with Figure~\ref{fig:crossingOrientation}, and increment the linking number by $+\nicefrac{1}{2}$ if it is positive, or $-\nicefrac{1}{2}$ if it is negative.

\paragraph{Gauss's Integral Form} Another mathematically equivalent method to compute the linking number, which doesn't require a projection plane or geometric overlap tests, is to compute the linking integral $\lambda$, first introduced by Gauss \cite{rolfsen1976knots}:
\begin{align}
    \label{eq:gaussintegral}
    \lambda(\gamma_1, \gamma_2) &= \frac{1}{4 \pi} \int_{\gamma_1} \int_{\gamma_2} \frac{\rv_1 - \rv_2}{|\rv_1 - \rv_2|^3} \cdot (\text{d}\rv_1 \times \text{d}\rv_2).
\end{align}
This integral form can be interpreted as an application of Ampere's law from magnetostatics for a current loop and a test loop\footnote{{\em Magnetostatics Interpretation:} If we imagine one loop has current flowing through it, the Gauss linking integral essentially applies Ampere's law on the other loop, using the magnetic field generated by the current loop (given to us by the Biot--Savart law), to compute the total current enclosed. Similar to the winding number computation in \cite{barill2018fast}, we note that the first term of the integrand is the gradient of the Laplace Green's function $G(\rv_1, \rv_2) = -1/(4 \pi | \rv_1 - \rv_2 |)$; in magnetostatics, the vector potential $\vec{A}$ of a monopole current source at $\rv_1$ with current $I$ and length vector $\vec{s}_1$ is $\mu_0 I \vec{s}_1 G(\rv_1, \rv_2)$, and the magnetic field at $\rv_2$ due to $\rv_1$ is $\vec{B} = \nabla_2 \times \vec{A} = -\mu_0 I \vec{s}_1 \times \nabla_2 G(\rv_1, \rv_2)$. Applying Ampere's law gives us the linking integral. This interpretation will later allow us to apply computational tools for multipole expansions.}.

\paragraph{Direct Summation Methods}
If the loops $\gamma_1, \gamma_2$ are discretized into line segments indexed by $j$ and $i$, and we denote the midpoints of $\gamma_1$ by $\vec{r}_j$ and $\gamma_2$ by $\vec{r}_i$, and the line segment length vectors by $\vec{s}_j$ and $\vec{s}_i$ respectively, then taking a midpoint quadrature, we have
\begin{align}
    \label{eq:midpointapproximation}
    \lambda(\gamma_1, \gamma_2) &\approx \frac{1}{4 \pi} \sum_{j, i} \frac{\vec{r}_j - \vec{r}_i}{|\vec{r}_j - \vec{r}_i|^3} \cdot (\vec{s}_j \times \vec{s}_i).
\end{align}
This midpoint approximation is only accurate when the segments are short compared to the distance between each pair. Fortunately, an exact expression \revisionp{for polyline loops} is known of the form
\begin{align}
    \label{eq:exactDEBackgrnd}
    \lambda(\gamma_1, \gamma_2) &= \sum_{j, i} \lambda_{ji}
\end{align}
where $\lambda_{ji}$ is the contribution from a pair of line segments \cite{berger2009topological, arai2013rigorous}.  If we denote the \revisionp{vertices (endpoints of the segments)} of each loop by $\vec{l}_j, \vec{k}_i$, the contribution from the segment pair $(j, i)$ is given in \cite{arai2013rigorous} (which uses the signed solid angle formula \cite{van1983solid}):
\begin{align}
\lambda_{ji} &= \frac{1}{2 \pi} \left(\atan\left( \frac{\vec{a} \cdot (\vec{b} \times \vec{c}) }{|\vec{a}| |\vec{b}| |\vec{c}| + (\vec{a} \cdot \vec{b}) |\vec{c}| + (\vec{c} \cdot \vec{a}) |\vec{b}|+(\vec{b} \cdot \vec{c}) |\vec{a}|}\right) \right. \nonumber \\
\label{eq:directarctan}
& \left. + \; \atan\left( \frac{\vec{c} \cdot (\vec{d} \times \vec{a}) }{|\vec{c}| |\vec{d}| |\vec{a}| + (\vec{c} \cdot \vec{d}) |\vec{a}| + (\vec{a} \cdot \vec{c}) |\vec{d}|+(\vec{d} \cdot \vec{a}) |\vec{c}|}\right)  \right),
\end{align}
where
\begin{align}
    \vec{a} &= \vec{l}_j - \vec{k}_i,  \;\; \vec{b} = \vec{l}_j - \vec{k}_{i+1}, \;\;  \vec{c} = \vec{l}_{j+1} - \vec{k}_{i+1},\nonumber \\
    \vec{d} &= \vec{l}_{j+1} - \vec{k}_i, \;\; \atan (y/x) = \text{atan2}(y, x). \nonumber
\end{align}
This computation requires two trigonometric operations per segment pair. More recently, \citet{bertolazzi2019efficient} used the $\arctan$ addition formula to eliminate all trigonometric operations\revisionp{.}

\paragraph{Fast Summation Methods} If we discretize the loops into $N$ samples, Gauss's double integral can be approximated using fast summation methods, such as the Barnes--Hut algorithm which uses a tree to compute $N$-body Gaussian summations in $O(N \log N)$ time. Alternately one can use the Fast Multipole Method (FMM); some FMM libraries provide off-the-shelf implementations of the Biot--Savart integral, which is the inner integral of the linking integral.

\subsection{Related Work}

Many works in the past have tackled the topic of robust collision processing. Epsilon geometry and robust predicates \revision{\cite{salesin1989epsilon, shewchuk97adaptive}} have been used to precisely answer \revision{``inside--outside'' questions, such as where a point lies with respect to a flat face (``vertex--face''),} or a straight line with respect to another (``edge--edge''). These determinant-based geometric tests (such as \cite{feito1997inclusion}) are also useful for detecting inversions in finite elements and preventing them from collapsing \cite{irving2004invertible, selle2008amass}. Some papers \cite{volino2006resolving, zhang2007generalized, selle2008amass, harmon2011interference, kim2019anisotropic} propose methods to repair interpenetrations between objects by minimizing a computable quantity, such as an intersection contour, a space--time intersection volume, an altitude spring, or another form of energy. For models with a large number of elements, a few works use hierarchical bounding volumes to compute energy-based collision verification certificates \cite{barbic2010subspace, zheng2012energy}, or to validate geometry in hair \cite{kaufman2014adaptive}, holey shapes \cite{bernstein2013putting}, or cloth \cite{baraff2003untangling}.

Many of the above methods either perform a check or repair a violation by answering \revision{discrete collision-detection} questions about a point vs.{} a flat face or a straight edge vs.{} another straight edge. \revision{To generalize these questions to arbitrary surfaces (and soups and clouds),} a few papers \cite{jacobson2013robust, barill2018fast} compute generalized winding numbers to answer the inside--outside question. The generalized winding number can be expressed as a Gaussian integral, and \citet{barill2018fast} uses Barnes--Hut style algorithms to speed up the computation.

\revision{Continuous collision detection (CCD) methods \cite{basch1999kinetic,bridson2002robust,harmon2011interference,brochu2012efficient,wang2021review} can
verify that a deformation trajectory is collision free. Such methods could be used to detect curve--curve crossings and thus topology changes.
In contrast, our approach of computing a topology verification checksum is a discrete before/after test: we verify the topology at a fixed state rather than the geometry along the entire deformation path. While our method cannot guarantee a deformation is collision free, it works even if the deformation path is ambiguous (e.g., in verifying model reparameterizations, which we will show in \S\ref{sec:results_reparam}), or, when the path is known, it can be performed as sparsely or frequently as the user requires.}

For simulations and deformations that are more suitable for 1D rod elements rather than volumetric elements, it can be difficult to apply \revision{discrete} inside--outside checks. For example, in yarn-level simulations of cloth, yarns are often modeled as cylindrical spline or rod segments \cite{kaldor2008simulating, kaldor2010efficient, leaf2018interactive, wu2020weavecraft}. While the inside--outside question \revisionp{applied to the cylindrical volumes} can determine interpenetration when detected, oftentimes an illegal or large simulation step can cause a ``pull-through,'' where the end state has no interpenetration but rather a topological change in the curve configuration. Pull-throughs can cause changes to the appearance and texture of the resulting relaxed pattern, yet remain undetected unless spotted by a human eye. Many of these simulations \cite{yuksel2012stitch, leaf2018interactive} prevent pull-throughs by using large penalty contact forces and limiting step sizes using a known guaranteed displacement limit. Some stitch meshing papers \cite{narayanan2018automatic, wu2018stitch, narayanan2019visualknit, wu2019knittable, guo2020representing} avoid closed loops as a result of making the model knittable or machine knittable. In contrast, many of the models from \cite{yuksel2012stitch} consist of yarns that form closed loops. We propose a method to automatically detect pull-throughs in closed-loop yarn models after they occur, so that the simulation can take more aggressive steps and backtrack or exit when the violation is detected. Other modeling applications in graphics that are more suitable for 1D elements include chains and chainmail \cite{bergou2008discrete, tasora2016chrono, mazhar2016ontheuse}, threads \cite{bergou2010discrete}, \revision{knots \cite{harmon2011interference,spillmann2008adaptive,bergou2008discrete,yu2020repulsive}}, hair \cite{chang2002practical,selle2008amass,kaufman2014adaptive}, and necklaces \cite{guibas2002collision}.

To detect pull-through violations of edge--edge contacts, one can locally verify the sign of the volume of the tetrahedron (computed as a determinant) spanned by the two edges. To generalize this local notion to arbitrary closed curves, one can compute the linking number between the two curves. Interestingly, Gauss's integral form of the linking number uses, in the integrand, this exact determinant, applied on differential vectors, but scaled so that it is the \revisionp{degree, or signed area of the image,} of the Gauss map \revisionp{of the link}.

In graphics, some works \cite{dey2009computing, dey2013efficient} compute the linking number, by projecting and computing crossings, to determine if a loop on an object's surface is a handle versus a tunnel, which is useful for topological repair and surface reparameterization. Another work \cite{edelsbrunner2001computing} computes the linking numbers of simplicial complexes within DNA filtrations, by using Seifert surface crossings (which are generated from the filtration process), to detect nontrivial tangling in biomolecules. In the computational DNA topology community, many papers \cite{fuller1978decomposition, klenin2000computation, clauvelin2012characterization, krajina2018active, sierzega2020wasp} use topological invariants such as the linking number, twist, and writhe to understand DNA ribbons; in particular, \citet{berger2009topological} summarizes how to compute these quantities, and \cite{arai2013rigorous, bertolazzi2019efficient} present more robust exact expressions for line segment curves using the signed solid angle formula \cite{van1983solid}. \citet{Peters2006ComputationalTO} also discusses how to subdivide spline curves for computational topology. A few works in animation and robotics \cite{ho2009character, ho2010controlling, zarubin2012hierarchical, pokorny2013grasping} also use the linking integral as a topological cost in policy optimization, e.g., for grasping. However, none of these works propose using acceleration structures, such as Barnes--Hut trees or the Fast Multipole Method \cite{greengard1987fast}, to compute the linking number between curves, mainly because the input size is sufficiently small or the problem has a special structure (e.g., simplicial complexes that come with surface crossings, or ribbons where the twist is easy to compute) that allows for a cheaper method. We propose using acceleration structures to speed up linking-number computation for general closed curves.

The linking integrand uses the Green's function for the Laplace problem in a Biot--Savart integral, and there are numerous Fast Multipole libraries for the Laplace problem, including FMM3D \cite{cheng1999fast} and FMMTL \cite{cecka2015fmmtl}. FMMTL provides an implementation of the Biot--Savart integral and has recently been used to simulate ferrofluids in graphics \cite{huang2019accurate}. Another library \cite{burtscher2011efficient} implements a very efficient, parallel Barnes--Hut algorithm for the N-Body Laplace problem on the GPU.

%% file: method.tex
\begin{figure*}[!htb]
    \centering
    \includegraphics[width=\textwidth]{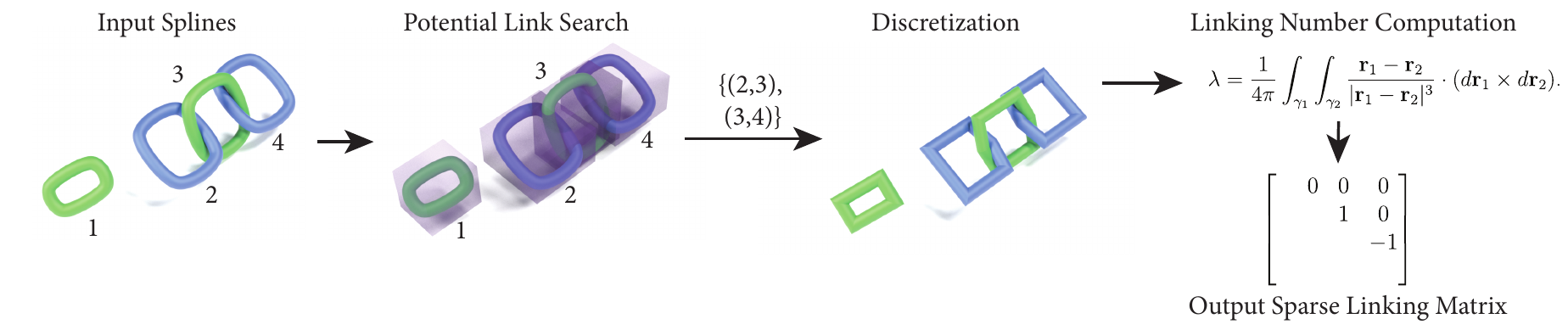}
    \caption{{\bf Method Overview:} Our method inputs a set of spline curves and outputs a sparse triangular linking number matrix as a topology invariant. First, (\S\ref{sec:methodspls}) we perform a potential link search to get a list of potentially linked loop pairs. Second, (\S\ref{sec:methodsdiscretization}) we discretize the input into a set of homotopically equivalent polylines. Third, (\S\ref{sec:methodsintegration}) we compute the linking number for each potentially linked pair, and output the sparse linking matrix. }
    \label{fig:overview}
\end{figure*}

\section{Fast Linking Number Methods}
\label{sec:method}

In this section, we describe several ways to compute linking numbers (based on Algorithm~\ref{alg:method_overview}) which all take an input model and generate the sparse upper-triangular linking matrix as a topology-invariant certificate for verification.

\begin{algorithm}[]
\caption{Method ComputeLinkingNumbers}
\label{alg:method_overview}
\SetAlgoLined
\SetKwInOut{Input}{Input}
\SetKwInOut{Output}{Output}
\Input{$\Lambda = \{\gamma_i \}$, a list of looped curves.}
\Output{$M$, a sparse triangular linking matrix of integers.}
\BlankLine
\SetKwFunction{ComputeLinkingNumbers}{ComputeLinkingNumbers}
\SetKwFunction{PotentialLinkSearch}{PotentialLinkSearch}
\SetKwFunction{Discretize}{Discretize}
\SetKwFunction{ComputeLink}{ComputeLink}
\SetKwProg{Fn}{Function}{:}{}
\Fn{\ComputeLinkingNumbers{$\Lambda$}}{
$M \leftarrow 0$ \;
$P \leftarrow$ \PotentialLinkSearch{$\Lambda$}\;
$\{\Gamma_i\} \leftarrow $ \Discretize{$\Lambda, P$}\;
\ForEach{$(i, j) \in P$}{
  $M_{ij} \leftarrow $\ComputeLink{$\Gamma_i, \Gamma_j$}\;
}
\textbf{return} $M$\;
}
\end{algorithm}

\subsection{Method Overview}
\label{sec:methodOverview}

The core method takes as input a model of closed loops defined either by line-segment endpoints or polynomial spline control points, and outputs a certificate consisting of a sparse triangular matrix of pairwise linking numbers between loops. Our goal is to robustly handle a wide range of inputs, such as those with a few very large intertwined loops, e.g., DNA, or many small loops, e.g., chainmail.

The main method ``ComputeLinkingNumbers'' is split into three stages (which are summarized in Figure~\ref{fig:overview} and described in pseudocode in Algorithm~\ref{alg:method_overview}):
\begin{enumerate}
\item Potential Link Search (PLS) (\S\ref{sec:methodspls}): This stage exploits the fact that any pair of loops with disjoint bounding volumes have no linkage. It takes a sequence of loops and produces a list of potentially linked loop pairs by using a bounding volume hierarchy (BVH).
\item Discretization (\S\ref{sec:methodsdiscretization}): This stage discretizes the input curves, if they are not in line segment form, into a sequence of line segments for each loop, taking care to ensure that the process is \revisionp{link} homotopic (i.e., if we continuously deform the original curves into the final line segments, no curves cross each other) and thus preserves linking numbers.
\item \revisionp{Linking Number Computation} (\S\ref{sec:methodsintegration}): This stage computes the linking matrix for the potentially linked loops, and can be performed with many different linking-number methods.
\end{enumerate}

\paragraph{Fast Verification} Verification of the linking matrix can be done exhaustively by computing and comparing the full linking matrix computed using Algorithm~\ref{alg:method_overview}. However, in some applications it is sufficient to know that any linkage failed, e.g., so that a simulation can be restarted with higher accuracy. In such cases, an ``early exit'' strategy can be used to quickly report topological failure \revision{by exiting when any linking number differs from the input}.

\paragraph{\revision{\revisionp{Bounding Volumes}}} \revision{\revisionp{All bounding volumes in our method are convex. The convexity provides guarantees in \S\ref{sec:methodspls} and \S\ref{sec:methodsdiscretization}.}}

\subsection{Potential Link Search (PLS)}
\label{sec:methodspls}

When the number of loops, $L$, is large, computing the linking integral between all loop pairs can be expensive, as there are $\left(L \atop 2\right) = L(L-1)/2$ pairs. For this stage of the method, we input a list of loops $\{ \gamma_i \}$, and output a list of potentially linked loop pairs, $P=\{ (i, j) \}$ (where $i<j$ are loop indices), which gets passed to the next stages (\S\ref{sec:methodsdiscretization} and \S\ref{sec:methodsintegration}). All other loop pairs must have $0$ linkage.

We exploit the fact that if two loops have a nonzero linkage, then their bounding volumes, \revisionp{which are convex,} must overlap. Therefore, we precompute an \revision{axis-aligned bounding box (AABB)} for each loop, and build an AABB-tree of loops. \citet{edelsbrunner2001computing} also use bounding boxes to cull the list of potentially linked loops. For any two loops with overlapping bounding volumes, we add the pair to a list of potentially linked loops. See Algorithm~\ref{alg:pls} for pseudocode, and Figure~\ref{fig:overview} for an illustration.

\begin{algorithm}[]
\caption{Potential Link Search}
\label{alg:pls}
\SetAlgoLined
\SetKwFunction{ComputeBoundingBoxOfLoop}{ComputeBoundingBoxOfLoop}
\SetKwFunction{BuildBVH}{BuildBVH}
\SetKwFunction{GetIntersectingBoxes}{GetIntersectingBoxes}
\SetKwInOut{Input}{Input}
\SetKwInOut{Output}{Output}

\Input{$\Lambda = \{\gamma_i \}$, a list of looped curves.}
\Output{$P = \{(i_k,j_k)\}$, a list of pairs of loop indices that potentially link.}
\BlankLine
\tcp{This function returns a list of pairs of loops that have overlapping bounding boxes.}
\SetKwFunction{PotentialLinkSearch}{PotentialLinkSearch}
\SetKwProg{Fn}{Function}{:}{}
\Fn{\PotentialLinkSearch{$\Lambda$}}{
Boxes $\leftarrow \{ \}$\;
\For{$i \leftarrow 0$ \KwTo $|\Lambda| - 1$}{
    Boxes[$i$] $\leftarrow$ \ComputeBoundingBoxOfLoop{$\gamma_i$}\;
}
Tree $\leftarrow$ \BuildBVH{\normalfont{Boxes}}\;
$I \leftarrow$ \GetIntersectingBoxes{\normalfont{Tree}, \normalfont{Tree}}\;
$P \leftarrow \{ (i,j) : (i,j) \in I \, \wedge \, (i < j) \}$\;
\textbf{return} $P$ \;
}
\end{algorithm}

\subsection{Discretization}
\label{sec:methodsdiscretization}

\begin{figure}[]
    \centering
    \includegraphics[]{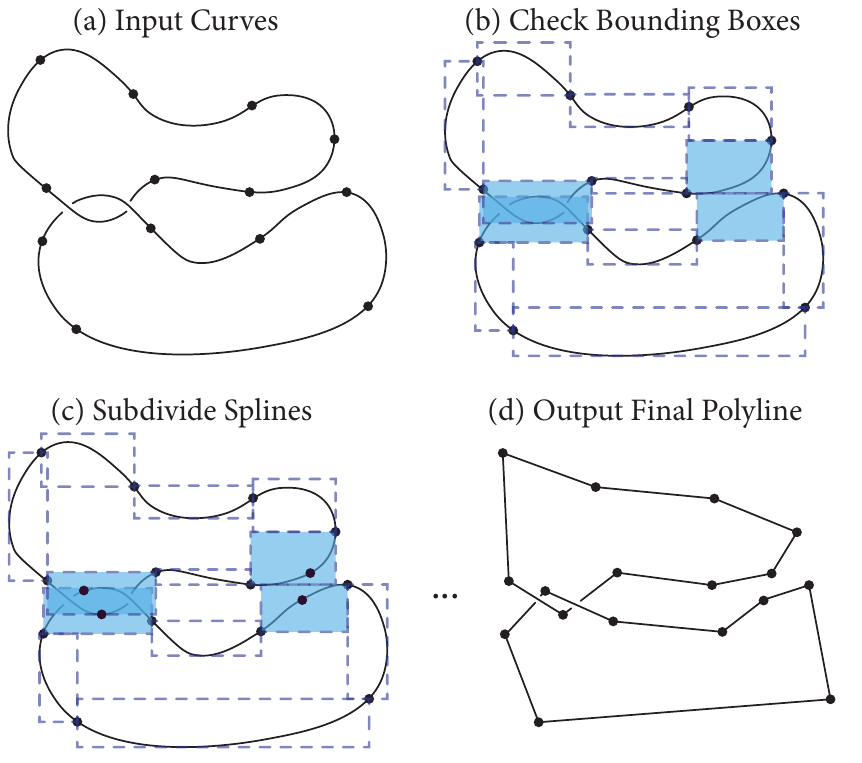}
    \caption{{\bf Discretization Illustration:} (a) Input curves with spline knots illustrated. (b) Given this input, we compute bounding boxes (in 3D) and find splines with overlapping boxes. (c) We refine each spline with an overlapping box by subdividing it into two splines, and we repeat steps (b) and (c) until no boxes overlap. (d) The result of discretization is a set of polylines homotopically equivalent to the input.}
    \label{fig:discretizeillustration}
\end{figure}

Given a set of input loops discretized into polynomial splines or line segments and a set of overlapping loop pairs, our discretization step computes a set of corresponding loops discretized into homotopically equivalent line segments, so that we can use the method in the next stage (Section~\ref{sec:methodsintegration}) to compute the exact linkage. See Figure~\ref{fig:discretizeillustration} for an illustration. While there are many ways to refine spline segments into line segments \cite{de1959outillages, MICCHELLI1989uniform}, many applications simply refine them to a fixed tolerance. Our method is spatially adaptive and refines the spline further only when curves from different loops pass closely by each other. The discretization routine takes advantage of the fact that if two convex volumes don't overlap, then two curves bounded by these volumes can be deformed into two line segments within these volumes without crossing each other. Example pseudocode is provided in Algorithm~\ref{alg:discretization}, and the routine is as follows.

\begin{algorithm}[]
\caption{Discretization}
\label{alg:discretization}
\SetAlgoLined
\SetKwFunction{BuildBVH}{BuildBVH}
\SetKwFunction{GetIntersectingBoxes}{GetIntersectingBoxes}
\SetKwFunction{DiameterOfBBox}{DiameterOfBBox}
\SetKwInOut{Input}{Input}
\SetKwInOut{Output}{Output}

\Input{$\Lambda = \{\gamma_i \}$, a list of looped curves, each of which can be a cubic spline or a polyline; \\
       $P = \{(i_k,j_k)\}$, a list of pairs of loops that potentially link.}
\Output{$\{\Gamma_i\}$, a list of polylines corresponding to the curve list.}
\BlankLine
\tcp{This function homotopically refines input curves into polylines.}
\SetKwFunction{Discretize}{Discretize}
\SetKwProg{Fn}{Function}{:}{}
\Fn{\Discretize{$\Lambda$, $P$}}{
L $\leftarrow |\Lambda|$\;
UnprocessedSegments $\leftarrow \{\{ \} _i\}$; \tcp{L sets of segments.}
$\xi \leftarrow$ the average coordinate magnitude\;
$\epsilon \leftarrow$ machine epsilon\;
\tcp{Ensure all segments have finite, nonzero length.}
\textbf{parallel }\For{$i \leftarrow 0$ \KwTo $L - 1$}{
    $\Gamma_i \leftarrow \{ \}$; UnprocessedSegments[$i$] $\leftarrow \{ \}$\;
    \ForEach{\normalfont Curve segment $s \in \gamma_i$}{
        $s$.ComputeBoundingBox()\;
        \If{\normalfont (\DiameterOfBBox{$s$} $< \epsilon \xi$)}{
          \textbf{return} Error(``Input has zero-length segments.'')\;
        }
        UnprocessedSegments[$i$].InsertSegment($s$)\;
    }
}

Trees $\leftarrow \{ \}$; \tcp{One tree for each curve.}
\While{\normalfont(At least one UnprocessedSegments[$i$] is not empty)}{
\textbf{parallel }\For{$i \leftarrow 0$ \KwTo $L - 1$}{
    \tcp{(\textsc{i}) Build an AABB-tree for each curve.}
    Trees[$i$]$\leftarrow$\BuildBVH{\normalfont UnprocessedSegments[$i$]}\;
}
\textbf{parallel }\For{$i \leftarrow 0$ \KwTo $L - 1$}{
    CurrentSegments $\leftarrow$ UnprocessedSegments[$i$]\;
    UnprocessedSegments[$i$] $\leftarrow \{ \}$\;
    \ForEach{\normalfont($j: ((i,j) \in P$ \textbf{or} $(j,i) \in P) $)}{
       \tcp{(\textsc{ii}) Traverse BVHs, and (\textsc{iii}-\textsc{v}) subdivide ``under-refined'' segments.}
       \ForEach{\normalfont($s :$ ($s \in $  CurrentSegments \textbf{and} $ s \in $ \GetIntersectingBoxes{\normalfont Trees[$i$], Trees[$j$]}))}{
          $r, t \leftarrow $ SubdivideSegmentIntoHalves($s$)\;
          $r$.ComputeBoundingBox()\;
          $t$.ComputeBoundingBox()\;
          \If{\normalfont (\DiameterOfBBox{$r$ \normalfont or $t$} $< \epsilon \xi$)} {
              \textbf{return} Error(``Curves $i$ and $j$ intersect.'')\;
          }
          UnprocessedSegments[$i$].InsertSegments($r, t$)\;
          CurrentSegments.Exclude($s$)\;
       }
    }
    \tcp{(\textsc{vi}-\textsc{viii}) Collect remaining segments into results.}
   $\Gamma_i \leftarrow \Gamma_i \; \cup$ CurrentSegments\;
}
}
\textbf{return} $\{ \Gamma_i \}$\;
}
\end{algorithm}

\revisionp{Suppose every spline segment is parameterized in $t$ as $\vec{p}(t)$.} We take multiple passes. At the beginning, all spline segments are unprocessed. In each pass, we start with a list of unprocessed spline segments in each loop. For each loop, we \revision{(\textsc{i})} build a BVH of its unprocessed spline segments. \revisionp{If any segment has almost zero length, we exit with an error, as this indicates that either the input is corrupt or the curves are nearly intersecting.} In our implementation we set the length threshold $\epsilon$ to the appropriate machine epsilon. For each loop pair, we \revision{(\textsc{ii})} traverse the loops' BVHs, and \revision{(\textsc{iii})} any two segments from the two different loops with overlapping bounding volumes are marked as ``under-refined.'' After all BVH pairs are traversed, we \revision{(\textsc{iv})} split every ``under-refined'' segment into two spline segments, by keeping the same coefficients but splitting the $t$ parameter's domain in halves, and \revision{(\textsc{v})} add them into a new list of unprocessed segments for the next pass.  Every segment not marked ``under-refined'' is \revision{(\textsc{vi})} converted into a line segment with the same endpoints, and \revision{(\textsc{vii})} added to a results list for the loop. Afterwards, we repeat this process until no segments are unprocessed. At the end, we \revision{(\textsc{viii})} sort the line segment list for each loop. In our implementation (Algorithm~\ref{alg:discretization}), all steps are parallelized across loops, and \revisionp{so steps \revision{(\textsc{ii--iii})} are actually ran twice for each loop pair: once from each loop in the pair}.

To ensure the algorithm terminates if two loops nearly intersect, we exit with an error if any segment length gets shorter than floating-point epsilon times the average coordinate. In our examples, this limit is reached in about 45 passes for double precision, and under 20 passes for single precision, with most of the later passes only processing a small number of primitives. \revisionp{Exiting at this limit ensures the next stage (\S\ref{sec:methodsintegration}) does not compute linking numbers on nearly intersecting curves.}

\subsubsection{Proof that discretization is homotopically equivalent}
\label{sec:proof}

For non-intersecting input curves with polynomial spline segments, we \revision{show that Algorithm~\ref{alg:discretization} terminates when using tight AABBs.} Also, we show that if this algorithm terminates, it produces a homotopically equivalent segmentation. These proofs can be found in Appendix~\ref{app:discretizationProofs}.

\subsection{Linking \revisionp{Number} Computation}
\label{sec:methodsintegration}

Given a set of loops discretized into line segments, and a set of loop pairs to check, we wish to compute their respective linkages and output a linking matrix. We present several classes of methods here based on different prior work, and our results and analysis section (\S\ref{sec:results}) will discuss the relative advantages of each method.

When running this stage on the CPU, we parallelize across the list of potentially linked loop pairs and have each thread compute its own linking number individually. We only use the parallelized version of these methods if we have very few or only one loop pair, which can arise when there are very few loops. For the GPU-accelerated methods, we just iterate through the list of loop pairs and launch the GPU kernels sequentially.

\subsubsection{Method 1: Count Crossings (CC$_{ann}$ and CC)}

\revisionp{The first approach is, given a pair of loops, find a regular projection and use it to count crossings. We modify the implementation provided with \cite{dey2013efficient} for computing the linking number, and evaluate two optimized methods (CC$_{ann}$ and CC).  Their projection code takes three steps to heuristically find a regular projection: first, it uses approximate nearest neighbors (ANN) to estimate a projection direction away from all line-segment directions among the two loops, defining an initial frame whose $z$ axis will be the direction of projection. Next, it corrects the frame by rotating about the $z$ axis into the direction such that all the projected segments would be oriented, at minimum, as far away as possible from vertical ($y$) or horizontal ($x$) directions. Thirdly, it rotates the frame about the $y$ axis so that all the projected segments would be, at minimum, the farthest from being degenerate (aligned with the $z$ axis). Each of these steps take $O(N \log N)$ time to sort the projected segment angles and perform the maximin. Afterwards, they project the segments onto the XY plane, and then perform an $O(N_k N_l)$ (where $N_k$, $N_l$ are the numbers of segments in each input loop) brute-force intersection check between all segment pairs.} \revision{The intersection check uses 2D exact line segment queries from CGAL \cite{cgal2d} and exits with an error when degeneracies are detected.}

While this procedure is sufficient for the loops \revision{generated} from the surface mesh analysis in \cite{dey2013efficient}, our linking number computation must support larger inputs. We accelerated the intersection search by building a BVH for one of the two loops when their average loop size exceeds \revision{750 segments}, and evaluating segments from the other loop against this tree in parallel. We refer to this modified method as CC$_{ann}$. \revisionp{The ANN call also} introduces a large overhead, so, after experimentation, we simply replace the initial frame with a randomized frame. This greatly speeds up performance for many examples, and we report the performance for this method (\revision{which} we call CC) as well as robustness tests in the results \S\ref{sec:resultsrobustness}. See Algorithm~\ref{alg:crossings} for pseudocode.

\begin{algorithm}[]
\caption{ComputeLink: Counting Crossings}
\label{alg:crossings}
\SetAlgoLined
\SetKwFunction{ComputeLink}{ComputeLink}
\SetKwFunction{GenerateRegularProjectionFrame}{GenerateRegularProjectionFrame}
\SetKwFunction{GenerateRandomOrthogonalFrame}{GenerateRandomOrthogonalFrame}
\SetKwFunction{RotateFrameAboutZToOptimalAngle}{RotateFrameAboutZToOptimalAngle}
\SetKwFunction{RotateFrameAboutYToOptimalAngle}{RotateFrameAboutYToOptimalAngle}
\SetKwFunction{BuildBVH}{BuildBVH}
\SetKwFunction{OrientationOfCrossing}{OrientationOfCrossing}
\SetKwInOut{Input}{Input}
\SetKwInOut{Output}{Output}

\Input{$\{\vec{l}_j\}$, $\{\vec{k}_i\}$, the two polyline loops, each represented as a list of vertices. Each vertex is a column vector.}
\Output{$\lambda$, the linking number}
\BlankLine

\tcp{This function computes the linking number between two closed polylines by counting crossings and uses a BVH.}
\SetKwProg{Fn}{Function}{:}{}
\Fn{\ComputeLink{$\{\vec{l}_j\}$, $\{\vec{k}_i\}$}}{
$\lambda \leftarrow 0$\;
$N_k \leftarrow |\{\vec{k}_i\}|$\;
\tcp{Find a regular projection frame.}
$F \leftarrow ( \vec{\hat{x}}, \vec{\hat{y}}, \vec{\hat{z}}) \leftarrow $ \GenerateRegularProjectionFrame{$\{\vec{l}_j\}$, $\{\vec{k}_i\}$}\;
\tcp{Rotate into the frame.}
$\{\vec{l}_j\} \leftarrow \{ F^T \vec{l}_j\}$; \,\, $\{\vec{k}_i\} \leftarrow \{ F^T \vec{k}_i\}$\;
\tcp{Build a tree from the projected segments of $\{\vec{l}_j\}$.}
tree $\leftarrow$ \BuildBVH{$\{ ((l_{jx}, l_{jy}), (l_{(j+1)x}, l_{(j+1)y}))\}$}\;
\tcp{Sum the orientations of all crossings}
\textbf{parallel} \For{$i \leftarrow 0$ \KwTo $N_k$}{
   ProjectedSegment $\leftarrow ((k_{ix}, k_{iy}), (k_{(i+1)x}, k_{(i+1)y}))$ \;
   \ForEach{$j \in $ \normalfont{ tree.Query(ProjectedSegment)}}{
     SegmentI, SegmentJ $ \leftarrow (\vec{k}_i, \vec{k}_{i+1}), (\vec{l}_j, \vec{l}_{j+1})$\;
     $\lambda \leftarrow \lambda + 0.5$ Orientation(SegmentI, SegmentJ)\;
   }
}
\textbf{return} $\lambda$ \;
}
\end{algorithm}

\subsubsection{Method 2: Direct Summation (DS)}

This approach simply uses the exact double-summation formula for Gauss's integral from \cite{arai2013rigorous} (which uses the signed solid angle formula \cite{van1983solid}), produced above as \eqref{eq:directarctan}. That is, for loops $\gamma_1, \gamma_2$, consisting of $N_l, N_k$ line segments enumerated by $j,i$ respectively,
\begin{align}
    \lambda(\gamma_1, \gamma_2) &= \sum_{i}^{N_k} \sum_j^{N_l} \lambda_{ji}.
\end{align}
See Algorithm~\ref{alg:de_simp} for a simple implementation. Unfortunately this approach computes $2 N_l N_k$ arctangents for loops of sizes $N_l$ and $N_k$, which is expensive for large loops.

When this is a single-threaded computation, the approach from \cite{bertolazzi2019efficient} removes all arctangents by using angle summations (counting negative $x$-axis crossings) instead. We use a modified version of their approach, and for robustness, we add a single $\arctan$ at the end to compute the remainder after the angle summation. We only compute one triple scalar product per segment pair because $\vec{a} \cdot (\vec{b} \times \vec{c}) = \vec{c} \cdot (\vec{d} \times \vec{a})$. See Appendix~\ref{sec:de_gpu} for details on how we multithread this on the CPU and the GPU.

\begin{algorithm}[]
\caption{ComputeLink: Direct Summation.}
\label{alg:de_simp}
\SetAlgoLined
\SetKwInOut{Input}{Input}
\SetKwInOut{Output}{Output}

\Input{$\{\vec{l}_j\}$, $\{\vec{k}_i\}$, the two polyline loops, each represented as a list of vertices. Each vertex is a column vector.}
\Output{$\lambda$, the linking number}
\BlankLine


\SetKwFunction{ComputeLink}{ComputeLink}
\SetKwFunction{Concatenate}{Concatenate}
\SetKwProg{Fn}{Function}{:}{}

\tcp{This function computes the linking number between two closed polylines using the \cite{arai2013rigorous} expression.}
\Fn{\ComputeLink{$\{\vec{l}_j\}$, $\{\vec{k}_i\}$}}{
$\lambda \leftarrow 0$\;
$N_l \leftarrow |\{\vec{l}_j\}|$; \,\, $N_k \leftarrow |\{\vec{k}_i\}|$\;
\ForEach{\normalfont{int} $(i,j)\in [0, N_k-1] \times [0, N_l-1]$}{
    $\vec{a} \leftarrow \vec{l}_j - \vec{k}_i$\;
    $\vec{b} \leftarrow \vec{l}_j - \vec{k}_{i+1}$\;
    $\vec{c} \leftarrow \vec{l}_{j+1} - \vec{k}_{i+1}$ \;
    $\vec{d} \leftarrow \vec{l}_{j+1} - \vec{k}_i$ \;
    $p \leftarrow \vec{a} \cdot (\vec{b} \times \vec{c})$\;
    $d_1 \leftarrow |\vec{a}| |\vec{b}| |\vec{c}| + \vec{a} \cdot \vec{b} |\vec{c}| + \vec{b} \cdot \vec{c} |\vec{a}| + \vec{c} \cdot \vec{a} |\vec{b}|$ \;
    $d_2 \leftarrow |\vec{a}| |\vec{d}| |\vec{c}| + \vec{a} \cdot \vec{d} |\vec{c}| + \vec{d} \cdot \vec{c} |\vec{a}| + \vec{c} \cdot \vec{a} |\vec{d}|$ \;
    $\lambda \leftarrow \lambda + (\text{atan2}(p, d_1) + \text{atan2}(p, d_2))/ (2 \pi)$\;
}
\textbf{return} $\lambda$\;
}
\end{algorithm}

\subsubsection{Method 3: Fast Multipole Method (FMM)}

We can also use external FMM libraries (provided by FMMTL \cite{cecka2015fmmtl} or FMM3D \cite{cheng1999fast}) to compute the linking integral using fast summation approximations. \revisionp{In particular, if we represent $\gamma_2$ as a source current, the Fast Multipole Method on the Biot--Savart integral will approximately evaluate, at each target point $\rv$, the field}
\begin{align}
    \vec{f}(\rv) &= \frac{1}{4 \pi}\int_{\gamma_2} \frac{\text{d} \rv_2 \times (\rv - \rv_2) }{|\rv - \rv_2|^3}.\label{eq:fmmeval}
\end{align}
\revisionp{The final linking number is then $\lambda(\gamma_1, \gamma_2) = \int_{\gamma_1} \vec{f}(\rv_1) \cdot \text{d}\rv_1$. In this notation, loop $\gamma_2$ is the ``source,'' and $\gamma_1$ is the ``target.''}

\revisionp{Because many of these libraries do not directly support finite line-segment inputs, we use segment midpoints for the source and target points. After evaluation by the library, we compute a set of finite-segment corrections for close segment pairs poorly approximated by midpoint samples, and add the correction. While we already know the exact evaluation expressions for line-segment pairs and could have modified libraries to use them, our goal is to make no modifications to existing FMM libraries that operate on point samples, so that our results can be easily reproduced and improvements to FMM libraries on newer hardware can be easily incorporated. This is why we perform post-evaluation correction rather than directly modify the FMM kernels and tree building structures.}

\revisionp{Given the aforementioned loop discretizations of $\gamma_1, \gamma_2$ in the direct evaluation section, denote the midpoint of each line segment as $\vec{r}_j$ or $\vec{r}_i$, respectively, and the displacement vector (difference of endpoints) as $\vec{s}_j$, $\vec{s}_i$, respectively. FMMTL directly provides a Biot--Savart implementation, so we simply have to pass in the line-segment displacements $\vec{s}_i$ as the sources, with positions $\rv_i$, and FMMTL computes the fields $\vec{f}_j = \vec{f}(\rv_j)$ at positions $\rv_j$.}

\revisionp{The linking number can then be computed from the field $\vec{f}_j$ using}
\begin{align}
    \lambda(\gamma_1, \gamma_2) &= \sum_j \vec{s}_j \cdot \vec{f}_j.
\end{align}

To reduce the duplicate work of building the FMM trees, we batch evaluation points for each loop. That is, given source loop $\gamma_i$, we concatenate the target evaluation points $r_j$ for all target loops to compute $F_j$ for all target loops, and then compute each linking-number sum individually.

When a library does not provide a native Biot--Savart implementation, we can \revisionp{alternately pass in a list of three-vector sources $\{\vec{s}_i\}$ into a Laplace FMM. The gradient output gives us Eq.~\eqref{eq:fmmeval}, except with an outer instead of a cross product}. The output is a $3\times3$ tensor $C_j = \sum_i  \vec{s}_i \otimes \nabla G(\vec{r}_j, \vec{r}_i) $ at each evaluation point $\vec{r}_j$, and we can simply grab the field $\vec{f}_j$ from the definition of cross product: 
\begin{align}
    \vec{f}_j &= ( C_{j23} - C_{j32} , C_{j31} - C_{j13} , C_{j12} - C_{j21}).
\end{align}

After the FMM library computation, we add a finite-segment correction, because we passed in our line segments as point sources and targets. We pick a distance ratio $\beta$ and apply this correction to all segment pairs that are less than $\beta (l_1 + l_2)/2$ apart, where $l_1, l_2$ are their segment lengths. Specifically, before the FMM calls we \revision{(\textsc{i})} build a bounding volume for each segment, and \revision{(\textsc{ii})} dilate each bounding volume by $\beta l/2 $, where $l$ is the segment length. We then \revision{(\textsc{iii})} build a BVH of segment bounding volumes for each loop. For any loop pair, \revision{(\textsc{iv})} traverse their BVH trees, and \revision{(\textsc{v})} accumulate the difference between expression \eqref{eq:directarctan} and its approximation \eqref{eq:midpointapproximation} for every pair of overlapping segment bounding volumes. We then \revision{(\textsc{vi})} add this correction to the original FMM result. In our implementation, we parallelize steps \revision{(\textsc{i--iii})} by loop and steps \revision{(\textsc{iv--v})} by loop pair, with a reduction at the end. \revision{We used a finite-segment correction because the correction is faster to compute in practice than using denser sample points along each segment for the FMM input; in our results in \S\ref{sec:results}, the correction takes a small fraction ($<\nicefrac{1}{8}$) of the runtime.}

\subsubsection{Method 4: Barnes--Hut (BH)}

Similar to \cite{barill2018fast}, simple Barnes--Hut trees \cite{barnes1986hierarchical} can be sufficient for \revision{approximately} evaluating the integral using fast summation without the overhead of the full FMM. \revisionp{For each loop pair, the FMM computes a field at every ``target'' point; however, we only need a single scalar overall. Since each loop can participate in many loop pairs, we can instead precompute a tree for each loop. Then for each loop pair, we traverse \emph{both trees} from the top down using Barnes--Hut; this can save effort especially when the loops do not hug each other closely and multiple loop pairs use the same target loop.} 

The Barnes--Hut algorithm takes advantage of a far-field multipole expansion. Given two curves $\gamma_1, \gamma_2$, parameterize $\gamma_1$ as $\vec{r}_1 \eq \vec{r}_1(s)$ and $\gamma_2$ as $\vec{r}_2 \eq \vec{r}_2(t)$, \revisionp{and denote $\nicefrac{d\rv_1}{ds}$ and $\nicefrac{d\rv_2}{dt}$ by $\rv_1'$ and $\rv_2'$}. Suppose also that we are only integrating the portions of $\gamma_1$ near $\vec{\tilde{r}}_1$, and the portions of $\gamma_2$ near $\vec{\tilde{r}}_2$. Also, let $\grad$ denote differentiation with respect to $\vec{r}_2$ \revisionp{(note that $\grad_{\rv_1} G$ is just $-\grad_{\rv_2} G$), $\otimes$ denote a tensor (outer) product, and $\cdot$ denote an inner product over all dimensions. Taylor expanding \eqref{eq:gaussintegral} with respect to $\rv_1$ and $\rv_2$, we have}
\begin{align}
    \lambda &= -\int \text{d}s\; \text{d}t \; (\vec{r}_1' \times \vec{r}_2') \cdot \grad G(\vec{r}_1, \vec{r}_2).\\
    \lambda &= -\int \text{d}s\; \text{d}t \; \left[ ( \vec{r}_1' \times \vec{r}_2' ) \cdot \grad G(\vec{\tilde{r}}_1, \vec{\tilde{r}}_2) \phantom{\frac{1}{2}}\right.\nonumber  \\
    & + ( (\vec{r}_1' \times \vec{r}_2' ) \otimes (-(\vec{r}_1 - \vec{\tilde{r}}_1) + (\vec{r}_2 - \vec{\tilde{r}}_2))) \cdot \grad^2  G(\vec{\tilde{r}}_1, \vec{\tilde{r}}_2) \nonumber \\
    & + \frac{1}{2}( (\vec{r}_1' \times \vec{r}_2' ) \otimes(-(\vec{r}_1 - \vec{\tilde{r}}_1) + (\vec{r}_2 - \vec{\tilde{r}}_2)) \otimes (-(\vec{r}_1 - \vec{\tilde{r}}_1) + (\vec{r}_2 - \vec{\tilde{r}}_2)))  \nonumber \\
    & \; \; \; \; \left. \phantom{\frac{1}{2}} \cdot \grad^3
     G(\vec{\tilde{r}}_1, \vec{\tilde{r}}_2)  + O(|\vec{\tilde{r}}_1- \vec{\tilde{r}}_2|^{-5}) \right].\label{eq:expansion}
\end{align}

Expanding \eqref{eq:expansion} and applying some algebra, every term can be split into a product of two separate integrals, one for the $\vec{r}_1$ factors and one for the $\vec{r}_2$ factors. These integrals, known as moments, can be precomputed when constructing the tree. For each node \revision{on both curves,} we precompute the following moments:
\begin{align}
    \vec{c}_M &= \int \vec{r}' \;\text{d}s,\label{eq:monopolemoment}\\
    C_D &= \int \vec{r}' (\vec{r} - \vec{\tilde{r}})^T \;\text{d}s,\label{eq:dipolemoment}\\
    C_Q &= \int \vec{r}' \otimes (\vec{r} - \vec{\tilde{r}})\otimes (\vec{r} - \vec{\tilde{r}}) \; \text{d}s.\label{eq:quadrupolemoment}
\end{align}

Having just the $\vec{c}_M$ moment at each node is sufficient to give us the first term in \eqref{eq:expansion}. Adding the $C_D$ and $C_Q$ moments gives us the second and third terms respectively. The derivatives of $G(\vec{\tilde{r}}_1, \vec{\tilde{r}}_2)$, how we compute these moments, and how we parallelize the Barnes--Hut on the CPU and GPU (using the Barnes--Hut N-Body implementation from \cite{burtscher2011efficient}) are in Appendix~\ref{sec:barneshutdipole}.

We set a $\beta$ tolerance parameter, similar to most Barnes--Hut algorithms, to distinguish between the near field and the far field: two nodes at locations $\vec{p}$ and $\vec{q}$ are in the far field of each other if \revision{$|\vec{p}-\vec{q}|$} is greater than $\beta$ times the sum of the bounding-box radii of the two nodes. Starting from both tree roots, we compare their bounding boxes against each other. If the nodes are in the far field, we use the precomputed moments and evaluate the far-field expansion; otherwise, we traverse down the tree of the larger node and check its children. If both nodes are leaves, we use the direct arctangent expression \eqref{eq:directarctan}. See Algorithm~\ref{alg:barneshut} for a summary.

\begin{algorithm}[]
\caption{ComputeLink: Barnes--Hut}
\label{alg:barneshut}
\SetAlgoLined
\SetKwFunction{ComputeLink}{ComputeLink}
\SetKwInOut{Input}{Input}
\SetKwInOut{Output}{Output}

\Input{tree1, tree2, the root nodes of the two trees, with $\rv$ the node's center, $R$ the bounding box radius, and $\vec{c}_M, C_D, C_Q$ the moments, precomputed at each node.}
\Output{$\lambda$, the linking number}
\BlankLine

\tcp{This function computes the linking number between two closed polylines using the Barnes--Hut algorithm with a quadrupole far-field expansion.}
\SetKwProg{Fn}{Function}{:}{}
\Fn{\ComputeLink{\normalfont{tree1, tree2}}}{
\uIf{$(|$\normalfont{tree1.}$\rv -$\normalfont{tree2}.$\rv| > \beta ($\normalfont{tree1.}$R + $\normalfont{tree2.}$R))$}{
   \textbf{return} FarFieldEvaluation(tree1, tree2); \tcp{Use \eqref{eq:expansion}}
}
\uElseIf{\normalfont{(tree1 and tree2 have no children)}} {
   \tcp{Use \eqref{eq:directarctan}}
   \textbf{return} DirectEvaluation(tree1.Segment, tree2.Segment)\;
}
\uElseIf{\normalfont{(tree1.}$R > $ \normalfont{tree2.}$R$ \normalfont{ and tree1 has children)}}{
   \textbf{return} $\sum_c$ \ComputeLink(tree1.child[$c$], tree2)\;
}
\Else{
   \textbf{return} $\sum_c$ \ComputeLink(tree1, tree2.child[$c$])\;
}
}
\end{algorithm}

In our implementation, we first run Barnes--Hut with the second-order expansion in \eqref{eq:expansion} (using all three terms) with $\beta_\text{init}=2$. Unlike other problems that use Barnes--Hut, the linking number certificate requires the same absolute error tolerance, regardless of the magnitude of the result. In large examples, some regions can induce a strong magnetic field from one loop on many dense curves from the other loop, resulting in a very large integrand. Our implementation therefore uses \revision{an error estimate based on the next-order expansion term} to set a new $\beta$ for a second evaluation if necessary. \revision{The estimate is accumulated} at each far-field node pair \revision{and reported after the evaluation}. The new $\beta_t$ is given by
\begin{align}
    E_\text{estimate} &= k\sum_{(i,j) \in \text{far-field node pairs} }  |\vec{r}|^{-5} (R_i|\vec{c}_{jM}| \|C_{iQ}\| \nonumber\\
    & \; \; \; + R_j|\vec{c}_{iM}| \|C_{jQ}\| + 3(\|C_{iD}\| \|C_{jQ}\| + \|C_{iQ}\| \| C_{jD} \|)),\\
    \beta_t &= \left(\frac{E_\text{estimate}}{E_\text{target} } \right)^{1/4} \beta_\text{init}.
\end{align}
Here $\vec{r} = \vec{\tilde{r}}_j - \vec{\tilde{r}}_i$, $R$ is the bounding-box radius at that node, and $k$ is a constant. If $\beta_t > \beta_\text{init}$, then it reruns Barnes--Hut using the second-order expansion and $\beta=\min(\beta_t,\beta_\text{max})$.

%% file: results.tex
\section{Results and Analysis}
\label{sec:results}

\begin{table*}[t]
    \centering
    \caption{{\bf Linkage Computation for Various Inputs:} In this table, $N_I$ is the total number of input curve segments, $L$ is the number of input loops, and $P$ is the computed number of potentially linked loop pairs. The first 11 models correspond to animated results shown in the supplemental video, and the rest are stress tests depicted in Fig.~\ref{fig:stresstests}. \revision{The sweater and glove models originate from \cite{yuksel2012stitch}, with ``Sweater'' referring to ``sweater flame ribbing.''} ``PLS Time'' is the Potential Link Search Time and ``Dscr Time'' is the Discretization time; DS is direct summation; CC$_{ann}$ and CC for counting crossings with ANN and randomized direction initialization, respectively; BH for Barnes--Hut, and FMM for Fast Multipole Method (using FMMTL with batched target points; FMM3D is slower in almost all cases). DSG and BHG run the DS and BH, respectively, on the GPU. A ``N/A'' indicates the run took longer than a half hour or did not fit in memory. $\lambda$ indicates the linkage between curves in a model, and $\nu$ is a woundball parameter explained in Fig.~\ref{fig:stresstests}.\\$^*$ For this example, FMMTL failed, and this is the FMM3D runtime instead.\\$^\dagger$ These used $\beta=40$, instead of $\beta=16$ normally used for large input. (The GPU BH only evaluates up to dipole moments, requiring high $\beta$ for large input.)}
    Linkage Computation for Various Input, Using Various Methods (All Times in [s])\\
    \setlength\tabcolsep{4pt}
    \begin{tabular}{|l|*{15}{c|}}
        \hline
        \multirow{2}{*}{Model} & \multirow{2}{*}{$N_I$} & \multirow{2}{*}{$L$} & \multirow{2}{*}{$P$} & PLS & Dscr. & \multicolumn{7}{c|}{\revision{Linking Number Matrix} Compute Times} & \multicolumn{3}{c|}{Abs. Error} \\
         \cline{7-16}
          & & & & Time & Time & DS & $\!\!$CC{\tiny $_{ann}$}$\!\!$ & CC & BH & FMM & DSG & BHG & BH & BHG & FMM \\
         \hline
        \revision{Alien Sweater (Initial)} & 139506 & 146 & 1335 & 0.013 & 0.059 & 11.1 & 12.6 & 0.34 & 0.32 & 1.09 & 0.63 & 0.27 & 3E-3 & 2E-3 & 1E-3 \\
         \hline
        \revision{Alien Sweater (Final)} & 139506 & 146 & 2426 & 0.017 & 0.050 & 13.9 & 22.4 & 0.38 & 1.18 & 1.61 & 1.01 & 0.63 & 3E-3 & 8E-4 & 1E-3 \\
         \hline
        \revision{Sheep Sweater} & 729628 & 549 & 4951 & 0.024 & 0.152 & 169 & 95.5 & 0.85 & 1.31 & 5.85 & 4.34 & 0.96 & 2E-3 & 5E-3 & 3E-3 \\
         \hline
        Sweater & 271902 & 253 & 2985 & 0.015 & 0.120 & 29.3 & 33.0 & 0.29 & 0.74 & 2.33 & 1.43 &  0.56 & 3E-3 & 2E-3 & 9E-4 \\
         \hline
        Glove & 58537 & 70 & 1022 & 0.012 & 0.046 & 2.71 & 8.49 & 0.10 & 0.14 & 0.56 & 0.35 & 0.16 & 2E-3 & 2E-3 & 4E-4 \\
         \hline
        Knit Tube \revision{(Initial)} & 18228 & 39 & 233 & 0.012 & 0.032 & 0.21 & 1.01 & 0.07 & 0.05 & 0.09 & 0.02 & 0.08 & 4E-3 & 9E-4 & 1E-3 \\
         \hline
        Knit Tube \revision{(Final)} & 18228 & 39 & 180 & 0.012 & 0.082 & 0.30 & 1.03 & 0.06 & 0.08 & 0.11 & 0.30 & 0.12 & 6E-3 & 3E-4 & 3E-3 \\
         \hline
        Chainmail \revision{(Initial)} & 211680 & 14112 & 18781 & 0.028 & 0.070 & 0.04 & 1.85 & 0.03 & 0.09 & 0.10 & 1.20 & N/A  & 1E-3 & N/A & 5E-4 \\
        \hline
        Chainmail \revision{(Final)} & 211680 & 14112 & 61417 & 0.056 & 0.086 & 0.11 & 6.30 & 0.07 & 0.15 & 0.17 & 3.91 & N/A & 3E-4 & N/A & 7E-4 \\
        \hline
        Chevron $3\times3$ & 11741 & 32 & 58 & 0.013 & 0.049 & 0.19 & 0.44 & 0.02 & 0.07 & 0.06 & 0.02 & 0.03 & 5E-3 & 2E-4 & 1E-3 \\
        \hline
        Rubber Bands & 51200 & 1024 & 20520 & 0.018 & 0.035 & 0.23 & 7.99 & 0.05 & 0.12 & 0.16 & 1.11 & 1.87 & 7E-4 & 2E-5 & 5E-4 \\
        \hline
        {\small \revision{Double} Helix Ribbon $\lambda\!=\!10$} & \revision{200000} & 2 & 1 & 0.018 & 0.076 & 177 & 5.60 & 0.06 & 0.15 & 0.34 & 0.48 & 0.02 & 7E-3 & 1E-3 & 1E-4  \\
        \hline
        {\small \revision{Double} Helix Ribbon $\lambda\!=\!10^3$} & \revision{200000} & 2 & 1 & 0.016 & 0.112 & 177 & 1.82 & 0.07 & 1.62 & 0.55 & 0.48 & 0.03 & 0.22 & 7E-2 & 0.37  \\
        \hline
        {\small \revision{Double} Helix Ribbon $\lambda\!=\!10$}
        & 2E7 & 2 & 1 & 0.465 & 6.816 & N/A & 131 & 8.62  & 11.8 & N/A & N/A & 2.63$^\dagger$ & 3E-2 & 5E-4 & N/A  \\
        \hline
        {\small \revision{Double} Helix Ribbon $\lambda\!=\!10^3$}
        & 2E7 & 2 & 1 & 0.441 & 6.816 & N/A & 118 & 9.51 & 31.3 & N/A & N/A & 5.30$^\dagger$ & 0.17 & 0.15 & N/A  \\
        \hline
        Thick Square Link & 500000 & 1250 & 4.1E5 & 0.054 & 0.146 & 161 & N/A & 31.1 & 6.83 & 37.6 & 26.4 & 23.2 & 1E-3 & 6E-4 & 1E-4 \\
        \hline
        Thick Square Link & 4E6 & 5000 & 6.4E6 & 0.531 & 1.874 & N/A & N/A & 105 & 145 & 830 & 1844 & N/A & 2E-3 & N/A & 2E-4 \\
        \hline
        Torus $\lambda\!=\!10^6$ & 2E7 & 2 & 1 & 0.456 & 7.027 & N/A & 100.3 & 9.30 & 13.9 & 22.2 & N/A & 4.94 & 5E-3 & 6E+2 & 9E-4 \\
        \hline
        Torus $\lambda\!=\!0$ & 4E7 & 4 & 6 & 0.519 & 7.698 & N/A & N/A & 55.1 & 18.3 & N/A & N/A & N/A & 3E-3 & N/A & N/A \\
        \hline
        Woundball $\nu\!=\!10^3$ & 1E6 & 2 & 1 & 0.048 & 0.387 & N/A & 6.93 & 0.55 & 0.68 & 1.32 & 12.0 & 0.14 & 6E-2 & 1E-2 & 2E-5 \\
        \hline
        Woundball $\nu\!=\!10^4$ & 2E6 & 2 & 1 & 0.056 & 0.756 & N/A & 43.9 & 7.96 & 15.3 & 5.19$^*$ & 47.7 & 90.0 & 1E-2 & 5E-4 & 7E-6 \\
        \hline
    \end{tabular}
    \label{tab:aggregateresults}
\end{table*}

\begin{figure}
    \centering
    \includegraphics[width=0.5\hsize]{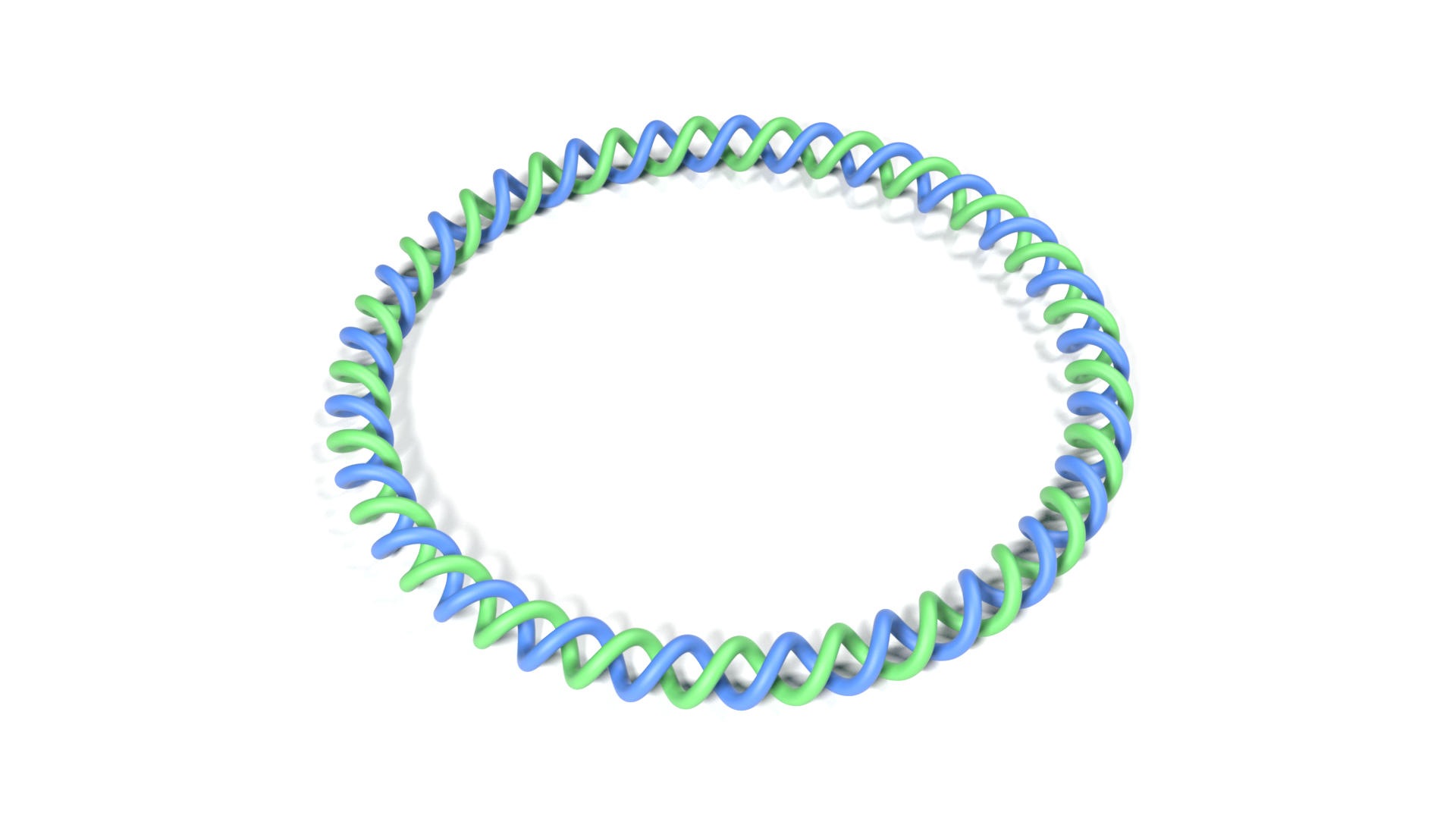}\includegraphics[width=0.5\hsize]{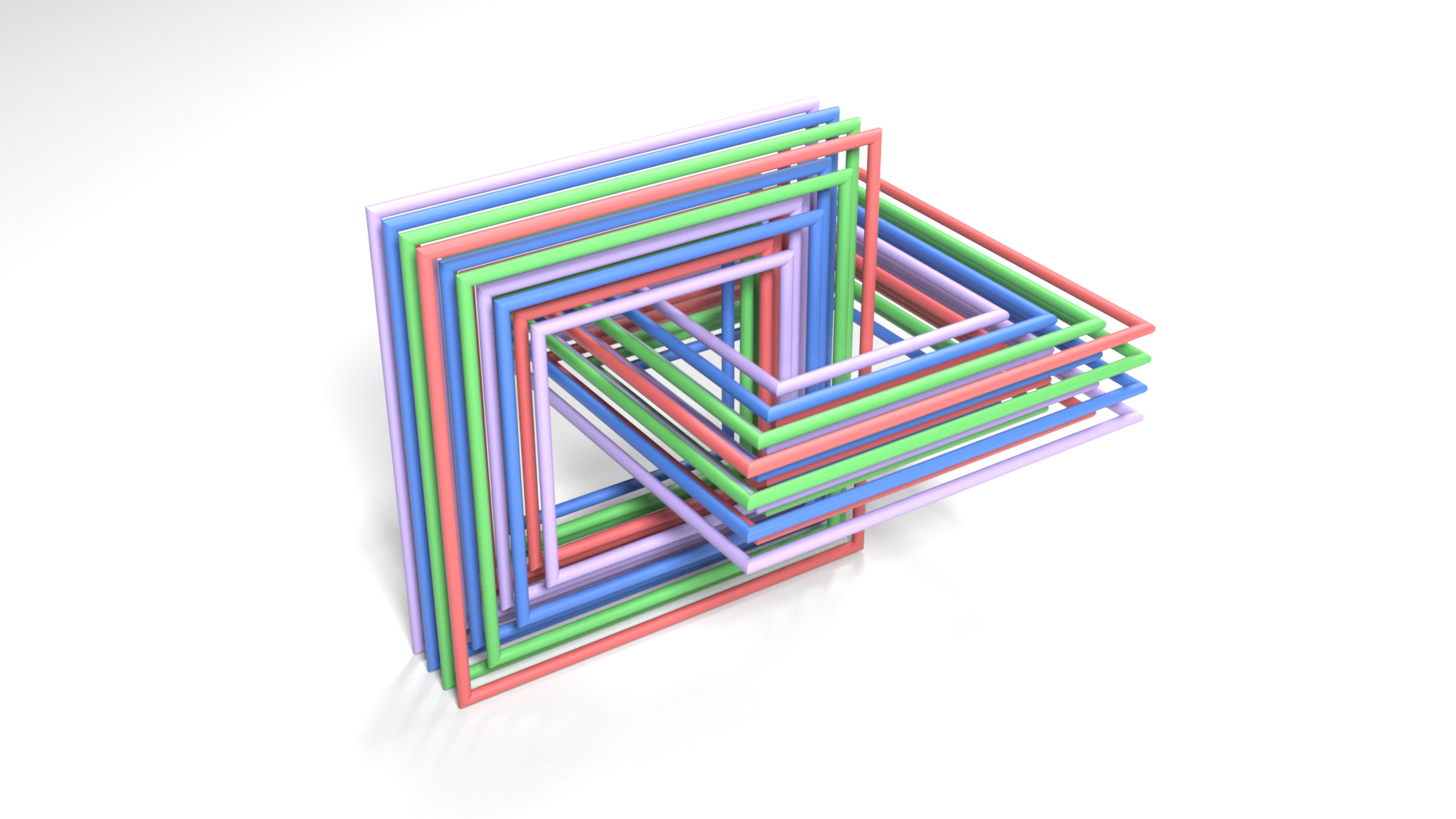} \\
    \includegraphics[width=0.5\hsize]{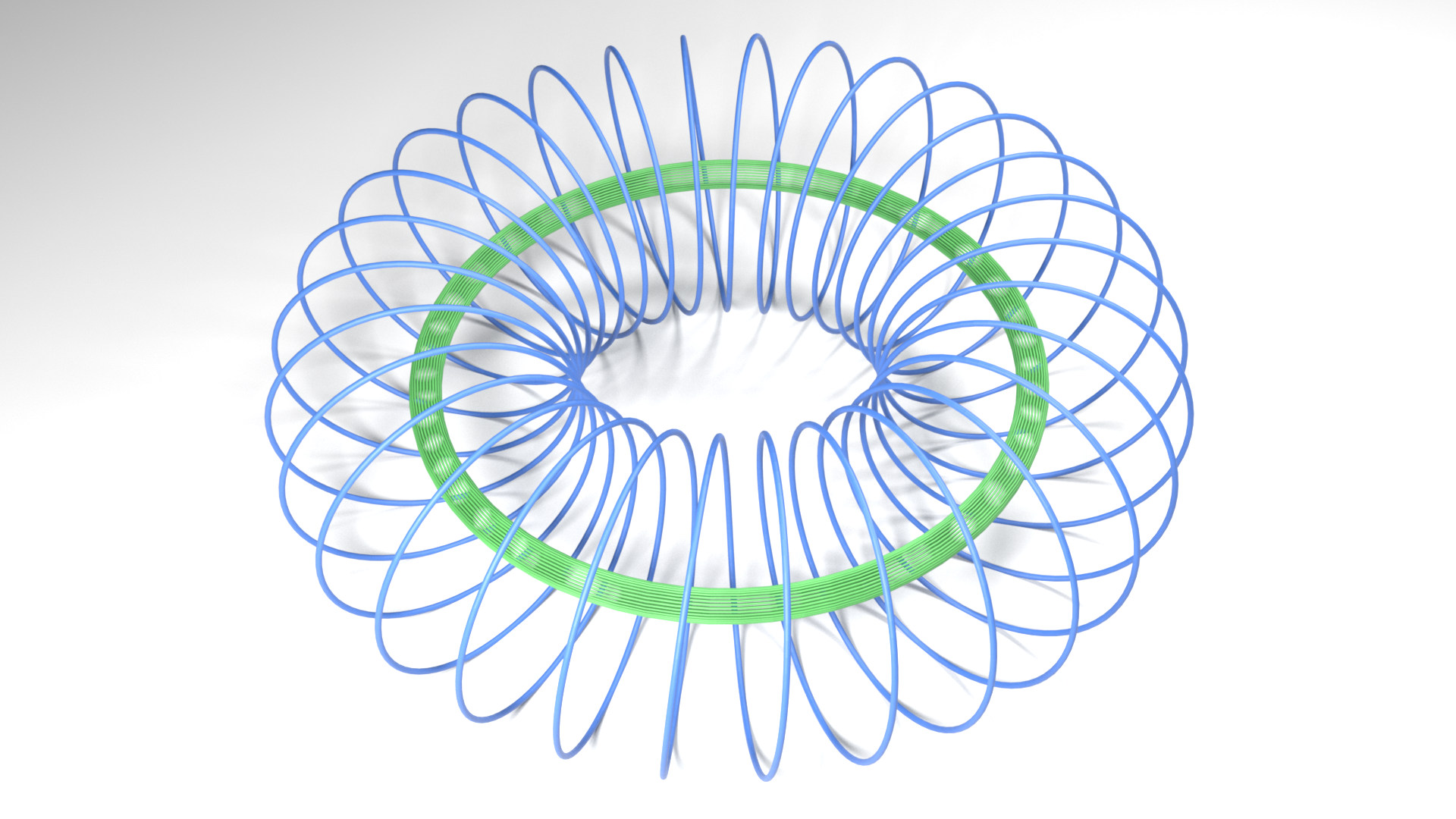}\includegraphics[width=0.5\hsize]{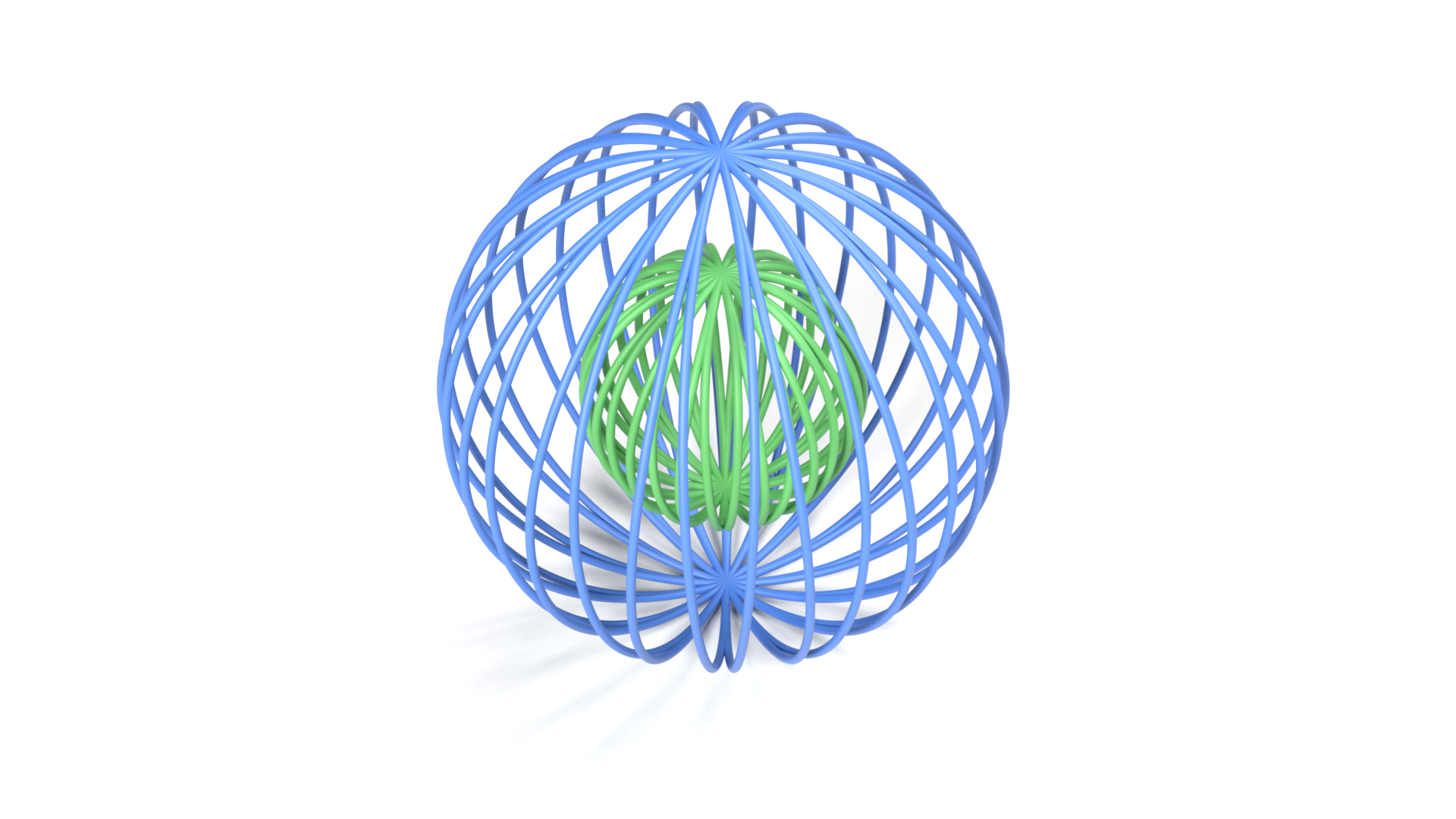}
    \caption{\revision{{\bf Large-Input Stress Tests:}} Synthetic examples were generated for (Top Left) DNA-like double-helix ribbons, (Top Right) thick square links, (Bottom Left) tori, and (Bottom Right) woundballs, of variable sizes. The ``thick square links'' example has $\nicefrac{L^2}{4}$ links each with $\lambda=1$, and the torus has a single $\lambda$ equal to the product of the number of toroidal periods in the green loop and poloidal periods in the blue loop. We also generated a $\lambda=0$ torus where both loops simply alternate directions every period, and it also has 2 more larger copies of the blue loop. The woundball has two similar concentric spherical curves, one winding a ball half the size of the other, and a parameter $\nu$ indicating the number of times each curve crosses the North pole.}
    \label{fig:stresstests}
\end{figure}

\begin{table}
    \centering
    \caption{{\bf Discretization Results:} This table shows, for the first 10 models in Table~\ref{tab:aggregateresults}, how many line segments each model was discretized into. All other models were input as line segments separated enough that Discretization did not change the segment count. Most of these 10 models ended with only a little more than one segment per control point, with the exception of the twisted knit tube and the chevron pattern.}
    Number of Discretized Segments for Various Inputs\\
    \begin{tabular}{|*{3}{c|}}
    \hline
    Model & $N_I$ & $N_D$ \\
    \hline
            \revision{Alien Sweater (Initial)} & 139506 & 158708 \\
         \hline
        \revision{Alien Sweater (Final)} & 139506 & 143228 \\
         \hline
        \revision{Sheep Sweater} & 729628 & 801447 \\
         \hline
        Sweater & 271902 & 296813 \\
         \hline
        Glove & 58537 & 65938 \\
         \hline
        Knit Tube \revision{(Initial)} & 18228 & 18941 \\
         \hline
        Knit Tube \revision{(Final)} & 18228 & 27625 \\
         \hline
        Chainmail \revision{(Initial)} & 211680 & 211680 \\
        \hline
        Chainmail \revision{(Final)} & 211680 & 212552 \\
        \hline
        Chevron $3\times3$ & 11741 & 32804 \\
        \hline
    \end{tabular}
    \label{tab:discretizationresults}
\end{table}

We now compare the many aforementioned methods on several benchmarks and applications. Please see the supplemental video for all animation results.

\subsection{Experiments and Verification}

We perform all our experiments on a single 18-core Intel\textsuperscript{\textregistered} Xeon\textsuperscript{\textregistered} CPU E5-2697 v4 @ 2.30GHz with 64GB of RAM, and the GPU versions are evaluated on an NVIDIA GeForce GTX 1080. In Table~\ref{tab:aggregateresults}, we evaluate the methods on inputs with varying numbers of loops, loop sizes, and loop shapes, and report runtimes and accuracies of the linkage values. For CPU Barnes--Hut, we used $\beta_\text{init}\eq 2$, $\beta_\text{max}\eq 10$, and $E_\text{target}\eq 0.2$. We chose $\beta_\text{init}$ so that the absolute errors were well below $10^{-2}$ for most examples. For Barnes--Hut on the GPU, since we only evaluated up to dipole moments with no error estimation, we used $\beta=4$ if $N_1 N_2 < 10^8$ and $\beta=16$ otherwise. We also included the runtime and accuracies achieved using direct summation, the FMMTL library \cite{cheng1999fast, cecka2015fmmtl}, and the modified count-crossings method from \cite{dey2013efficient} accelerated with parallelism and a BVH. \revisionp{The number of discretized line segments for each input is listed in Table~\ref{tab:discretizationresults}. In most cases, Discretization did not significantly increase the segment count (The exception is the chevron $3 \times 3$, which uses a special procedure defined in \S\ref{sec:openCurveVerification} where each curve participates in two virtual loops.).}

In general, the Counting Crossings (CC) and the two Barnes--Hut (BH) implementations give the fastest runtimes, with \revision{the CC, BH, and BHG all taking under 2 seconds} for all examples shown in the video (first 11 rows of Table~\ref{tab:aggregateresults}). \revision{As a result, we believe either a CC or a BH method is performant enough for most applications.}

The GPU implementations perform best when they have very few, or one, pairs of large loops to evaluate, and this can be seen in the ribbon, $\lambda=10^6$ torus, and low-frequency (low $\nu$) woundball. In particular, the smaller double-helix ribbons ($N_I$=\revision{200000}) examples illustrate that the GPU Direct Summation (DS) is a $370\times$ speedup over the CPU DS, and both DS implementations have very little error (mostly under \revision{$10^{-5}$} for the single-precision GPU implementation). The Barnes--Hut CPU implementation benefits from traversing two trees, a quadrupole expansion, as well as error estimation, allowing it to perform almost as well as the BH GPU on a wide range of examples that involve more than one loop pair. However, both Barnes--Hut implementations suffer when the segment lengths are relatively long (compared to the distance between curves), which is true in the higher-$\nu$ woundball (a higher $\nu$ has longer segments in order to traverse the ball more times), because their limited far-field expansion requires a relatively high $\beta$. This effect is worse for the BH GPU on the high-$\nu$ woundball, which is slower than the BH CPU here because our GPU version only implements monopole and dipole terms and thus requires an even higher $\beta$ to achieve the same accuracy. In contrast, the FMM excels on the high-$\nu$ woundball, giving the fastest performance, because it uses arbitrary-order expansion terms as needed. Both BH and FMM suffer from hard-to-predict absolute errors, especially in examples with high $\lambda$. In general, CC appears to perform the best in the most scenarios, and because it tabulates an integer, it has 0 error in the runs we observed.

\subsubsection{Barnes--Hut $\beta$ behavior:} 
\begin{figure}[t]
    \centering
    Average Barnes--Hut Error Versus $\beta$\\ \includegraphics[]{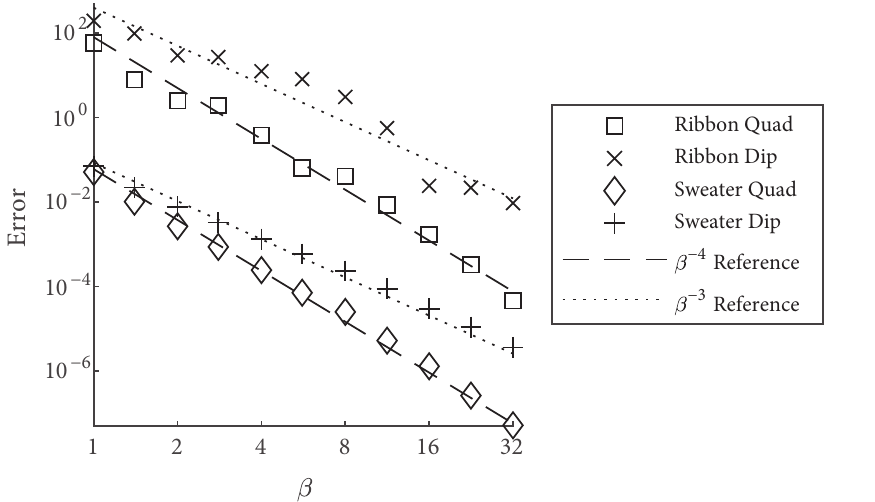}
    \caption{{\bf Barnes--Hut Accuracy Versus $\beta$:} We plot the (L1) average error versus the Barnes--Hut $\beta$ parameter for the double-helix DNA ribbon and the alien sweater examples. We also show this with only dipole (``Dip,'' first-order) terms versus having the quadrupole (``Quad,'' second-order) terms as well. \revision{The error appears to} drop off as $\beta^{-3}$ using the dipole expansion and as $\beta^{-4}$ for the quadrupole expansion.}
    \label{fig:betas}
\end{figure}

We illustrate how varying $\beta$ can impact accuracy in Figure~\ref{fig:betas}. \revisionp{\revision{The error appears to behave as $\beta^{-3}$ for the first-order (dipole) expansion, whereas for the second-order (quadrupole) expansion the error appears to behave as $\beta^{-4}$.}}

\subsubsection{Numerical conditioning and robustness:}\label{sec:resultsrobustness} For direct summation (and lowest-level evaluations of the tree algorithms), we use the expression from \citet{arai2013rigorous} (reproduced in \S\ref{sec:bgComputingLN} as \eqref{eq:directarctan}), because it is well defined over the widest range of input. In particular, it is only singular when the two line segments intersect or collinearly overlap, or if either has zero length. Our method already detects and flags these conditions in the Discretization step. Furthermore, the signed solid angle formula is widely used due to its simplicity and stability \cite{van1983solid}.

All methods compute using double precision when implemented on the CPU, while the GPU implementations use single precision.

To learn more about the robustness of the projection in the count-crossings (CC) method, we tested it, using random initial frames before the rotation step, 50,000 times on the compressed chainmail \revision{(final)}, and 5,000 times on the alien sweater \revision{(initial)}, and, based on their $P$ values in Table~\ref{tab:aggregateresults}, this sums up to 3.08 billion projections. \revisionp{\revision{If our implementation encounters a degeneracy, the method would fail and exit.}} For each loop pair, we recomputed the linking number with a new projection every trial, and it completed with the correct result every time. We also tested it with very large inputs such as the dense tori and woundballs (Fig.~\ref{fig:stresstests}), which span a large range of angles, and the linking number was successfully computed, albeit the crossings computation was very slow for the unlinked $\lambda=0$ torus. This robustness is likely because degeneracies are unlikely to appear with this input size in double precision. That said, we do not guarantee with our implementation that it is able to find a non-degenerate, regular projection on every input. If this is a concern, the Gauss summation implementations such as the Barnes--Hut algorithm, which do not require a projection, can also be used on most inputs with comparable speed.

\subsection{Performance}
\begin{figure}
    \centering
    Linking Number Computation Runtimes vs. Discretized Model Size\\
    \includegraphics[]{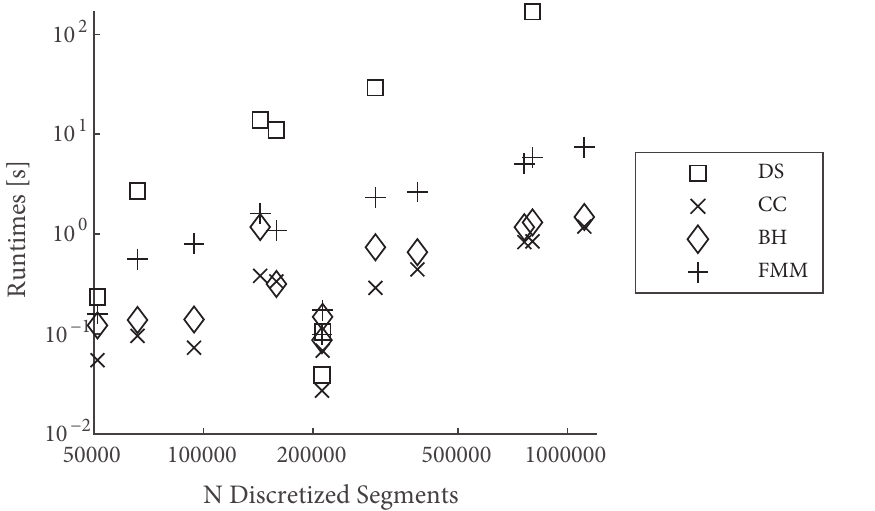}
    \caption{{\bf Runtimes for Linking Number Computation:} This plots the runtime in ms against the discretized model size, for the examples shown in the video (first 11 rows of Table~\ref{tab:aggregateresults} plus a few reparameterized models). Because the models vary in shape complexity, number of loops, and number of segments per loop, the input size cannot be used to perfectly predict the runtime.}
    \label{fig:runtimes}
\end{figure}

For all results shown in the video (first 11 rows of Table~\ref{tab:aggregateresults}), PLS and Discretization took under 200~ms combined. For the third stage, we plot the runtimes versus discretized model size in Figure~\ref{fig:runtimes}. Both the count-crossings and Barnes--Hut methods significantly outperform direct summation, and also outperform FMM in most cases. We also tested four classes of large input as stress tests; see Fig.~\ref{fig:stresstests} for an illustration.

\subsubsection{Usual-case theoretical estimates:} Similar to many tree-based methods, the performance depends on the distribution of input, and certain inputs can severely degrade performance. For inputs with relatively ``short'' segments (i.e. segments are short compared to the spacing between curves) and uniformly sized loops, we expect PLS to take $O(L \log L)$ time and Discretization to take $O(N \log N )$ time, where $L$ is the number of loops and $N$ is the number of input segments. The Barnes--Hut, Count-Crossings, and FMM \revisionp{computations} are all expected to take $O(P N_L \log N_L)$ time, \revisionp{where $P$ is the number of potentially linked pairs and $N_L$ approximates the number of segments per loop} (FMM is not linear in $N_L$ because we compute the finite-segment correction, although for all the inputs we reported in Table~\ref{tab:aggregateresults}, the finite-segment correction took under $\nicefrac{1}{8}$ of the FMM runtime). For chainmail, the total simplifies to $O(C \log C)$ where $C$ is the number of rings, and for most stitch-mesh knit patterns \revisionp{where each row interacts with a small number of other rows}, this simplifies to $O(R \log R + N \log N_L)$ where $R$ is the number of rows.

However, if segments are extremely long compared to the distance between curves, which can happen after a simulation has destabilized, every algorithm can slow down to $O(N^2)$ performance to finish. This is where the early-exit approach can be a huge gain; furthermore, if our method runs periodically during a simulation, topology violations can be detected long before this occurs, as we will show in the Knit Tube Simulation (Figure~\ref{fig:tubesim}).

\subsection{Applications}
In this section we demonstrate the versatility of our method across several applications. In most scenarios (except Example~\ref{sec:cleanupinputyarn}) we first compute the linking matrix of the initial model, and then compare it against the linking matrices of the deformed models.

\subsubsection{Yarn model analysis:} \label{sec:cleanupinputyarn}
\begin{figure}
    \centering
    \includegraphics[width=\hsize]{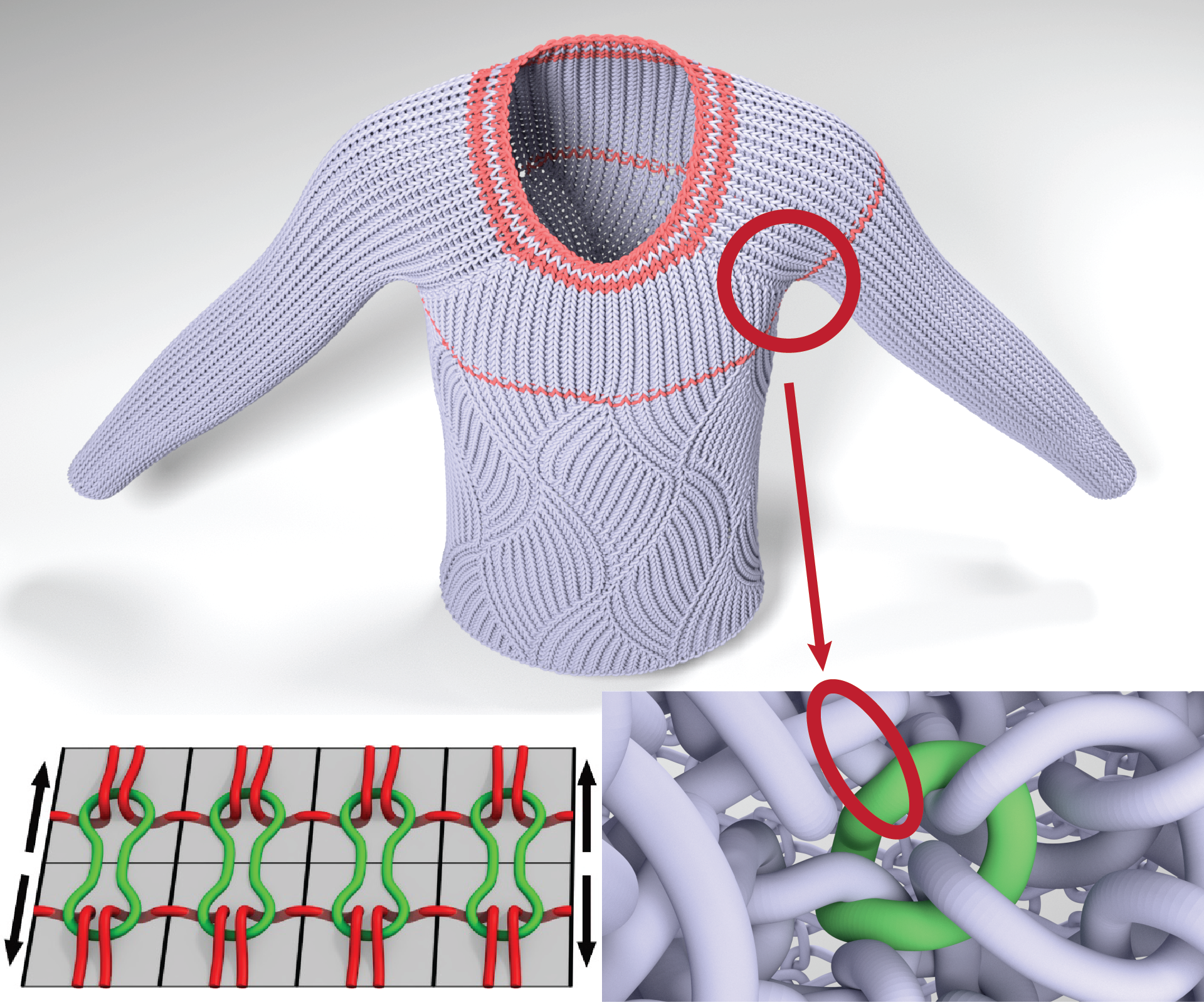}
    \caption{\revision{\bf Yarn Model Analysis}: We examine the ``sweater flame ribbing'' yarn model from Stitch Mesh \cite{yuksel2012stitch, wu2017real}, which was provided in relaxed form. This model contains small yarn loops to join rows with mismatched wale direction (Bottom Left, reproduced from Figure 8 in \cite{yuksel2012stitch}). Even with these loops, the entire stitch mesh should have zero pairwise linkage between its yarn rows. We verify the relaxed yarn model using our algorithm, and mark linkages in red (Top). From this, we zoom in on one of the linkages, and find that the small green loop (Bottom Right), which should topologically match the small green loops in the bottom left figure, has only 3, rather than 4, curve segments passing through it, indicating there was yarn pull-through during relaxation.}
    \label{fig:inputyarnanalysis}
\end{figure}

Many yarn-level models in graphics and animation are represented as a set of closed curves; a major example is the models generated from StitchMesh \cite{yuksel2012stitch}. Implicit (and/or implicit constraint direction) yarn simulations solve smaller systems when the model uses closed curves, and the model stays intact without the requirement of pins. Our method can aid in the topology validation of these models, by computing a linking matrix certificate before and after a deformation, whether the deformation is a simulation, relaxation, reparameterization, or compression. In addition, we can examine the integrity of input models from prior work. In our examples we analyze four models from \cite{yuksel2012stitch,wu2017real}\footnote{\revision{Models downloaded from \url{http://www.cemyuksel.com/research/yarnmodels}}}: the ``alien sweater,'' ``sweater flame ribbing,'' ``glove,'' and the ``sheep sweater,'' which are provided after a first-order implicit yarn-level relaxation. Stitches between rows introduce zero pairwise linkage, and our method verifies whether the relaxed models have valid stitches. We found that the ``glove'' and the ``alien sweater'' have no pairwise linkages, indicating no obvious issues, while the ``sheep sweater'' and ``sweater flame ribbing'' contain linkage violations. See Figure~\ref{fig:inputyarnanalysis} for a violation in the ``sweater flame ribbing'' model.

\subsubsection{Reparameterizing and compressing \revision{yarn-level} models:}\label{sec:results_reparam} 
\begin{figure}
    \centering
    \includegraphics[width=\hsize]{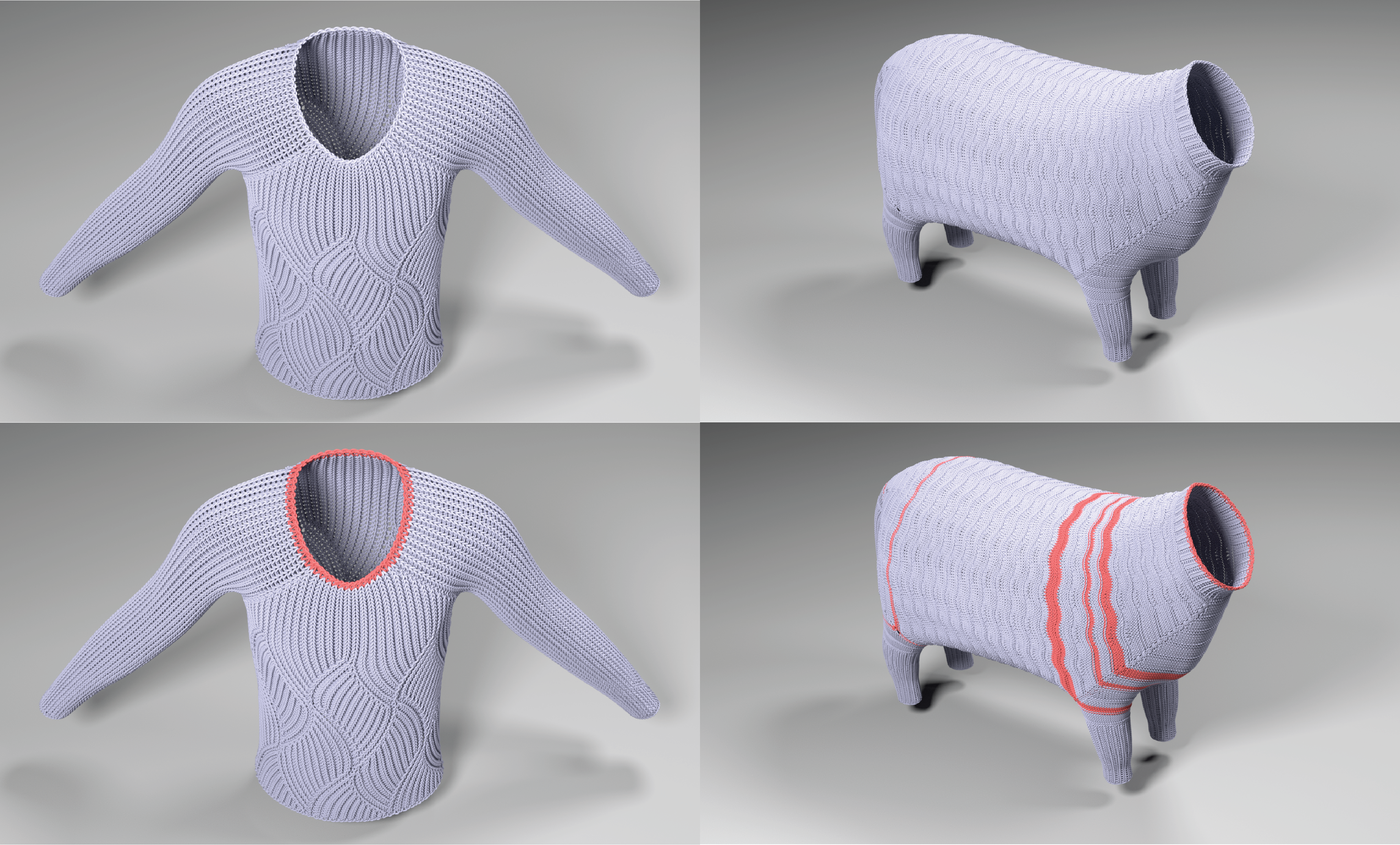}
    \begin{tabular}{cc}
      Compression & \hspace{1.5cm} Reparameterization
    \end{tabular}
    \caption{{\bf Compression and Reparameterization of Detailed Yarn Models:} (Left) We compress a sweater model down to 16-bit (Top Left) and 12-bit (Bottom Left) precision by quantizing coordinates. Our algorithm verifies that the 16-bit compression is valid, but highlights the loops that contain topology violations in the 12-bit compression. (Right) If we aggresively resample the Catmull--Rom splines of the sheep sweater model to reduce the number of control points from 729628 (Top Right) to 547264 (Bottom Right), our method highlights violations in the latter.}
    \label{fig:compressionandreparam}
\end{figure}

Our methods can be used to verify the topology of yarn-level cloth models following common processing operations. Spline reparameterization can be used to reduce the number of spline control points prior to simulation, or afterwards for model export. Furthermore, it is common to compress control point coordinates using quantization for storage and transmission. Unfortunately such operations can introduce topological errors, which limits the amount of reduction that can occur. Our method can be used to verify these reparameterization and compression processes. Since large modifications of the model may introduce multiple linking errors that could cancel each other out, we leverage our methods' speed to analyze a sweep of the reparameterization/compression process to better detect errors. See Figure~\ref{fig:compressionandreparam} for results.

\subsubsection{Embedded deformation of yarn-level models:} It is computationally appealing to animate and pose yarn-level models using cheaper deformers than expensive yarn-level physics, but also retain yarn topology for visual and physical merits. To this end, we animated the yarn-level glove model using embedded deformation based on a tetrahedral mesh simulated using Houdini 18.5 Vellum. For the modest deformations shown in Figure~\ref{fig:embeddedDefoGlove} the yarn-level model retains the correct topology throughout the animation, and therefore can be used for subsequent yarn-level processing.

\newcommand{\glovy}[1]{\includegraphics[width=0.33\hsize]{#1} \hspace{-4mm} } 
\begin{figure}
  \centering
  \begin{tabular}{ccc}
    \hspace{-3mm}
    \glovy{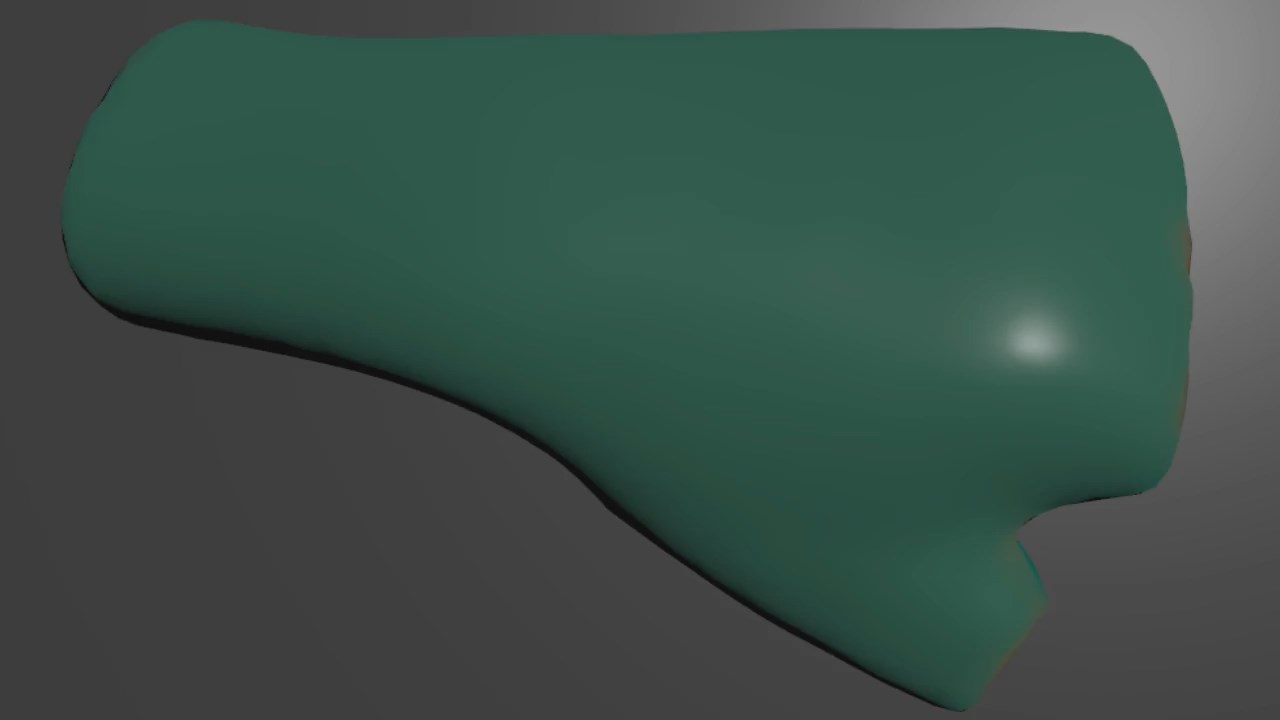} & 
    \glovy{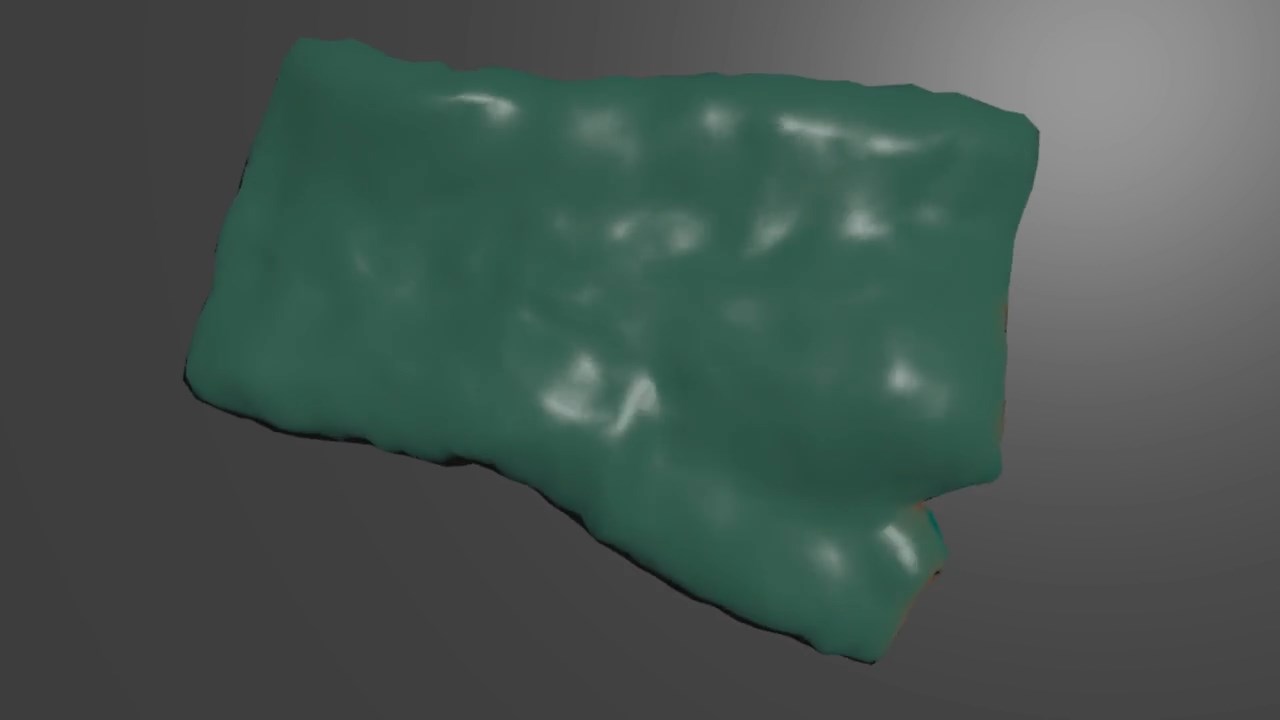} & 
    \glovy{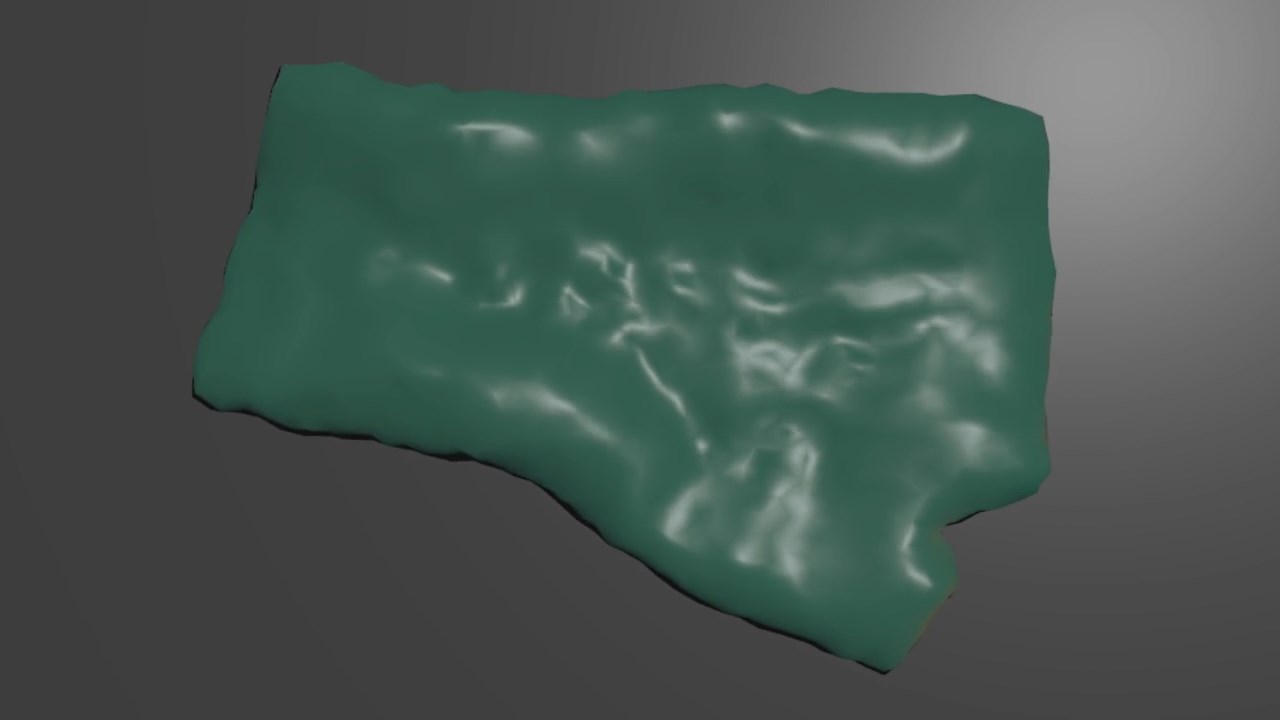} \\
    \hspace{-3mm}
    \glovy{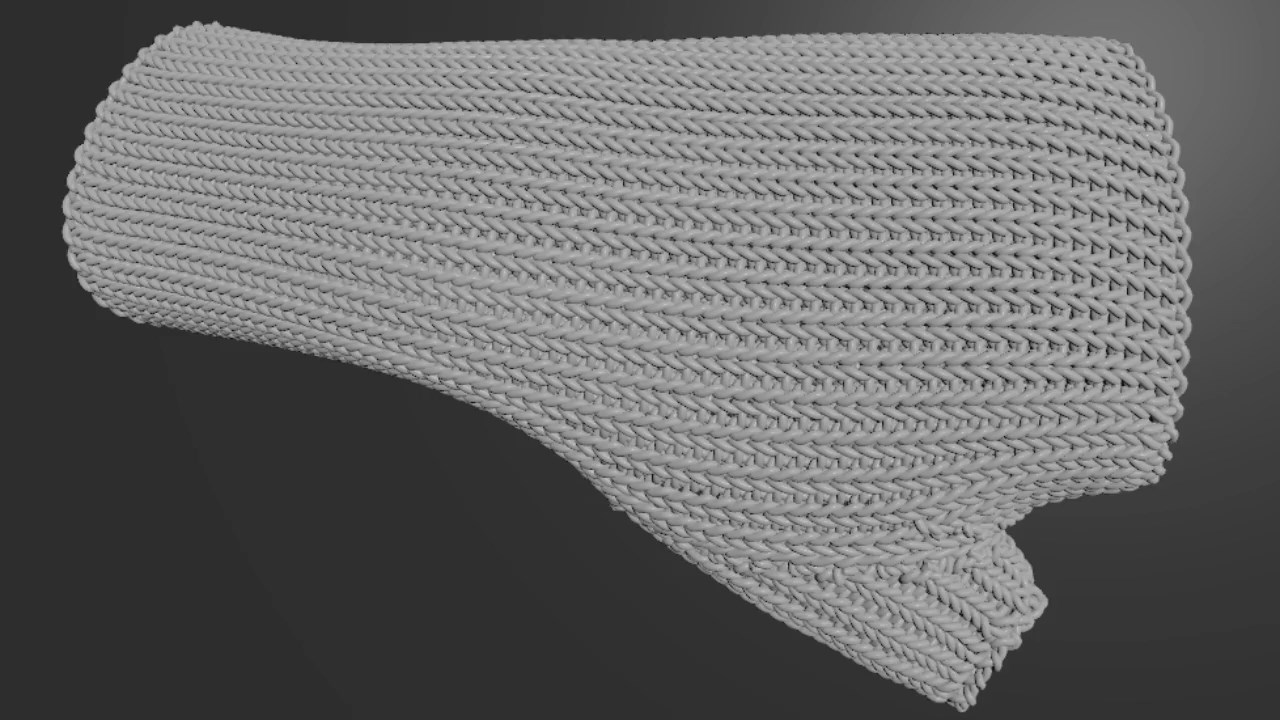} & 
    \glovy{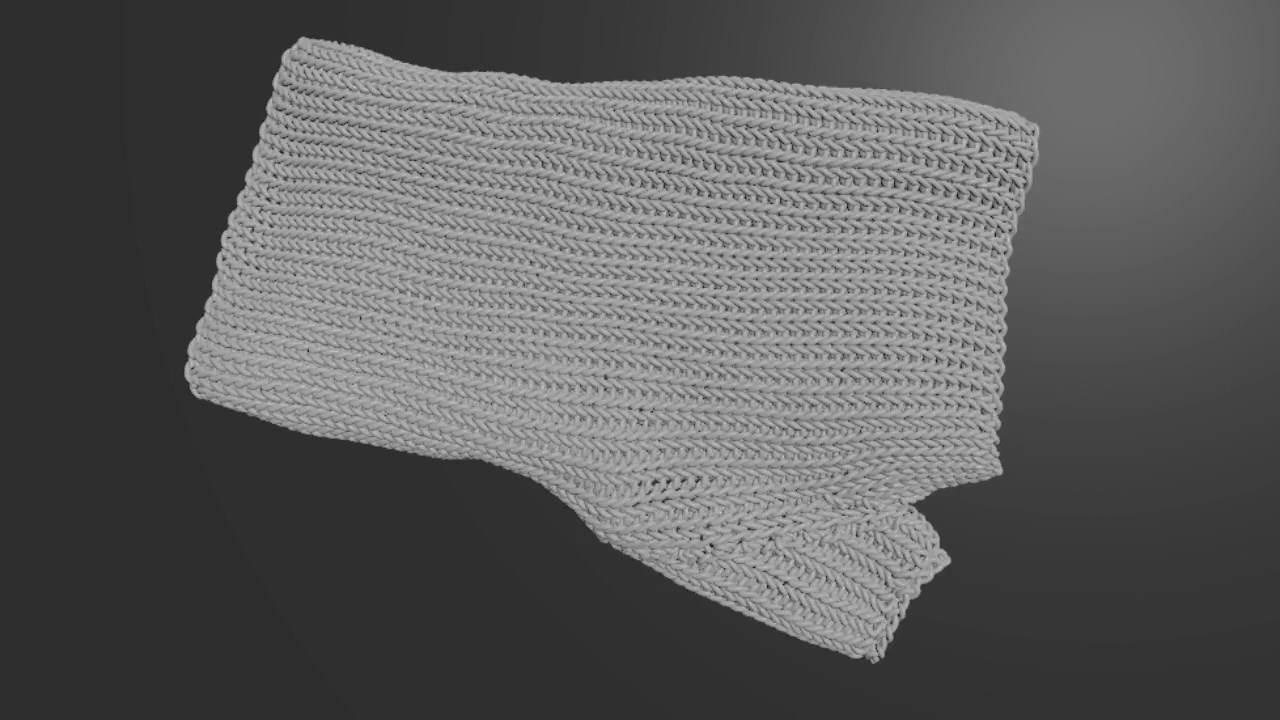} & 
    \glovy{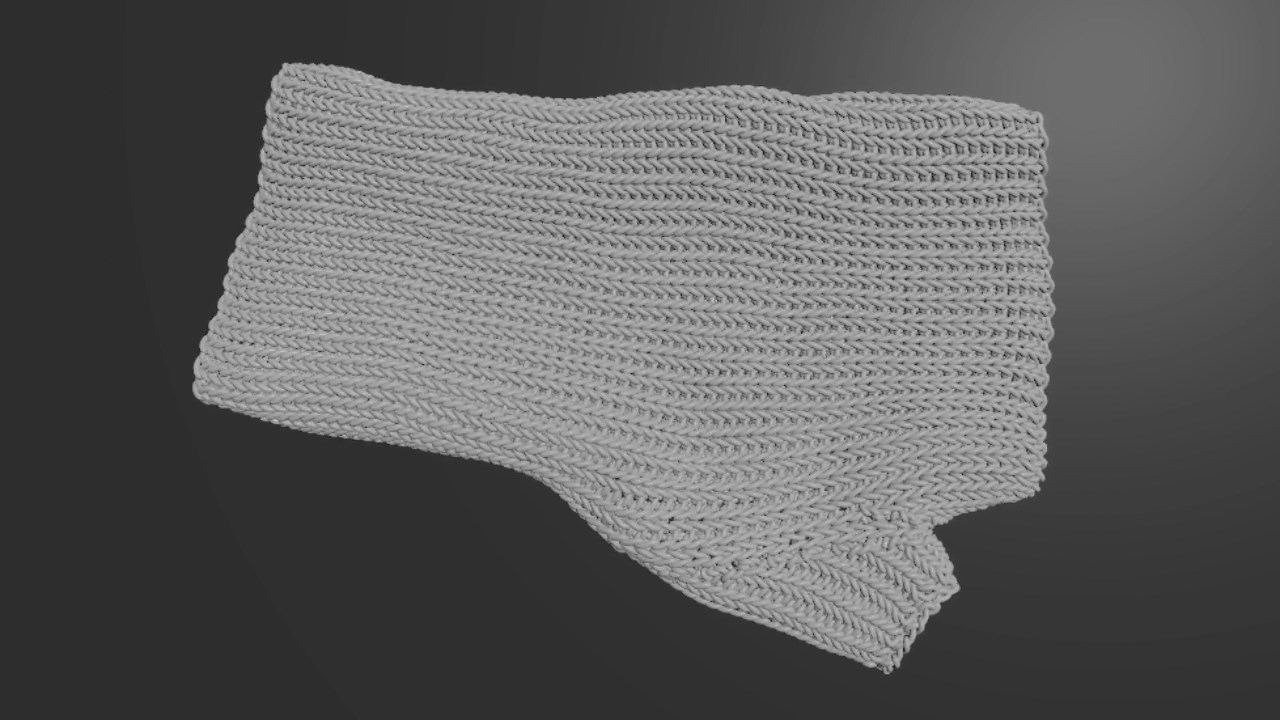}
  \end{tabular}
  \caption{{\bf Embedded Deformation of Yarn-Level Models:} A coarse single-layer
    tetrahedral mesh (simulated by Houdini's Vellum solver) is used to place a yarn-level
    glove model on a surface using embedded deformation.  We verify that no topology
    changes are introduced by this deformation.
    \label{fig:embeddedDefoGlove}}
\end{figure}  

\begin{figure*}
    \centering
    \includegraphics[width=0.5\hsize]{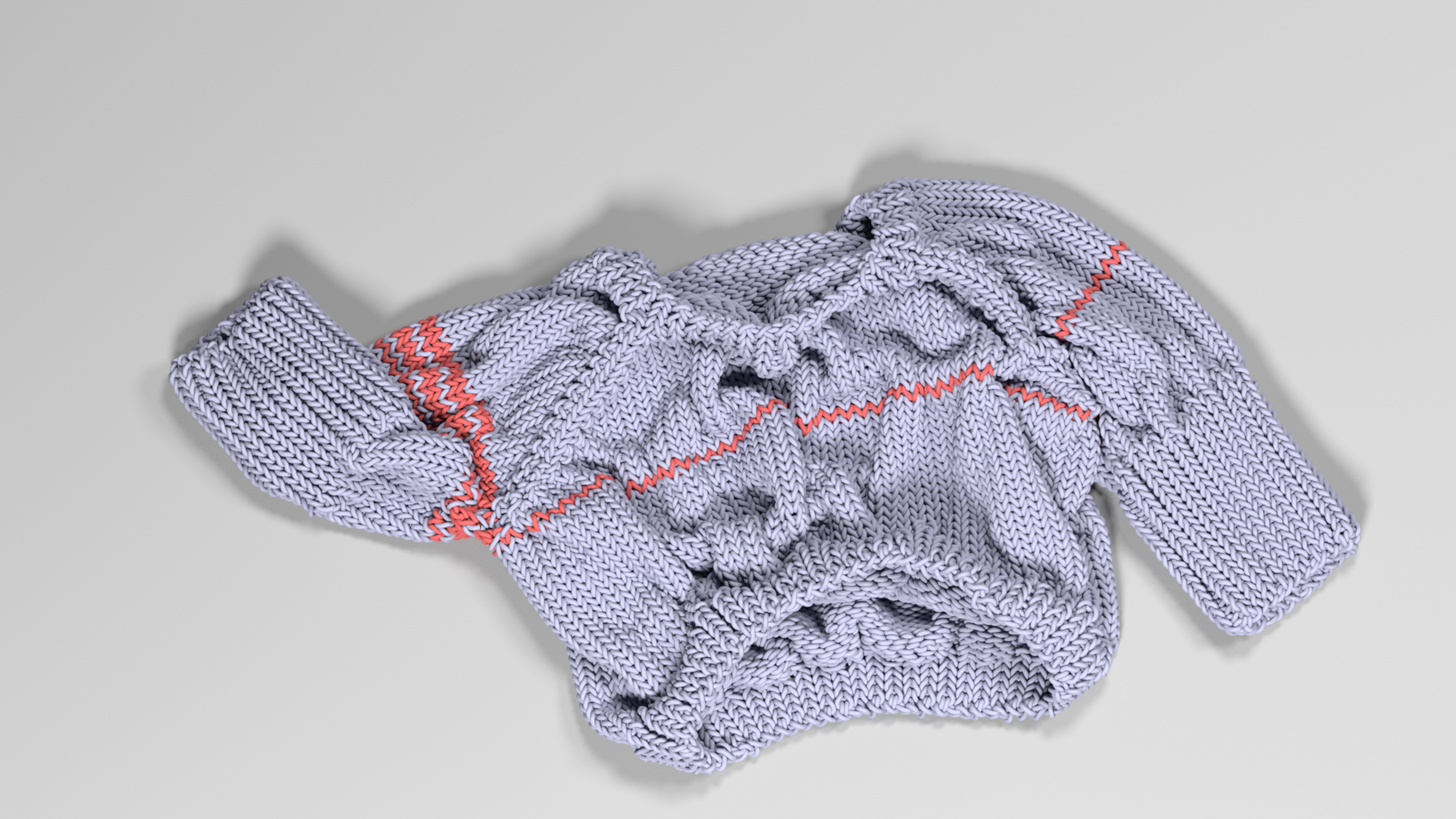}\includegraphics[width=0.5\hsize]{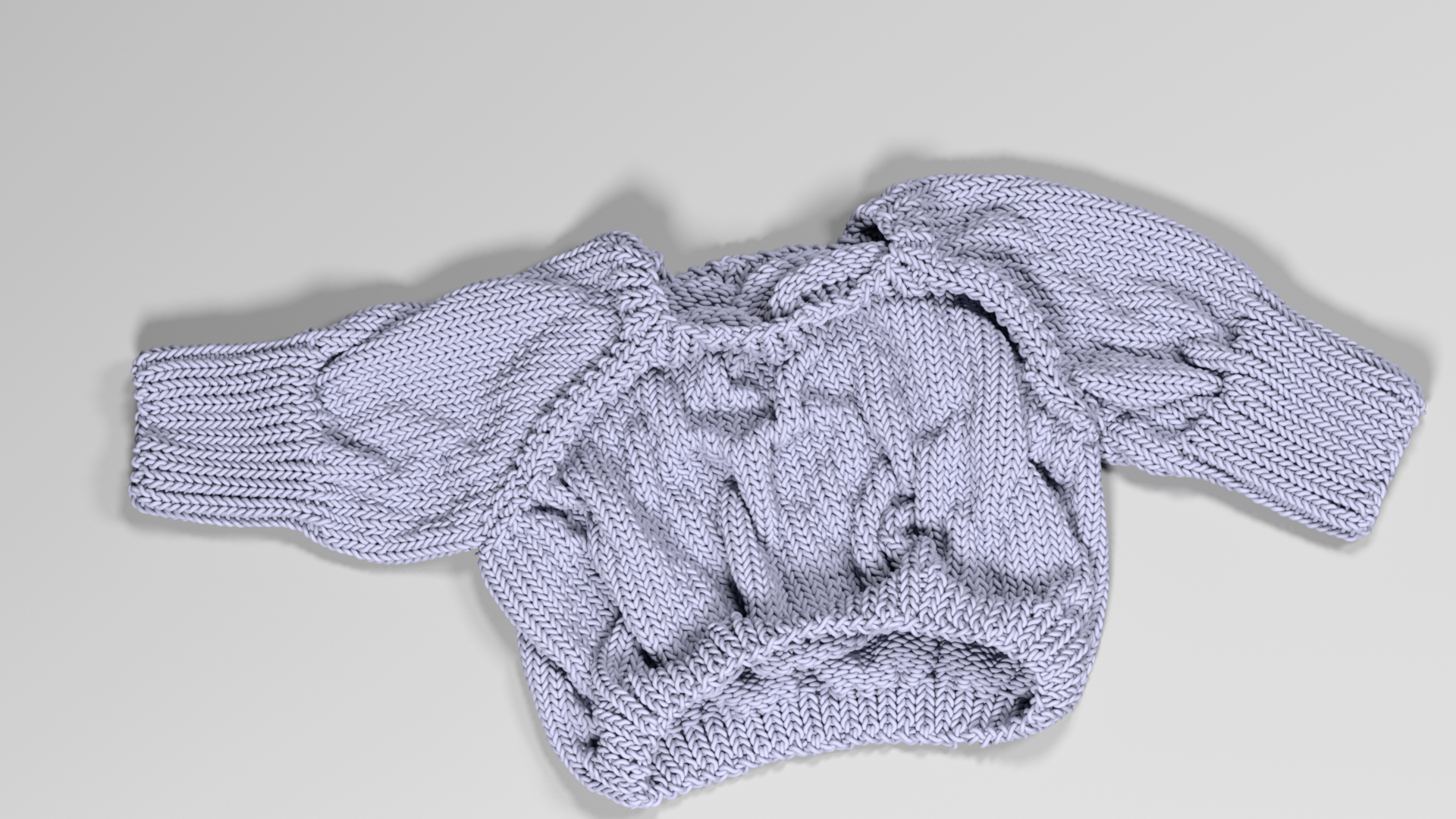}
    \caption{{\bf Verification of Sweater Drop Simulation:} Our method detected that (Left) a preview implicit simulation with larger step sizes failed to preserve yarn topology, while (Right) another simulation with smaller steps maintained it.}
    \label{fig:sweatersim}
\end{figure*}

\begin{figure}
    \centering
    \includegraphics[width=\hsize]{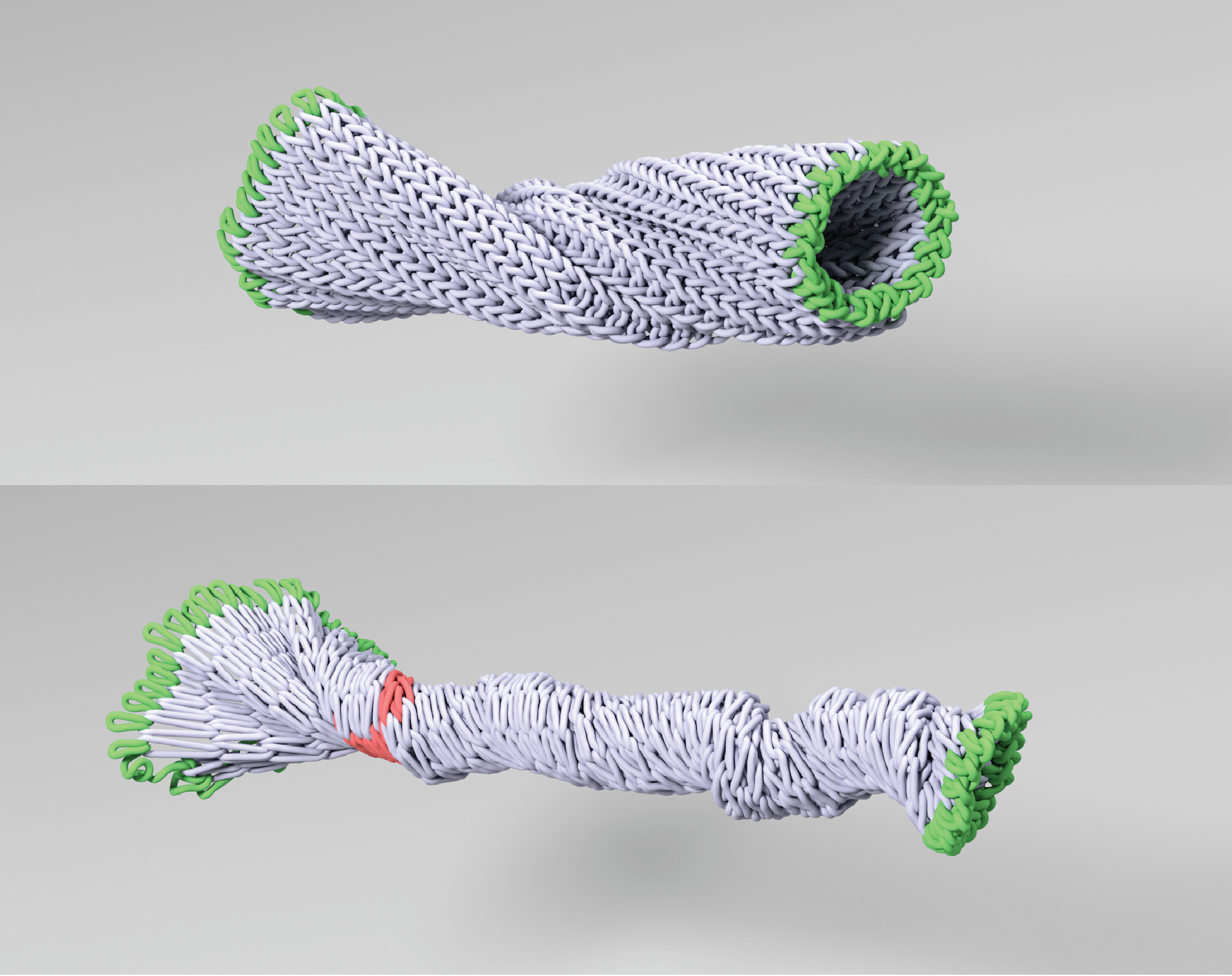}
    \caption{{\bf Verification of Tube Twist Simulation:} Twisting cloth is a common stress test for a simulator, and in this image, the end rows (green) are twisted using a stiff spring until breakage. Our method is able to cheaply detect topology violations in the model (red) over a thousand steps before the simulation visually explodes.}
    \label{fig:tubesim}
\end{figure}

\subsubsection{Implicit yarn-level simulation:} Simulation of yarn, as in \cite{kaldor2008simulating}, is challenging because of the large number of thin and stiff yarns in contact, and the strict requirement to preserve topology (by avoiding yarn--yarn ``pull through''), both of which can necessitate small time steps. For large models, these relaxations or simulations can take hours or days, and there may not be an expert user in the loop to visually monitor the simulations everywhere. While sometimes penalty-based contact forces will cause the simulation to blow up, alerting the user to restart, at other times pull-through failures can be silent and lead to incorrect results.

Our method is useful for validating yarn models during yarn-level relaxation, when large forces can cause topological changes, as well as large-step, preview implicit simulations of yarn-level models. We use implicit integration \cite{baraff1998large, kaldor2008simulating, leaf2018interactive, bergou2010discrete} to step through the stiff forces, which enables larger timesteps at the risk of introducing topology errors; our PCG implementation uses Eigen \cite{eigenweb} and falls back to the SYM-ILDL package \cite{greif15symildl} when the energy gets too nonconvex \cite{kim2020finite}. In Figure~\ref{fig:sweatersim}, we simulate the alien sweater model from \cite{yuksel2012stitch} using larger implicit steps (1/2400 s) for a preview simulation as well as smaller implicit--explicit step sizes (1/6600 s). Our method detects and marks the yarn loops that have engaged in pull through violations in the former, and validates that the latter maintains its topology. In Figure~\ref{fig:tubesim}, we perform a common stress test where we grab a knitted cylindrical tube by the ends and wring (twist) the ends in opposing directions until the simulation breaks. Even in this simulation with an obvious failure, our method is still useful because it detects the first linkage violations over a thousand steps before the simulation destabilizes. In all of these simulations, the runtime of our method is small compared to the runtime of the simulation: we only run our method once every animation frame, whereas the simulation takes dozens of substeps per frame.

\subsubsection{Open-curve verification: chevron $3\times3$ stitch pattern relaxation:}
\label{sec:openCurveVerification} Our method can be applied not only on closed loopy structures, but also on finite, open curves in specific circumstances, to detect topology violations between different curves. In particular, this notion is well defined when the curves form a braid, that is, when the curves connect fixed points on two rigid ends, and the rigid ends remain separated. This can be useful for verifying the relaxation of a yarn stitch pattern, if the top and bottom rows are either cast on or bound off, and the ends of each row are pinned to two rigid ends, turning the stitch pattern into a braid.

\begin{figure}
    \centering
    \includegraphics[width=\hsize]{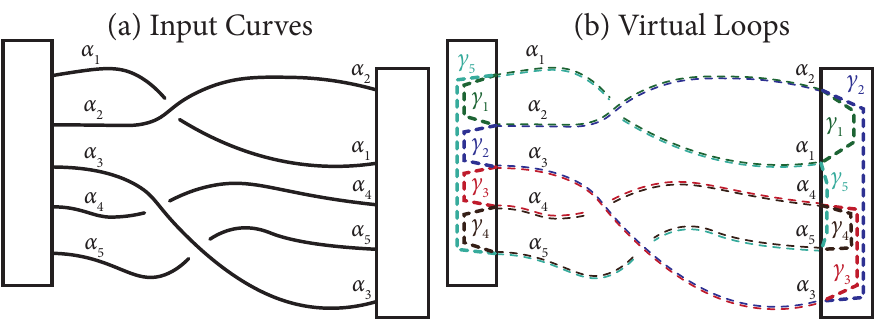}
    \caption{{\bf Verification of Open Curves Attached to Rigid Ends:} We can automatically validate curve topology for open curves with ends that are fixed with respect to each other. In this illustration, when we have five open curves, $\alpha_1$ to $\alpha_5$, we can form five virtual closed loops, $\gamma_1$ to $\gamma_5$ by attaching every two curves into a loop, with a virtual connection inside each end volume: $\gamma_i \eq \alpha_i \cup \alpha_{i+1}$ for $i \eq 1,2,3,4$ and $\gamma_5 \eq \alpha_5 \cup \alpha_1$. As every curve participates in two loops, we modify PLS to exclude pairs of loops that share curves, and then generate the certificate. During deformation, as long as no curve crosses the virtual end connections and no virtual connection crosses each other (this is guaranteed by rigid ends), the certificate will detect pull throughs. See the chevron stitch pattern (Fig.~\ref{fig:chevron}), for example.}
    \label{fig:chevronillustration}
\end{figure}

\begin{figure}
    \centering
    \includegraphics[width=\hsize]{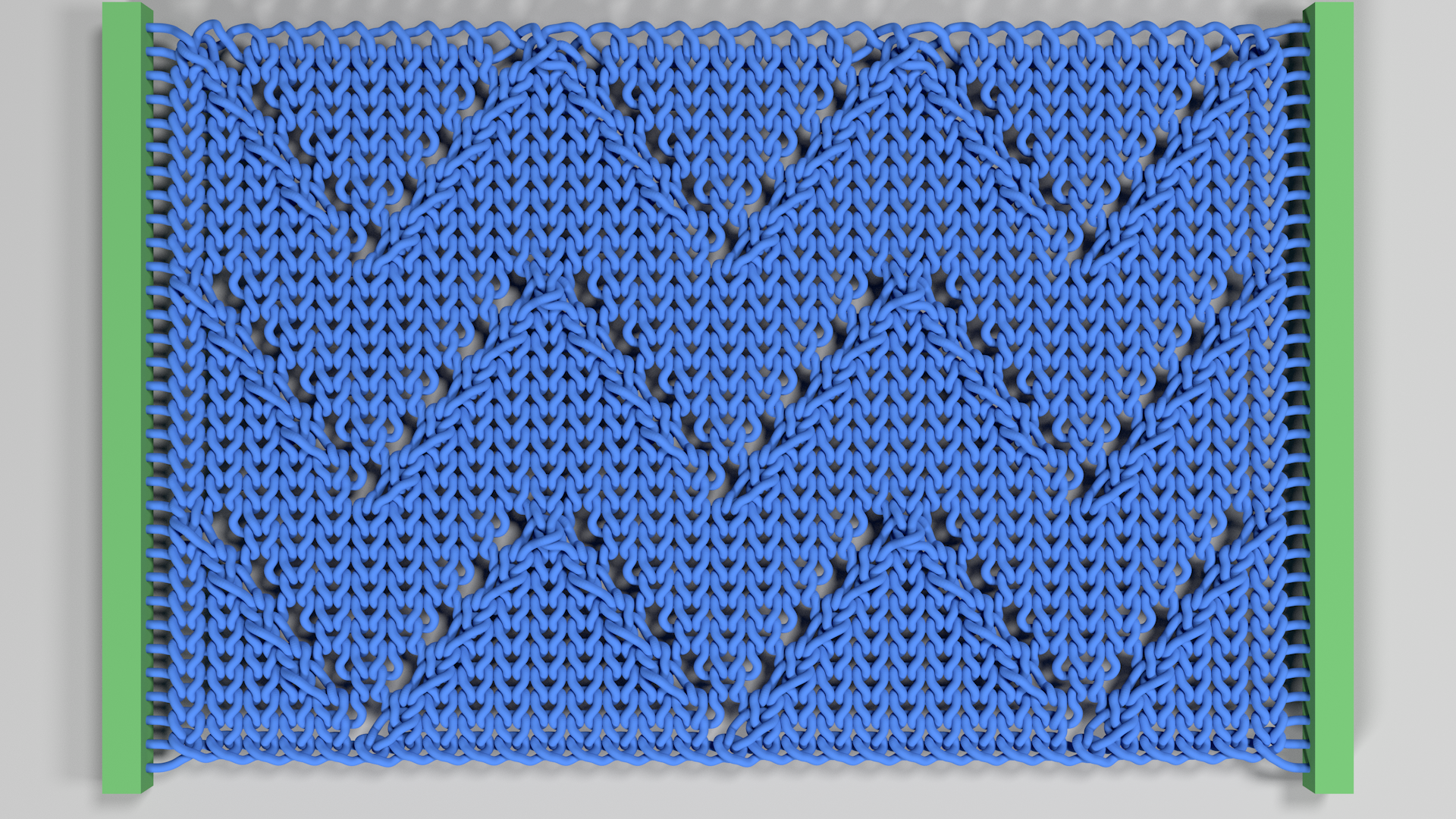}
    \includegraphics[width=\hsize]{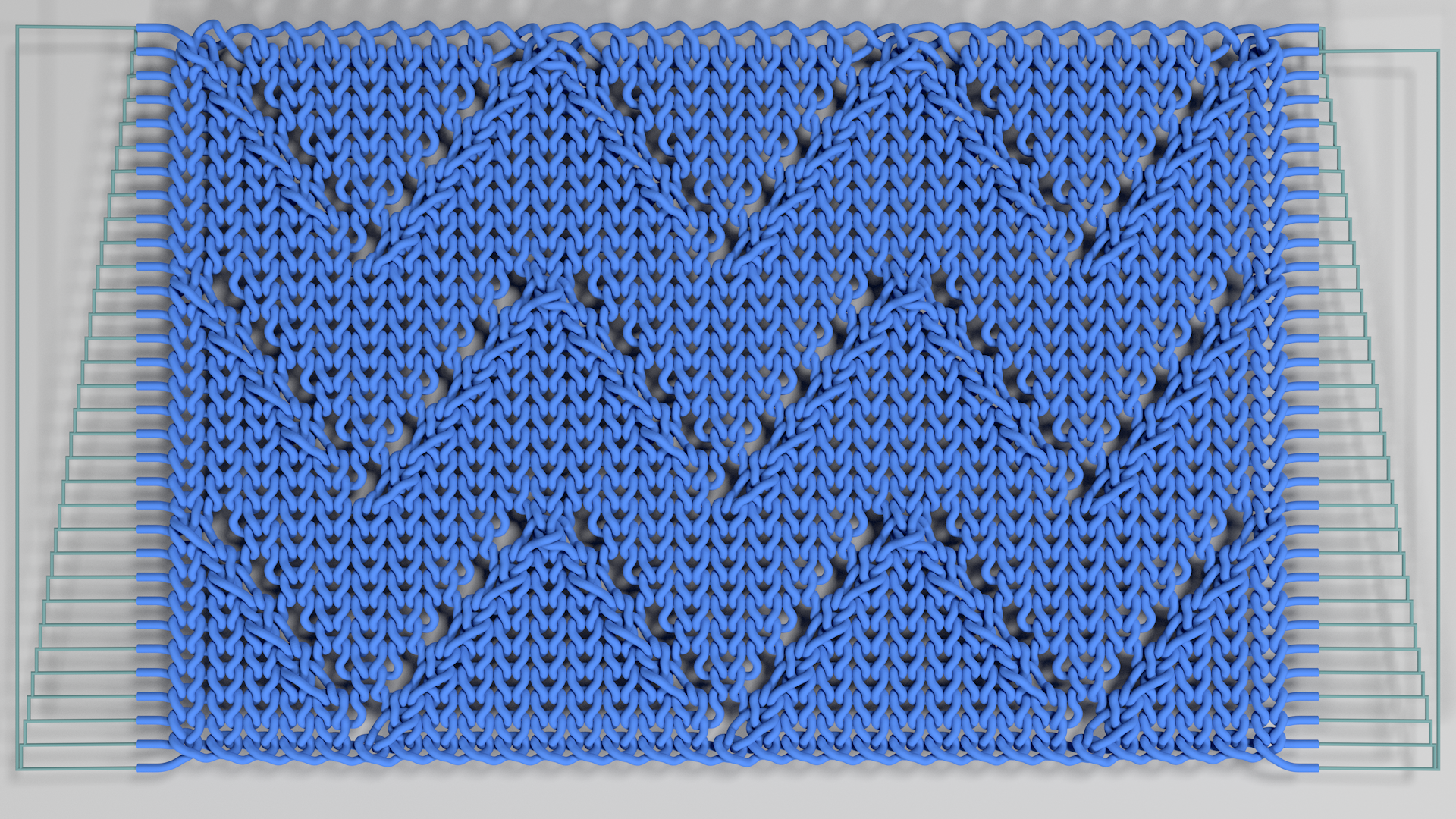}
    \caption{{\bf Verification of Chevron Stitch Pattern During Relaxation:} (Top) We attach a $3\times3$ chevron stitch pattern between two rigid bars and relax the pattern while slowly pulling the bars apart. Using our procedure to verify open curves attached to rigid ends, we verify that the relaxation preserves stitch topology. (Bottom) In this image, the virtual connections are shown, exaggerated outside their volumes, as green wires extending to the left and right. See the supplemental video for an alternate relaxation with an early pull through detected in the process. The violation there is detected in the first few steps, while the pull through only becomes visually apparent after hundreds of steps.}
    \label{fig:chevron}
\end{figure}

When the curves form a braid, we can connect them using virtual connections to form virtual closed loops (see Figure~\ref{fig:chevronillustration}). We propose a method to automatically form virtual connections which we include in Appendix~\ref{app:openCurveVCAlg}. To test this, we use a stitch pattern that repeats a chevron three times horizontally and vertically ($3 \times 3$). We apply our proposed method to the chevron $3 \times 3$ to verify the relaxation of the stitch pattern (see Figure~\ref{fig:chevron}). In the video, we also detect an early pull through in another relaxation with larger steps, hundreds of steps before it becomes visually apparent.

\subsubsection{Chainmail simulation:}
In Figure~\ref{fig:teaser}, we simulated a fitted Kusari 4-in-1 chainmail garment with 14112 curved rings and 18752 links.  Inspired by the 3D printing example from \cite{tasora2016chrono, mazhar2016ontheuse}, we simulated packing the garment into a small box for 3D printing using the Bullet rigid-body solver in Houdini 18.5. Without exceptionally small timesteps, the solver destroys links and creates spurious others as shown in Figure~\ref{fig:teaser} and supplemental animations. By using varying numbers of substeps, we could illustrate linkage violations in Figure~\ref{fig:teaser} (b) and (c), and certify that the result in (d) maintains its topological integrity, which would be otherwise difficult to ascertain until after expensive 3D printing.

\subsubsection{Rubber band simulation:} Errors in collision and contact handling can result in previously unlinked loops becoming entangled, and thus necessitate more accurate but expensive simulation parameters. We demonstrate this scenario by the simulation of 1024 unlinked rubber bands dropped onto the ground in Figure~\ref{fig:rubberBands}.  Our methods can be used to monitor and detect topological errors introduced by low-quality simulation settings (too few substeps), and inform practitioners when to increase simulation fidelity.

\newcommand{\band}[1]{\includegraphics[width=0.255\hsize]{#1} \hspace{-6mm} }
\begin{figure*}
  \begin{center}
    \begin{tabular}{cccc}
      \hspace{-4mm}
      \band{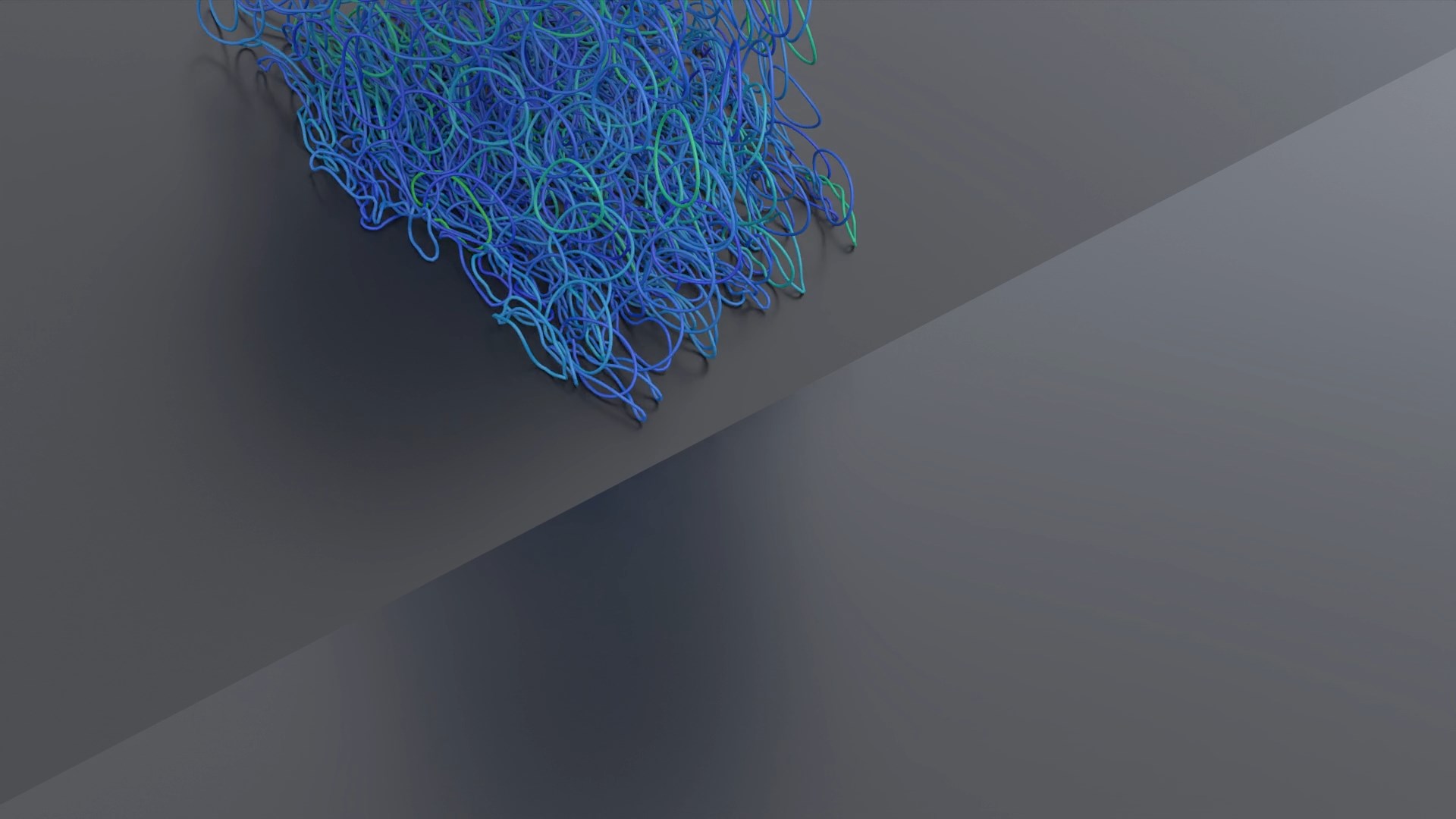} & 
      \band{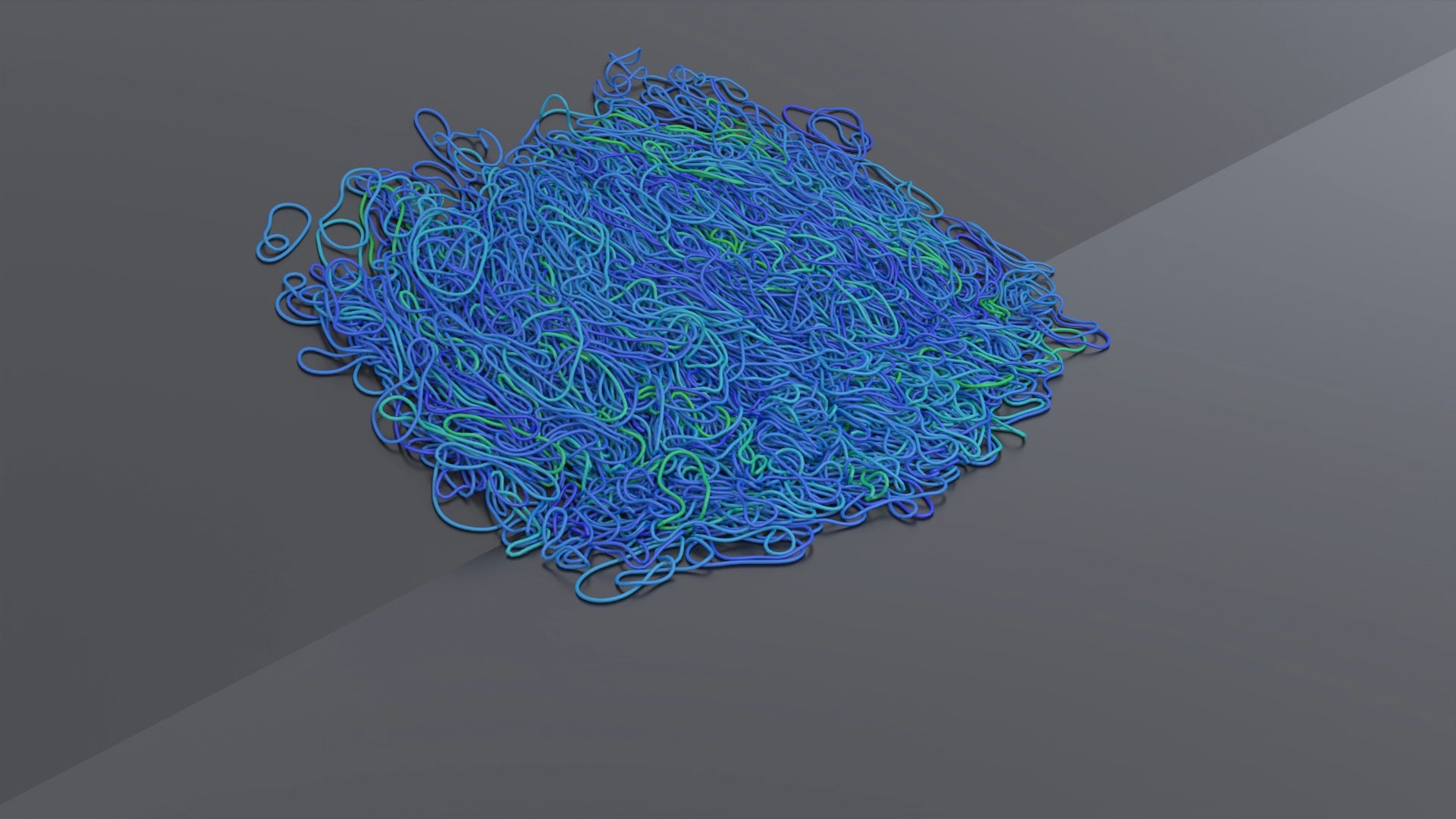} & 
      \band{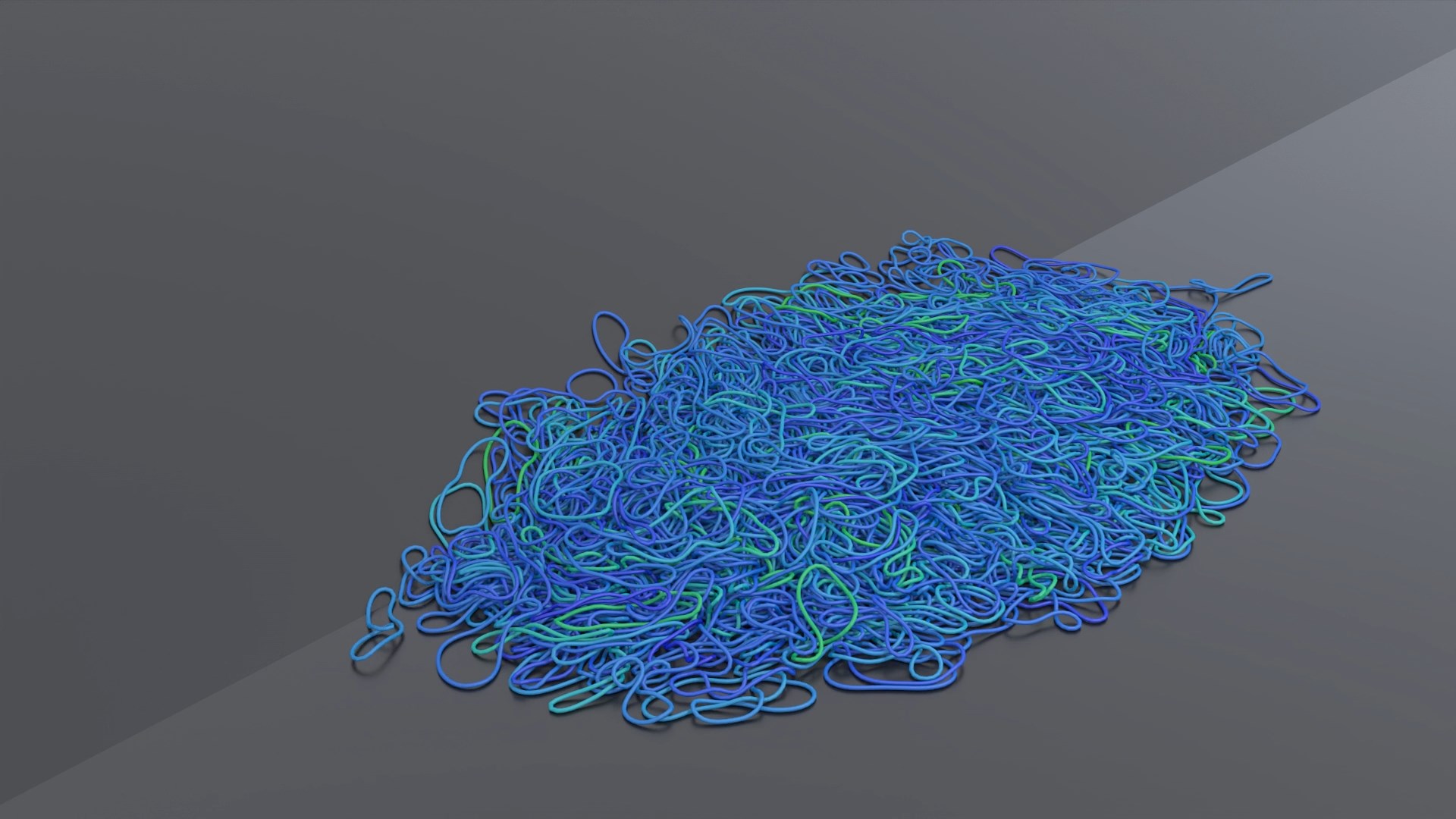} &
      \band{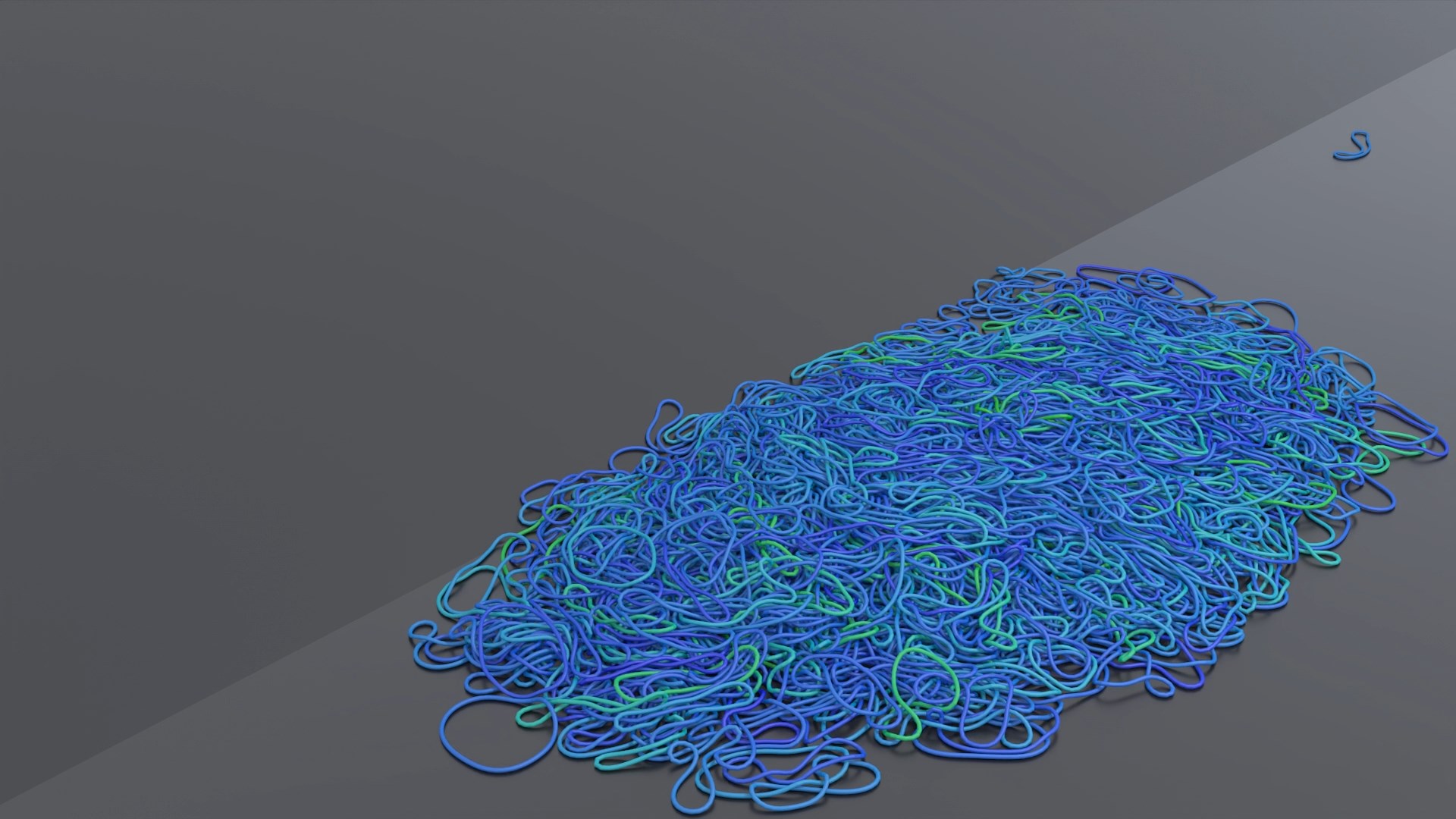} \\
      \hspace{-4mm}
      \band{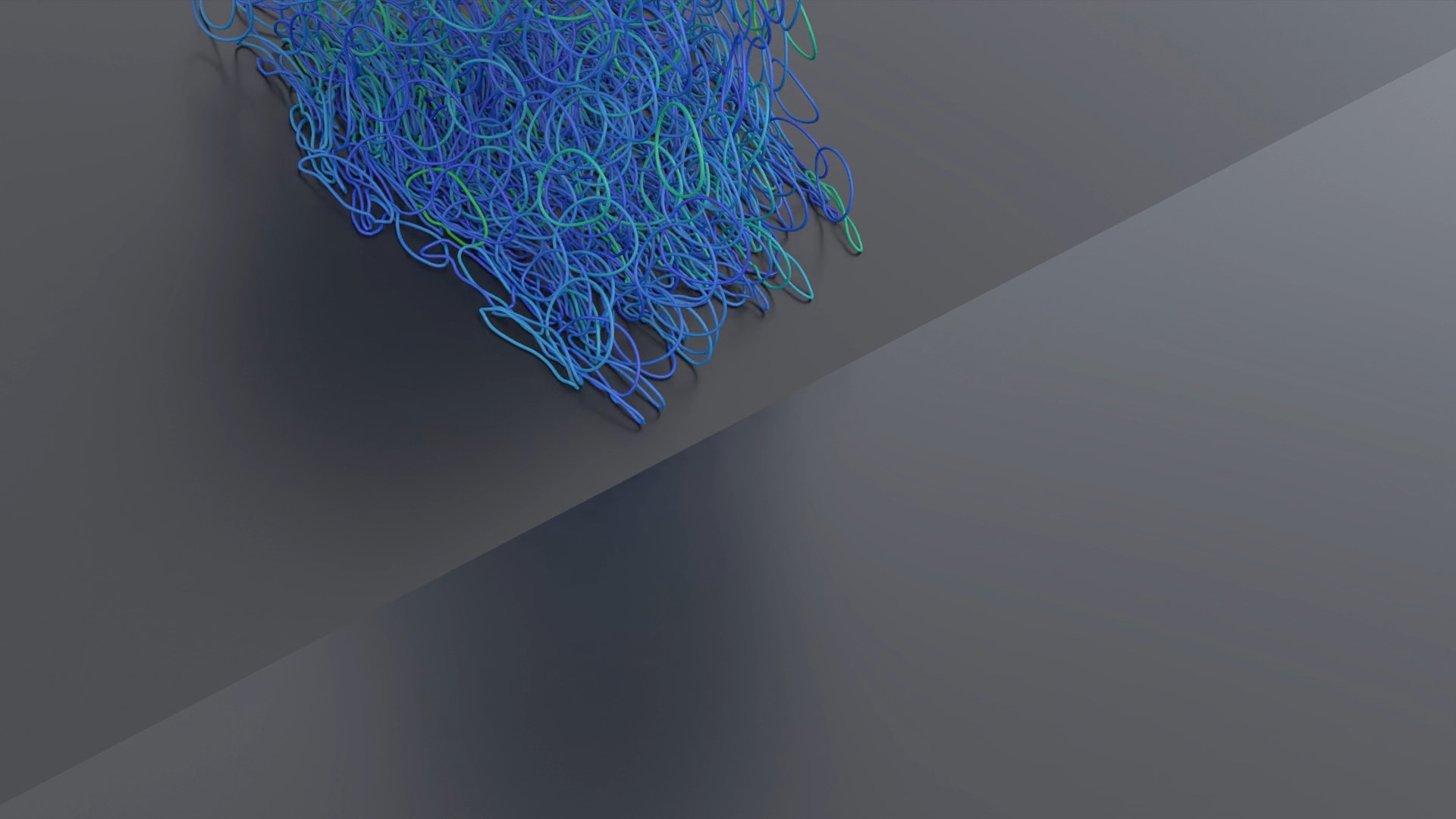} & 
      \band{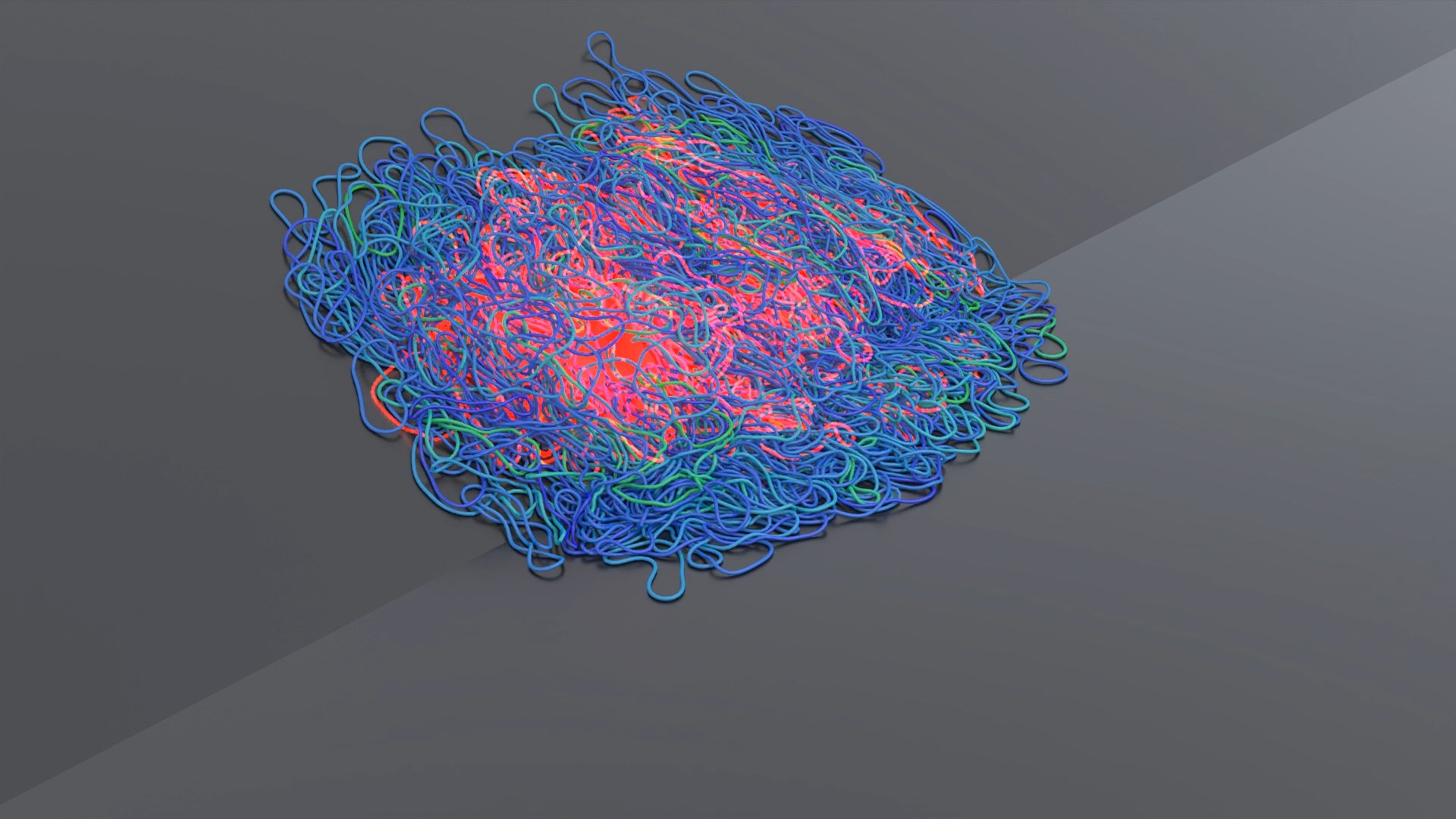} & 
      \band{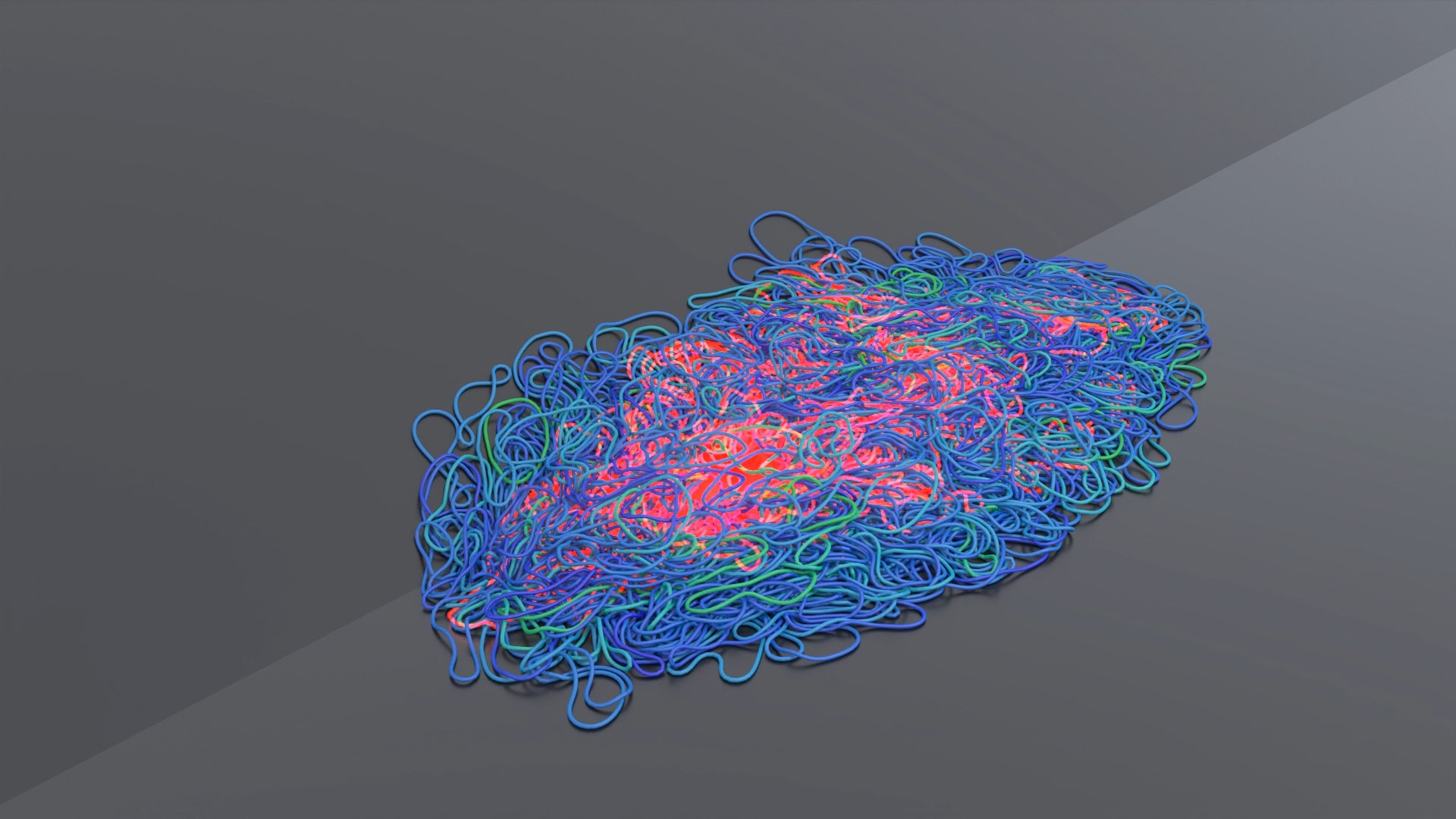} & 
      \band{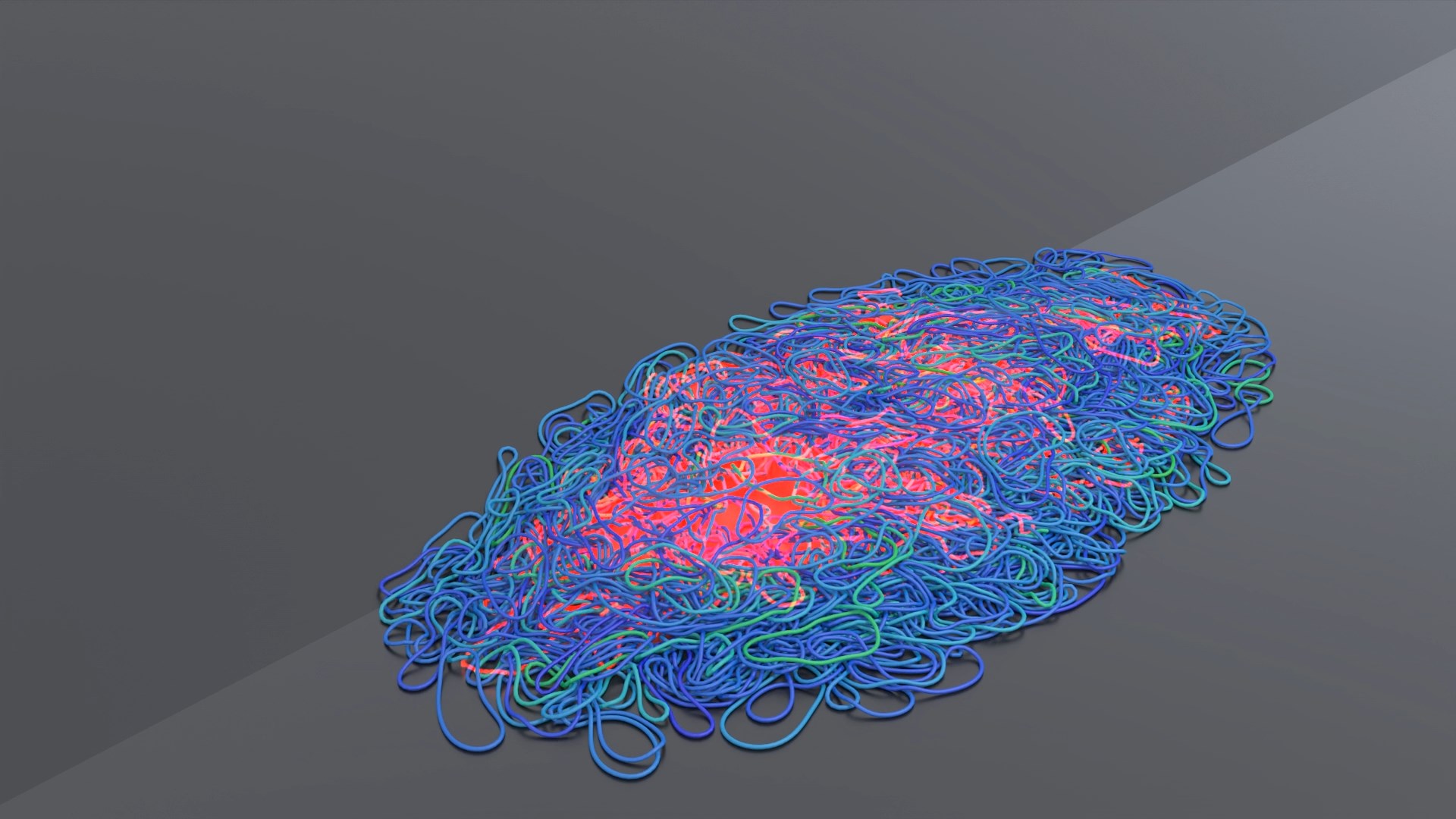} 
    \end{tabular}
  \end{center}
  \caption{{\bf Rubber Band Test:} An initially unlinked set of 1024 rubber bands falls onto an incline then slides onto the floor, as simulated with hair-like strands in Houdini's Vellum solver. We verify that (for typical contact/constraint iterations) (Top) the loops avoid spurious linking using 20 substeps per frame, whereas (Bottom) only 4 substeps per frame resulted in 322 linked rubber bands (shown in red).
    \label{fig:rubberBands}}
\end{figure*}  

\subsubsection{Ribbons and DNA:} Some applications in computational biology require the computation of linking numbers (and related quantities such as writhe) for extremely long DNA strands, which can change their linking numbers as a result of topoisomerase, thereby producing interesting effects such as super-coiling~\cite{clauvelin2012characterization}. For computational purposes, the strands are essentially ribbons with two twisted double-helix curves along each edge. While existing approaches typically only consider direct evaluation methods for relatively small models, e.g.,~\cite{sierzega2020wasp}, we demonstrate performance on very long and closely separated closed ribbons (akin to circular DNA) depicted in the top left of Fig.~\ref{fig:stresstests}. Results shown in Table~\ref{tab:aggregateresults} illustrate excellent performance by several of the algorithms; most notably, the GPU Barnes--Hut approach finishes 200,000-segment ribbons in under 30 milliseconds and, with sufficiently high $\beta$, finishes 20-million-segment ribbons in 5.3 seconds or faster; combined with the first two stages, our method takes up to 157 ms to compute the linking matrix for the former ribbon and 12.6 seconds for the larger ribbon.

%% file: conclusions.tex
\section{Conclusions}

We have explored practical algorithms and implementations for computing linking numbers for efficient monitoring and verification of the topology of loopy structures in computer graphics and animation. Linking-number certificates provide necessary (but not sufficient) conditions \revisionp{to ensure the absence} of topology errors in simulation or processing that would go undetected previously. We hope that the community can benefit from these improved correctness monitoring techniques especially given the many immediate uses in graphics applications, e.g., yarn-level cloth simulation, that have previously relied on linkage correctness but lacked practical tools for verification.

By comparing multiple exact and approximate linking-number numerical methods, and CPU and GPU implementations, we discovered that several methods give accurate and fast results for many types of problems, but alas there are no clear winners. Direct summation (DS) using exact formulae, although popular outside graphics for its simplicity and accuracy, was uncompetitive with the other methods for speed except for very small examples; GPU acceleration of the double summation was competitive in some cases. Our CPU parallel implementation of counting crossings (CC) was perhaps the simplest and fastest method for many problems; it relies on finding a suitable regular projection which is often cited as a weakness and potential source of numerical problems, but we were unable to generate incorrect results with it in double precision. CPU Barnes--Hut was a strong performer and competitive with CC in most cases, and GPU-accelerated Barnes--Hut was the fastest on several of the very largest examples, most notably for DNA ribbons. In some cases the GPU method struggles with accuracy due to having only dipole terms and single precision, and our GPU hardware and implementation were unable to store Barnes--Hut trees for large numbers of small loops, e.g., chainmail, for which CC or \revision{DS} did best.  The CPU-based Fast Multipole Method (FMM) was able to achieve high-accuracy fast summations due to the increased number of multipole terms, e.g., relative to Barnes--Hut, but it did not translate to beating CPU Barnes--Hut in general.

\subsection{Limitations and Future Work}

There are several limitations of our approaches, and opportunities for future work.  While changes in the linking number between two curves imply a change in topology, the same linking number does not rule out intermediate topological events, e.g., two knit yarn loops with $\lambda$ zero could ``pull through'' twice during a simulation and still have zero linkage at the end. \revision{Furthermore, collision violations can occur without a topology change: for example, given a key ring with three rigid keys, it is physically impossible for two adjacent keys to swap positions, yet the topology would still be maintained after a swap.} Therefore, our approach can flag linking number changes, \revisionp{but a passing certificate does not preclude other errors.}

We use AABB-trees to find overlapping loop AABBs in Potential Link Search (PLS) and overlapping spline-segment AABBs in Discretization, but tighter bounds than AABBs might reduce the number of linking-number calculations and discretized segments enough to provide further speedups.

We have considered linking numbers between closed loops and virtually closed loops for fixed patterns, but many loopy structures involve open loops, e.g., knittable yarn models~\cite{narayanan2018automatic, wu2018stitch, narayanan2019visualknit, wu2019knittable, guo2020representing}, periodic yarn-level cloth patterns~\cite{leaf2018interactive}, \revision{and hair}, and it would be useful to generalize these ideas for open and ``nearly closed'' loops, e.g., to reason about ``pull through'' between adjacent yarn rows.

Topology checks could be closely integrated with simulations and geometry processing to encourage topologically correct results and enable roll-back/restart modes. Our proposed \revision{method with the ``early exit'' strategy} could be better optimized to aggressively verify links before finishing PLS and Discretization on the entire model.

Finally, our GPU implementations and comparisons could be improved by specialized implementations for linking-number \revision{computations}, all-at-once loop--loop processing instead of serial, better memory management for many-loop scenarios, e.g., chainmail, and also more recent hardware.

%% file: acknowledgments.tex
\begin{acks}
\revision{We thank the anonymous reviewers for the constructive and detailed feedback. We thank Jonathan Leaf, Xinru Hua, Paul Liu, Gilbert Bernstein, and Madeleine Yip for helpful discussions, and Steve Marschner, Purvi Goel, and Jiayi Eris Zhang for proofreading the final manuscript. We also thank Cem Yuksel for the stitch mesh source code. Ante Qu's research was supported in part by the National Science Foundation (DGE-1656518). We acknowledge SideFX for donated Houdini licenses and Google Cloud Platform for donated compute resources. Mitsuba Renderer was also used to produce renders. Any opinions, findings, and conclusions or recommendations expressed in this material are those of the authors and do not necessarily reflect the views of the National Science Foundation.}
\end{acks}

%% file: appendix.tex
\appendix

\section{Discretization Proofs}
\label{app:discretizationProofs}

These proofs accompany the Discretization discussion in \S\ref{sec:proof}.

\begin{theorem}
If all input loops have only polynomial spline segments, and no two loops intersect, then \revisionp{Algorithm~\ref{alg:discretization} terminates in finite time if it uses tight AABBs as the bounding volumes in} its overlap check\revision{.}
\end{theorem}
\begin{proof}
Write each input spline segment using a parameter $t$ that has a domain of $(0,1)$. Since there is a finite number of input spline segments and each segment is polynomial over $t$ with a finite domain, the derivative along the curve with respect to $t$ has a global maximum, $M$. Since the loops don't intersect, let $d$ be the minimum separation between two loops. After pass $P$ of the algorithm, the parameter $t$ of any ``under-refined'' spline segment now spans a domain of length $2^{-P}$. Because polynomial spline segments are continuous, their arc lengths must be at most $2^{-P} M$. Their bounding AABBs must therefore have diameters that are also at most $2^{-P} M$. If the minimum separation between any two loops is $d$, then after $ P  > \log_2(2M/d)$ passes, no two bounding boxes of different loops can overlap, and so all splines segments must have been discretized.
\end{proof}

\begin{theorem}
If \revision{Algorithm~\ref{alg:discretization} terminates}, then it produces a line segment discretization that is homotopically equivalent to the input loops.
\end{theorem}
\begin{proof}
We can deform the input following the steps of the algorithm. At each pass, let us deform one loop, $\xi$, at a time. \revisionp{All the spline segments of this loop, $\xi$, that are ``processed'' during the pass can be deformed continuously to their line segments (which share the same endpoints) by parameterizing the line segment to have the same $t$ domain and performing a point-wise straight-line deformation \revision{at} each $t$. \revision{Because our bounding boxes are convex,} this deformation sweeps a surface that is entirely within the bounding box. The box overlaps with no boxes from other loops, so loop $\xi$} does not cross any other loop during the deformation. Notice, however, that we do not guarantee that a loop does not cross itself during its deformation; this is allowed in \revision{link homotopy} and does not affect the linking number. After all the passes, every loop has been deformed into its output polyline, and we are done. 
\end{proof}

\section{Implementation details of linking number computation routines}
\subsection{Direct \revision{Summation}}
\label{sec:de_gpu}
We reproduce the direct sum expression \eqref{eq:directarctan} here for reference. For loops $\gamma_1, \gamma_2$, consisting of $N_l, N_k$ line segments respectively, with $j,i$ enumerating the segments respectively, we want to compute
\begin{align}
    \lambda(\gamma_1, \gamma_2) &= \sum_{i}^{N_k} \sum_j^{N_l} \lambda_{ji}.\\
    \lambda_{ji} &= \frac{1}{2 \pi} \left(\atan\left( \frac{\vec{a} \cdot (\vec{b} \times \vec{c}) }{|\vec{a}| |\vec{b}| |\vec{c}| + (\vec{a} \cdot \vec{b}) |\vec{c}| + (\vec{c} \cdot \vec{a}) |\vec{b}|+(\vec{b} \cdot \vec{c}) |\vec{a}|}\right) \right. \nonumber \\
& \left. + \; \atan\left( \frac{\vec{c} \cdot (\vec{d} \times \vec{a}) }{|\vec{c}| |\vec{d}| |\vec{a}| + (\vec{c} \cdot \vec{d}) |\vec{a}| + (\vec{a} \cdot \vec{c}) |\vec{d}|+(\vec{d} \cdot \vec{a}) |\vec{c}|}\right)  \right).
\end{align}
In our implementation, both on the CPU and for large examples on the GPU, we multithread the outer loop of $i$, and perform angle summations \revision{(using the arctan addition formula)} in the inner loop with a single arctangent at the end of each loop, and perform a sum reduction on the total. This results in $N_k$ arctangents. \revision{When performing the inner loop, we split the $j$ loop into batches, and each threadblock fetches the entire $\vec{l}_j$ batch in parallel into shared memory before each $i$ thread iterates through the batch. See Alg. \ref{alg:de_angle_sum} for more details.} To maximize GPU usage, for small examples ($N_l N_k < 1.4 \times 10^6$), we parallelize both for-loops, and within each iteration we perform one arctangent\revision{.}

\begin{algorithm}[]
\SetAlgoLined
\SetKwInOut{Input}{Input}
\SetKwInOut{Output}{Output}

\Input{$\{\vec{l}_j\}$, $\{\vec{k}_i\}$, the two polyline loops}
\Output{$\lambda$, the linking number}
\BlankLine
\SetKwFunction{ComputeLink}{ComputeLinkDS}
\SetKwProg{Fn}{Function}{:}{}

  \tcp{This function computes the linking number between two closed polylines based on \cite{bertolazzi2019efficient}. The function $s(x,y)$ outputs $+1$ if (($y>0$) or ($y=0$ and $x<0$)), $-1$ otherwise.}
\Fn{\ComputeLink{$\{\vec{l}_j\}$, $\{\vec{k}_i\}$}}{

$\lambda \leftarrow 0$\;
$N_l \leftarrow |\{\vec{l}_j\}|$; \,\, $N_k \leftarrow |\{\vec{k}_i\}|$\;
\textbf{parallel }\For{$i\in [0, N_k-1]$}{
  $x_S \leftarrow 1$; \,\, $y_S \leftarrow 0$ \;
  $S \leftarrow -1$; \,\, $\lambda_i \leftarrow 0$ \;
  \tcp{If this is the GPU, preload a batch of $\vec{l}_j$ values here into shared memory, and then add a threadfence.}
  \For{$j\in [0, N_l-1]$}{
    $\vec{a} \leftarrow \vec{l}_j - \vec{k}_i$\;
    $\vec{b} \leftarrow \vec{l}_j - \vec{k}_{i+1}$\;
    $\vec{c} \leftarrow \vec{l}_{j+1} - \vec{k}_{i+1}$ \;
    $\vec{d} \leftarrow \vec{l}_{j+1} - \vec{k}_i$ \;
    $p \leftarrow \vec{a} \cdot (\vec{b} \times \vec{c})$\;
    $d_1 \leftarrow |\vec{a}| |\vec{b}| |\vec{c}| + \vec{a} \cdot \vec{b} |\vec{c}| + \vec{b} \cdot \vec{c} |\vec{a}| + \vec{c} \cdot \vec{a} |\vec{b}|$ \;
    $d_2 \leftarrow |\vec{a}| |\vec{d}| |\vec{c}| + \vec{a} \cdot \vec{d} |\vec{c}| + \vec{d} \cdot \vec{c} |\vec{a}| + \vec{c} \cdot \vec{a} |\vec{d}|$ \;
    $x' \leftarrow d_1 d_2 - p^2$ \;
    $y' \leftarrow p (d_1 + d_2)$ \;
    \If{\normalfont(($s(d_1, p) s(d_2,p) > 0$) \textbf{and} ($s(d_1, p) s(x',y') < 0$))}{
      $\lambda_i \leftarrow \lambda_i + s(d_1, p)$\;
    }
    $x'' \leftarrow x_S x' - y_S y'$\;
    $y'' \leftarrow x_S y' + y_S x'$\;
    \If{\normalfont(($s(x',y') \, S > 0$) \textbf{and} ($s(x'',y'') \, S < 0$))}{
      $\lambda_i \leftarrow \lambda_i + S$\;
    }
    $S \leftarrow s(x'',y'')$\; 
    $x_S \leftarrow x''/ \max(|x''|, |y''|)$\;
    $y_S \leftarrow y''/ \max(|x''|, |y''|)$\;
  }
  $\lambda_i \leftarrow \lambda_i + \text{atan2}(y_S, x_S) / (2 \pi)$\;
  $\lambda \leftarrow \lambda + \lambda_i$; \tcp{Reduce this summation.}
}
\textbf{return} $\lambda$\;
}

\caption{\revision{ComputeLink:} Direct Summation via Angle Summation, CPU or Large Input on the GPU}
\label{alg:de_angle_sum}
\end{algorithm}

\subsection{Barnes--Hut: Derivatives of $G(\vec{\tilde{r}}_1, \vec{\tilde{r}}_2)$, Moment Computation, and Parallel Implementation}
\label{sec:barneshutdipole}

For reference, here is the Barnes--Hut far-field expansion (Eq.~\eqref{eq:expansion}) again, expanded about $(\vec{\tilde{r}}_1, \vec{\tilde{r}}_2)$:
\begin{align*}
    \lambda &= -\int \text{d}s\; \text{d}t \; (\vec{r}_1' \times \vec{r}_2') \cdot \grad G(\vec{r}_1, \vec{r}_2).\\
    \lambda &= -\int \text{d}s\; \text{d}t \; \left[ ( \vec{r}_1' \times \vec{r}_2' ) \cdot \grad G(\vec{\tilde{r}}_1, \vec{\tilde{r}}_2) \phantom{\frac{1}{2}}\right.\nonumber  \\
    & + ( (\vec{r}_1' \times \vec{r}_2' ) \otimes (-(\vec{r}_1 - \vec{\tilde{r}}_1) + (\vec{r}_2 - \vec{\tilde{r}}_2))) \cdot \grad^2  G(\vec{\tilde{r}}_1, \vec{\tilde{r}}_2) \nonumber \\
    & + \frac{1}{2}( (\vec{r}_1' \times \vec{r}_2' ) \otimes(-(\vec{r}_1 - \vec{\tilde{r}}_1) + (\vec{r}_2 - \vec{\tilde{r}}_2)) \otimes (-(\vec{r}_1 - \vec{\tilde{r}}_1) + (\vec{r}_2 - \vec{\tilde{r}}_2)))  \nonumber \\
    & \; \; \; \; \left. \phantom{\frac{1}{2}} \cdot \grad^3
     G(\vec{\tilde{r}}_1, \vec{\tilde{r}}_2)  + O(|\vec{\tilde{r}}_1- \vec{\tilde{r}}_2|^{-5}) \right].
\end{align*}
Now here are the derivatives of $G(\vec{\tilde{r}}_1, \vec{\tilde{r}}_2)$. Let $\vec{r} = \vec{\tilde{r}}_2 - \vec{\tilde{r}}_1$. Then
\begin{align}
    \grad G(\vec{\tilde{r}}_1, \vec{\tilde{r}}_2) &=  \frac{\vec{r}}{4 \pi |\vec{r}|^3}.\\
    \grad^2 G(\vec{\tilde{r}}_1, \vec{\tilde{r}}_2) &= \frac{I_{3\times3}| \vec{r}|^2 - 3 \vec{r} \otimes \vec{r}}{4 \pi |\vec{r}|^5}.\\
    \grad^3 G(\vec{\tilde{r}}_1, \vec{\tilde{r}}_2) &= \frac{-3\sum_i(\vec{r} \otimes \vec{e}_i \otimes \vec{e}_i  + \vec{e}_i \otimes \vec{r}\otimes \vec{e}_i+ \vec{e}_i \otimes \vec{e}_i \otimes \vec{r})}{4 \pi |\vec{r}|^5} \nonumber \\
    & \; \; \; \; + \frac{15 \vec{r} \otimes \vec{r} \otimes \vec{r}}{4 \pi |\vec{r}|^7},
\end{align}
where $\vec{e}_i$ is the $i$-th basis element and repeated indices are summed. Note that these are identical to the basis functions in Appendix A of \cite{barill2018fast}, except the third derivative there (Eq. 24 in \cite{barill2018fast}) has a misprint.

These expressions can be expanded and rearranged so that for each term, the $s$ and $t$ integrals can be separated; this allows us to use the multipole moments directly.

Now let's discuss how to build the moment tree. For reference, here are the moments (from Eqs. \eqref{eq:monopolemoment}, \eqref{eq:dipolemoment}, and \eqref{eq:quadrupolemoment}) again:
\begin{align*}
    \vec{c}_M &= \int \vec{r}' \;\text{d}s,\\
    C_D &= \int \vec{r}' (\vec{r} - \vec{\tilde{r}})^T \;\text{d}s, \\
    C_Q &= \int \vec{r}' \otimes (\vec{r} - \vec{\tilde{r}})\otimes (\vec{r} - \vec{\tilde{r}}) \; \text{d}s.
\end{align*}

For a line-segment element, we set \revision{$\vec{\tilde{r}}$} to the midpoint, and, evaluating the integrals over a line segment, we get that $\vec{c}_M$ is the displacement between the endpoints, $C_D=0$, and $C_Q = \frac{1}{12} \vec{c}_M \otimes \vec{c}_M \otimes \vec{c}_M$.

When we combine bounding boxes into their parents, we need to sum and shift moments. Let $\vec{r}_c$ be the displacement \revision{$\vec{\tilde{r}}_{c} - \vec{\tilde{r}}_{p}$} (from the parent node, indexed by $p$, to the child node, $c$), and simply use
\begin{align}
    \vec{c}_{Mp} &= \sum_c \vec{c}_{Mc},\\
    C_{Dp} &= \sum_c C_{Dc} + \vec{c}_{Mc} \vec{r}_c^T,\\
    C_{Qpijk} &= \sum_c C_{Qcijk} + C_{Dcij} r_{ck} + C_{Dc ik} r_{cj} + c_{Mci} r_{cj} r_{ck},
\end{align}
to compute the parent moments ($\vec{c}_{Mp}$, $C_{Dp}$, and $C_{Qp}$).

On the CPU, we parallelized the two-tree method by gathering the list of child node pairs to evaluate after each recursive pass, and proceeding to children only after a pass is finished (breadth-first). We first run breadth-first passes single-threaded until the list has over 1000 node pairs. We then multithread the evaluation with breadth-first passes when it has between 1000 and 500,000 node pairs to evaluate. Beyond the 500,000-node-pair upper limit, memory allocation overhead (for the node-pair list) dominates, and we simply call the depth-first recursive algorithm in parallel for the 500,000+ node pairs.

On the GPU, we use only a first-order far-field expansion, dropping the quadrupole terms. Although we use the parallel, two-trees algorithm on the CPU, we parallelize a one-tree algorithm when evaluating on the GPU by basing it on the GPU Barnes--Hut N-Body implementation from \cite{burtscher2011efficient}. In particular, we use Kernels 1 through 5. We first build a tree (Kernels 1 through 4) for each loop, then for each loop pair, we evaluate Kernel 5 using the source tree from one loop against target points from the other loop. We modified Kernel 3 to gather and compute the monopole and dipole moments of each node and also propagate the max segment length, $l_M$, from its children. Because segments are only inserted at their midpoints, we replace the $\beta$ far-field condition in Alg.~\ref{alg:barneshut} by checking whether $|\vec{r}_1 - $ tree$.\vec{r}| > \beta $\,(tree.$R$ + tree.$\nicefrac{l_M}{2}$ + $\nicefrac{l_1}{2}$), where $l_1$ is the length of the segment at $\vec{r}_1$. We modified Kernel 5 to evaluate a first-order (dipole) far-field expansion when the $\beta$ far-field condition is met by all threads; otherwise the arctan expression. However, one key difference from \cite{burtscher2011efficient} is that because the source and target trees are different and our input is already spatially coherent, Kernel 5 is actually slower if it evaluates the points in the sorted order from Kernel 4 than if it evaluates the points in the input order. So we keep Kernel 4 for sorting the individual source trees, but we evaluate Kernel 5 on the target evaluation points in the input order instead of the sorted order.

\section{Forming virtual closed loops in braids}
\label{app:openCurveVCAlg}

This section describes the formation of virtual closed loops for Open-Curve Verification \S\ref{sec:openCurveVerification}. The key is that every curve in the braid participates in two overlapping virtual closed loops, and every virtual closed loop contains two curves from the braid. Then in the Potential Link Search (PLS, \S\ref{sec:methodspls}), we exclude loop pairs that share a common curve.

Let $\alpha_1, \cdots , \alpha_L$ denote the open curves, and let one rigid end be the ``left'' end and the other be the ``right'' end. Furthermore, define a small rigid end volume attached to each end, such that no curve passes through it. Then follow this procedure, also illustrated in Figure \ref{fig:chevronillustration}:
\begin{enumerate}
    \item To ease bookkeeping, parameterize the curves so that they are oriented left to right.
    \item Form $L$ total closed loops, $\gamma_i$. For each $i \in [1,L-1]$, form $\gamma_i$ from $\alpha_i$ and $\alpha_{i+1}$ by doing the following:
    \begin{enumerate}
        \item Start with a copy of $\alpha_i$,
        \item Append a virtual connection, entirely within the right end volume, from the right end of $\alpha_i$ to the right end of $\alpha_{i+1}$, carefully making sure it intersects no other virtual curve,
        \item Append $\alpha_{i+1}$ in the reverse direction, and then
        \item Append a virtual connection, entirely within the left end volume, from the left end of $\alpha_{i+1}$ to the left end of $\alpha_i$, taking the same care to ensure it intersects no other curve.
    \end{enumerate}
    \item Also form $\gamma_L$ using $\alpha_L$ and $\alpha_1$ the same way as the last step. At this point, every curve $\alpha$ should be part of two closed loops $\gamma$, and every loop $\gamma$ should contain two curves from the model.
    \item Run our method from \S\ref{sec:method} using the input loops $\{\gamma_i\}$, with the difference that, at the end of the PLS stage, we exclude pairs $(i,j) \in P$, where $\gamma_i$ and $\gamma_j$ share an $\alpha$ as a component (which is when $i+1=j$ or ($i=1$ and $j=L$)), from $P$. The rest of the method proceeds as usual, computing a sparse linking matrix as a certificate.
    \item When this procedure is used again on a deformed model, the virtual connections at each end \emph{must be appended with the same exact paths}, up to a rigid transformation (this is possible because of the rigid ends), ensuring that the virtual connections don't pull through each other.
\end{enumerate}
We also require that $L \ge 4$.

The certificate meaningfully verifies the topology between open curves in the model, as long as no real curve enters the rigid end volumes and the end volumes don't mutually interpenetrate (ensuring no real or virtual curve moves through a virtual connection). To see how, suppose a single pull through occurs, between curves $\alpha_i$ and $\alpha_j$, $i<j$. Then this certificate must change, for these reasons:
\begin{itemize}
     \item Call indices ``cyclically adjacent'' if they are either adjacent (such as $i$ and $i+1$) or the first and last (such as $1$ and $L$). If $(i,j)$ are not cyclically adjacent, then the linkage between $\gamma_i$ and $\gamma_j$ will be computed (because they don't share a curve), and found to be different, since the former contains $\alpha_i$, and the latter contains $\alpha_j$.
     \item Otherwise if $(i, j)$ are cyclically adjacent, then the linkage between $\gamma_{i-1 \; (\text{modulo } L)}$ and $\gamma_j$ will be calculated (because $L \ge 4$, they don't share a curve) and found to be different, because the former contains $\alpha_i$, and the latter contains $\alpha_j$.
 \end{itemize}
 Conversely, if a linking number changes, then there must have been at least one topology violation among the real curves, because no virtual connection participated in any pull through.

Furthermore, this procedure can be applied independently on many interacting braids as long as each braid has at least 4 curves.